\documentclass[preprint,authoryear,12pt]{elsarticle}

\usepackage{amssymb}
\usepackage{color}
\usepackage{xcolor}
\usepackage{amsmath}
\usepackage{amsthm}

\usepackage{graphicx}
\usepackage{graphics}
\usepackage{booktabs}
\usepackage{threeparttable}
\usepackage{appendix}
\usepackage{tabularx}
\usepackage{array}
\usepackage{multirow}
\usepackage{pifont}
\usepackage{makecell}

\usepackage{algorithm}
\usepackage{algorithmic}

\usepackage{subfig}

\usepackage{geometry}
\usepackage[displaymath]{lineno}
\usepackage{setspace}
\usepackage[colorlinks,linkcolor=red,anchorcolor=blue,citecolor=green]{hyperref}
\usepackage{soul}

\usepackage{mathpazo}

\journal{ }

\begin{document}
\begin{spacing}{1.3}

\begin{frontmatter}

\title{Platooning Connected, Autonomous, and Human-Driven Vehicles: A Deep Reinforcement Learning-based Approach}

\author[rvt1]{Zhen Qin}
\author[rvt1]{Dong-Fan Xie\corref{cor1}}
\ead{dfxie@bjtu.edu.cn}

\author[rvt1]{Heng Ma\corref{cor1}}
\ead{HengMa1024@outlook.com}

\author[rvt1]{Xiaomei Zhao}

\author[rvt2]{Zhengbing He\corref{cor1}}
\ead{he.zb@hotmail.com}

\address[rvt1]{School of Systems Science, Beijing Jiaotong University, Beijing 100044, China}
\address[rvt2]{Faculty of Science and Engineering, University of Nottingham Ningbo China}

\cortext[cor1]{Corresponding authors}

\begin{abstract}
Conventionally, existing vehicle platooning approaches are designed for connected vehicles, typically including connected autonomous vehicles and connected human-driven vehicles. 
Non-connected vehicles, such as non-connected autonomous or human-driven vehicles, are not incorporated.
As a result, these platooning approaches may not properly reflect real-world mixed traffic conditions at the current stage.
To address this limitation, this study proposes a hybrid platooning pattern that conditionally permits non-connected vehicles to join platoons, thereby enhancing platooning diversity and flexibility.
However, it was found that the unregulated integration of non-connected vehicles can trigger rapid platoon expansion, significantly amplifying the risk of disturbance propagation in traffic flow.
This, in turn, exacerbates the inherent conflict between traffic throughput and stability.
To mitigate these challenges, this paper further develops a hybrid platooning control strategy based on deep reinforcement learning (DRL).
This strategy integrates vehicle dynamics, platoon topology, and traffic flow states through a multi-level state representation network, enabling a dynamic trade-off between traffic capacity and stability.
Numerical simulations demonstrate that the proposed strategy effectively suppresses velocity disturbance propagation by dynamically optimizing platoon structures, thereby significantly enhancing the stability and safety of mixed traffic while reducing fuel consumption and emissions.
\end{abstract}


\end{frontmatter}

% \newpage

% \tableofcontents

%\linenumbers

\newpage

\section{Introduction}\label{sec:Introduction}
 
The rapid progress of new technologies such as the Internet of Vehicles (IoV) and Artificial Intelligence (AI) has facilitated the development of vehicles.
Many existing studies categorize these vehicles into the following four types: 
(i) {\bf Human-driven Vehicles (HVs)}, i.e., traditional vehicles without automation or connection.
(ii) {\bf Autonomous Vehicles (AVs)}, which provide driving automation but no connection.
(iii) {\bf Connected Human-driven Vehicles (CHVs)}, which offer communication capabilities but no automation.
(iv) {\bf Connected Autonomous Vehicles (CAVs)}, which integrate both automation and connection.
Obviously, HVs/AVs are non-connected vehicles, while CHVs/CAVs are connected vehicles.
Those connected and non-connected vehicles form a new mixed traffic flow that is expected to become a defining feature of future traffic systems.

In mixed traffic, CHVs/CAVs can form platoons to enhance traffic throughput by enabling coordination among vehicles \citep{huang2018path,chang2020analysis,xin2021modeling}.
Under optimal platooning conditions, road capacity can exceed even 12,000 vehicles per hour per lane, representing a 5.3-fold increase compared with scenarios without platooning \citep{sala2020macroscopic}.
Furthermore, reduced inter-vehicle spacing within platoons decreases aerodynamic drag, thereby lowering fuel consumption and emissions \citep{shi_2025_Mixed_Vehicle_Platoon}.

Nevertheless, most existing platooning strategies rely on homogeneous and continuously distributed CAVs, which limits platooning scalability \citep{guan2023markov}.
Notably, CAVs are equipped with multi-source perception systems, such as LiDAR, radar, and vision sensors, enabling them to accurately perceive the states of HVs/AVs.
By broadcasting processed environmental information through communication networks, CAVs can incorporate HVs/AVs into platoons, thereby relaxing the requirement for continuous CAV distribution and enabling scalable platoon deployment.

In hybrid platooning strategies, CAVs assume dual roles as information relays and motion coordinators.
Optimizing the vehicle interaction network through topological restructuring enables the conditional integration of HVs/AVs into platoons.
This approach enhances platoon scenario diversity and flexibility, thereby better aligning with real-world traffic demands.
However, the unregulated incorporation of HVs/AVs can cause rapid expansion of platoon size, resulting in excessively large formations.
This substantially amplifies the risk of traffic flow disturbance propagation and increases the likelihood of congestion arising from systemic instability.
Moreover, conventional maximum platoon length constraints, originally developed for homogeneous platoons, are no longer adequate to accommodate the requirements of these emerging platooning strategies.

To address the aforementioned challenges, this study introduces a hybrid platooning pattern that permits the integration of HVs/AVs into platoons.
Building on this framework, we further develop a Deep Reinforcement Learning (DRL)-based platooning strategy to optimize multiple objectives in such mixed traffic flow.
The proposed strategy employs a multi-tier state representation network to integrate vehicle dynamics, platoon topology, and traffic flow states, enabling a dynamic trade-off between traffic throughput and flow stability.
Furthermore, the strategy adopts the Proximal Policy Optimization (PPO) algorithm within a Centralized Training with Decentralized Execution (CTDE) architecture, which helps improve policy optimization while enabling decentralized platooning decisions based on local observations.
Simulation results demonstrate that the proposed DRL-based hybrid platooning strategy effectively improves mixed traffic flow stability by dynamically optimizing platoon structures and suppressing velocity disturbance propagation. 
Evaluation of the trained policy confirms its robustness across different CHV/CAV penetration rates, with a $51.08\%$ improvement in the stability discriminant when $P_{\mathrm{CAV}}>0.1$. 
The trained policy also reduces peak average TTCI by $52.95\%$ under low CAV penetration and lowers fuel consumption and CO$_2$ and NO$_x$ emissions in disturbance-propagation scenarios.

%This study aims to leverage the technological potential of vehicle platooning for traffic flow benefits, with the following contributions.

% \begin{itemize}

% \item This study introduces a novel hybrid platooning pattern that conditionally integrates non-connected HVs/AVs into platoons.
% This integration enhances platooning diversity and operational flexibility, thereby expanding the design space for optimized platooning strategies.

% \item This study develops a DRL-based hybrid platooning strategy to dynamically optimize heterogeneous platoon structures.
% The proposed online optimization capability overcomes the static limitations inherent in conventional platooning strategies.
% Moreover, a multi-objective reward mechanism is designed to effectively balance traffic efficiency and flow stability.

% \item Simulation results demonstrate that the proposed DRL-based hybrid platooning strategy effectively suppresses velocity disturbance propagation through dynamic topology optimization, significantly enhancing mixed traffic flow stability.
% These improvements translate into tangible benefits, including enhanced traffic safety, reduced fuel consumption, and lower emissions of $CO_2$ and $NO_x$.
% In a representative scenario with $50\%$ CHVs/CAVs penetration, the proposed strategy achieves a $27.9\%$ increase in the traffic stability discriminant value and a $28.3\%$ reduction in maximum platoon length compared with conventional hybrid platooning strategies.

% \end{itemize}

The remainder of this study is organized as follows.
Section \ref{sec:Literature_review} presents a systematic review of existing platooning strategies, including both homogeneous and heterogeneous approaches.
Section \ref{sec:Platooning_scenarios} describes the mixed traffic environment and introduces the proposed hybrid platooning pattern incorporating HVs/AVs.
Section \ref{sec:PS_DRL} develops the DRL-based hybrid platooning strategy and details its training and testing procedures.
Section \ref{sec:Performance_evaluation} evaluates the proposed strategy through comprehensive simulations, assessing its generalization capability, traffic stability, safety, energy consumption, and emissions.
Finally, Section \ref{sec:Conclusions} summarizes the main findings and outlines potential directions for future research.

\section{Literature review}
\label{sec:Literature_review}

%The effectiveness of vehicle platooning in mixed traffic depends not only on cooperative control strategies but also on structural and environmental factors. Previous studies have developed a wide range of platooning strategies, including rule-based approaches, model predictive control (MPC)-based methods, data-driven techniques, and reinforcement learning frameworks. However, most existing control strategies are designed for homogeneous CAV platoons and do not explicitly account for vehicle heterogeneity. Other studies have investigated the impacts of factors such as platoon structure and pattern, maximum platoon length, and CAV clustering intensity on platooning performance.

\subsection{Platooning strategies}

Although extensive research has been conducted on cooperative control strategies to exploit the benefits of vehicle platooning, many existing studies primarily focus on platooning strategies for a single vehicle type.  
While advanced optimization- and learning-based algorithms have been widely explored, rule-based platooning strategies remain among the simplest and most computationally efficient solutions.  
For example, \cite{wu_2024_Multi_Lane_Unsignalized} and \cite{MAITI_2023_Ad_hoc_platoon} introduced rule-based platoon control strategies that offer high computational efficiency but exhibit limited adaptability in complex mixed traffic environments.

Vehicle platooning fundamentally relies on the sharing of vehicle state information and aims to enhance collective performance, which naturally motivates its formulation as an optimal control problem.  
Within this framework, optimal control techniques, such as Model Predictive Control (MPC) and Linear Quadratic Regulator (LQR), have been extensively adopted due to their effectiveness in handling complex and dynamic traffic conditions \citep{Wu2023DistributedDataDrivenModel, Wang2025DistributedNonlinearModel, Yang2023DistributedModelPredictive, SUN2024}.  
Specifically, \cite{Lan_2023_Data_Driven_Robust} investigated cooperative adaptive cruise control (CACC) for mixed traffic platoons by developing a data-driven MPC algorithm that explicitly accounts for noisy measurements.  
\cite{jin2020analysis} proposed a continuum model for mixed traffic flow and provided analytical methodologies for the design of cooperative platoon control strategies.  
\cite{Zhang2025PlatoonFormation} developed a self-organized cooperative control strategy for platoon formation based on distributed MPC.  
\cite{Zong2025He} introduced an adaptive control framework that allows CAVs to dynamically switch between platooning and individual operation, improving traffic stability and reducing emissions, while incorporating human driver behavior through transposition prediction.  
In addition, \cite{Liu2025ZhengWang} proposed a cooperative control strategy that combines an LQR controller with an active disturbance rejection controller (ADRC) to manage platoon state transitions under varying time headways.  
From a decentralized perspective, \cite{Wijnbergen2025NonlinearSystems} developed a nonlinear spacing policy that guarantees string stability when each vehicle follows its immediate predecessor.

As platooning research has expanded from control under prescribed vehicle formations to coordination problems involving mixed traffic, heterogeneous vehicle behaviors, and more flexible platoon interactions, it has become increasingly difficult to rely solely on handcrafted rules or explicitly formulated optimization models. This shift has motivated the use of RL and DRL, which offer a practical way to derive adaptive control strategies from traffic interactions. 
For example, \cite{Song2026VSL}integrated global perception prediction for CHVs/CAVs with lane-specific reinforcement-learning-based variable speed limit control, showing how richer traffic-state information can support adaptive decision-making in mixed traffic environments. 
\cite{Han2025Model-freeAdaptiveControl} proposed a data-driven finite-time adaptive controller that achieves model-free tracking of vehicle position and velocity.  
\cite{Lin2024Multi-AgentRLPlatoon} presented an enhanced state representation for multi-agent reinforcement learning to improve car-following performance.  
By integrating heuristic learning with distributed modeling, \cite{Du2024heuristicRL} developed a multilayer predictive control framework for autonomous vehicle platoons.  
For heterogeneous nonlinear platoons, \cite{Ma2024ModelFree} proposed a model-free distributed control strategy that relies only on local information from neighboring vehicles and guarantees stability via linear matrix inequalities without pre-training.  
Addressing communication efficiency in multi-agent reinforcement learning, \cite{Hua2025Multi-agentDRL} developed a fully decentralized CACC framework that incorporates quantified stochastic gradient descent and binary differential consensus mechanisms.

More closely related to this study, recent works have extended platoon control to mixed or heterogeneous vehicle settings. 
These studies typically take the platoon composition as a given setting and concentrate on the associated control problem, such as robust control of mixed platoons under HV prediction uncertainty \citep{feng_2021_Robust_Platoon_Control}, distributed MPC coordination of heterogeneous platoons with spacing constraints \citep{Qiang2023DMPC}, and hierarchical cooperative control of mixed platoons for oscillation mitigation \citep{Liang2025Jun}. 
This tendency is also evident in recent DRL-based studies.
\cite{Shi2023DRLDistributedControl} developed a distributed DRL strategy for CAV longitudinal control in mixed platoons by aggregating consecutive HVs and incorporating their traffic effects into the state representation. 
\cite{Yang2025MixedPlatoonRL} compared single-agent and multi-agent RL methods for eco-CACC in predefined mixed-platoon scenarios, where the learned policies are used to determine CAV acceleration actions.
\cite{Liu2026StringStability} considered mixed platoons containing HVs under a prescribed CAV--HV--CAV subplatoon structure and proposed a DRL-based cooperative controller with guaranteed string stability. 
\cite{Shi2025MixedVehiclePlatoonForming} moved one step closer to platoon organization by studying mixed-vehicle platoon forming; however, their framework first generates a feasible target formation through rule-based design and then applies MARL to control CAV acceleration, deceleration, cruising, and lane-changing actions to realize that formation. 
Overall, these studies mainly address how vehicles should be controlled once the platoon composition or target formation has been assumed, rather than how heterogeneous vehicles, including HVs and AVs, should be organized into feasible platoons in mixed traffic. 
This leaves an important gap between within-platoon vehicle control and hybrid platooning strategy design, which is the focus of this study.

%%%%%%%%%%%%%%%%%%%%%%%%%%%%%%%%%%%%%%%%%%%%%%%%%%%%%%%%%%%%%%%%%%%%%%%%%%%%%%%%%%%%%%%%%%%%%%%%%%%%%%
\subsection{Impacting factors of vehicle platooning on traffic flow}

The construction of platoons in mixed traffic requires more than identifying which vehicles are technically able to connect. 
It demands a clear understanding of what impacts different platoon arrangements may bring. 
Existing studies suggest that platoon formation pattern, maximum platoon length, and CAV clustering intensity are among the factors most closely related to these impacts.

Platoon formation pattern and CAV clustering intensity are closely related to how vehicles are spatially organized in mixed traffic. 
\cite{JIN_2021_Spatial_Distribution} evaluated different platoon formation patterns and showed that the arrangement of platoons can affect traffic efficiency and vehicle energy consumption. 
Similarly, \cite{li_2023_Cooperative_Formation} compared uniform, random, and platoon-based vehicle distributions from a set-function optimization perspective. 
Beyond formation patterns, CAV clustering intensity has also been used to characterize the spatial concentration of CAVs in mixed traffic. 
Related studies suggest that clustering intensity is associated with platooning opportunities, macroscopic traffic characteristics, and safety performance \citep{guan2023markov, zhou2021analytical, jiang2023platoon, ghiasi2017mixed, ghiasi2020lane}.

Maximum platoon length is another key consideration, as platoon expansion may bring both traffic benefits and operational risks. 
\cite{van2006impact} pointed out that unrestricted platoon growth may adversely affect traffic safety, while \cite{liu2018impact} argued that platoon splitting becomes necessary once platoon length exceeds an allowable threshold. 
From a traffic-flow perspective, \cite{jin2020analysis} showed that road capacity can improve with platoon formation, but may decline when platoon size exceeds a critical level. 
Considering capacity, safety, and environmental impacts jointly, \cite{seraj2018modeling} and \cite{zhou2021analytical} further showed that larger platoons may improve throughput and environmental performance, although these gains can be accompanied by safety or stability concerns. 
Accordingly, recommended platoon lengths vary across studies depending on the target objective \citep{liu2018modeling, seraj2018modeling, zhou2021impact, yao2022fundamental}.

\section{Platooning scenarios in mixed traffic}\label{sec:Platooning_scenarios}

For clarity, a platoon is defined as a coordinated group of adjacent vehicles traveling longitudinally with a shared platooning status. 
A hybrid platoon may include HVs and AVs if their states can be sensed by adjacent CAVs and made available to CHVs and CAVs, subject to the restriction that HVs and AVs cannot act as platoon leaders or tail vehicles.

\subsection{Mixed traffic scenario}

The mixed traffic flow considered in this study consists of four categories of vehicles: CAVs, CHVs, AVs, HVs.  
The vehicle type set is defined as:
\begin{equation}
\mathcal{V} = \{\text{CAV}, \text{CHV}, \text{AV}, \text{HV}\}
\label{eq:chap2_vehicle_set}
\end{equation}

From a communication perspective, CHVs/CAVs are capable of vehicle-to-vehicle (V2V) communication through dedicated short-range communication (DSRC) or cellular vehicle-to-everything (C-V2X) technologies.  
In contrast, HVs/AVs do not possess communication capabilities and therefore cannot exchange information with other vehicles.  
Accordingly, vehicles can be classified into two categories based on their communication capability: CHVs/CAVs ($\mathcal{V}_{C}$) and HVs/AVs ($\mathcal{V}_{H}$):
\begin{equation}
\begin{cases}
\mathcal{V}_{C} = \{\text{CAV}, \text{CHV}\},\\
\mathcal{V}_{H} = \{\text{AV}, \text{HV}\}.
\end{cases}
\label{eq:chap2_vc_vh}
\end{equation}

While CHVs and HVs are operated by human drivers, decision-making and control in CAVs and AVs are fully executed by autonomous driving systems, which typically employ control algorithms based on frameworks such as MPC or DRL.

This study focuses on mixed traffic scenarios on straight road segments, excluding external influences such as on- and off-ramps and lane-changing maneuvers.  
Vehicles are assumed to be spatially distributed with randomized type configurations.  
The penetration rates of CAVs, CHVs, AVs, and HVs are denoted by $P_{\text{CAV}}$, $P_{\text{CHV}}$, $P_{\text{AV}}$, and $P_{\text{HV}}$, respectively, and satisfy:
\begin{equation}
P_\text{CAV} + P_\text{CHV} + P_\text{AV} + P_\text{HV} = 1
\label{eq:chap2_sum_p_equal_1}
\end{equation}

Based on the above description of the mixed traffic scenario, the following assumptions are made:
(1) Vehicles move on a single-lane road without considering lane-changing behavior;
(2) The four types of vehicles, i.e., CAVs, CHVs, AVs, HVs, are randomly distributed along the road;
(3) V2V communication delays and sensor data processing latencies are neglected;
(4) CAVs and AVs operate fully under automated control without human intervention, whereas CHVs and HVs rely entirely on human drivers;
(5) CAVs are regarded as either  
(i) communication-enhanced AVs, namely, AVs equipped with additional V2V communication capabilities for state information sharing; or  
(ii) control-enhanced CHVs, namely, CHVs equipped with autonomous driving systems that replace human drivers;
(6) Vehicle platoon formation and dissolution are assumed to occur instantaneously, without considering transient effects induced by dynamic reorganization processes;
(7) All vehicles unconditionally participate in platooning operations whenever platoon size constraints and communication connectivity conditions are satisfied.

\subsection{Heterogeneous platooning behavior}

Based on whether HVs/AVs are permitted to participate in platoons, this study investigates vehicle platooning behaviors under two distinct scenarios.

\subsubsection{Platooning scenario for CHVs and CAVs}

In the traditional platooning scenario, only CHVs/CAVs, are allowed to form platoons.

When a CHV/CAV attempts to join a platoon, it first identifies a communication target vehicle (CTV) traveling ahead and seeks to establish a vehicle-to-vehicle (V2V) communication link. The vehicle can join the CTV’s platoon if and only if the inter-vehicle distance is within the maximum communication range of both vehicles.

Although any member of the target platoon could theoretically serve as a CTV, this study focuses on the most practical and commonly adopted case in which vehicles communicate exclusively with their immediately preceding neighbors. As illustrated in Fig.~\ref{fig:pla_scenarios} (a), all platoon members, except for the leader, maintain communication solely with their direct predecessor. For notational simplicity, this platooning scheme is referred to as $\text{PS}1$.

The probability that an ego vehicle successfully joins the platoon of its immediately preceding vehicle is denoted as
$P_{\text{pla},Y_\text{ego},Y_\text{front},u}^{\text{PS1}}$,
where $u$ represents the position index of the preceding vehicle within the platoon, and $Y_\text{ego}$ and $Y_\text{front}$ denote the vehicle types of the ego vehicle and the communication target vehicle, respectively.
Based on the steady-state velocity $v$ and the time headway $\Delta T$, the corresponding probability model is given by:

\begin{equation}
\begin{aligned}
& P^{\mathrm{PS1}}_{\mathrm{pla}}\!\left(Y_{\mathrm{ego}},Y_{\mathrm{front}},u\right)
\\
& =
\begin{cases}
\Pr\!\Big(
v\,\Delta T_{\mathrm{ego}}
<
\min\{ d_{Y_{\mathrm{ego}}},\, d_{Y_{\mathrm{front}}} \}
\Big),
& u \le l_{\max}-1,\\
0, & u = l_{\max}.
\end{cases}
\end{aligned}
\label{eq:chap2_pm1_eq}
\end{equation}
where $d_{Y_\mathrm{ego}}$ and $d_{Y_\mathrm{front}}$ denote the maximum effective communication ranges of the ego vehicle and the communication target vehicle, respectively. This study assumes identical maximum communication ranges for CHVs/CAVs. Since HVs/AVs lack communication capabilities, their maximum effective communication ranges are set to zero.
$\Delta T_{\mathrm{ego}}$ represents the time headway after the ego vehicle joins the platoon, and $l_{\max}$ denotes the maximum allowable platoon size, i.e., the maximum number of vehicles that a single platoon can accommodate.

\subsubsection{Platooning scenario for all vehicles}

Conventional homogeneous platooning requires all vehicles to possess communication capabilities. However, under randomly distributed mixed traffic, HVs/AVs that lack communication modules impose stringent constraints, as platoon formation becomes strictly dependent on the contiguous spatial distribution of CHVs/CAVs. This dependence severely limits the scalability of platooning technology.

To address this limitation, this study proposes a heterogeneous platooning strategy. The approach comprises the following two steps.
(1) Multi-sensor perception: CAVs utilize onboard multimodal sensors (lidar, mmWave radar, and cameras) to acquire real-time states (velocity, acceleration, and spacing) of adjacent HVs/AVs.
(2) Information propagation: CAVs broadcast the perceived state information of HVs/AVs via V2X communication, establishing an intra-platoon communication topology. This renders the states of all HVs/AVs observable to every CHV/CAV within the platoon.

This strategy enforces strict platoon integrity conditions: if the state of any vehicle becomes unobservable to a connected platoon member, immediate platoon dissolution and dynamic reorganization are triggered.

This platooning strategy conditionally allows HVs/AVs to join platoons, forming heterogeneous structures such as CAV--HV--HV--CAV. This enables more diverse and flexible platooning scenarios that better align with real-world traffic conditions. For subsequent reference, this strategy is referred to as platooning scheme $\text{PS}2$.

Notably, this study stipulates that HVs/AVs are not allowed to serve as platoon leaders or tail vehicles. Due to the absence of communication capabilities, HVs/AVs cannot obtain platoon-level information nor disseminate control or status signals to define platoon boundaries. Moreover, their car-following behaviors remain identical regardless of whether they are inside or outside a platoon. Consequently, positioning HVs/AVs at the head or tail of a platoon fails to establish effective platoon boundaries and does not contribute to improvements in traffic efficiency or stability through cooperative driving.
The following sections systematically analyze and elaborate on all feasible platooning scenarios under the proposed framework.

\begin{figure}[htbp]
    \centering \includegraphics[width=0.70\linewidth]{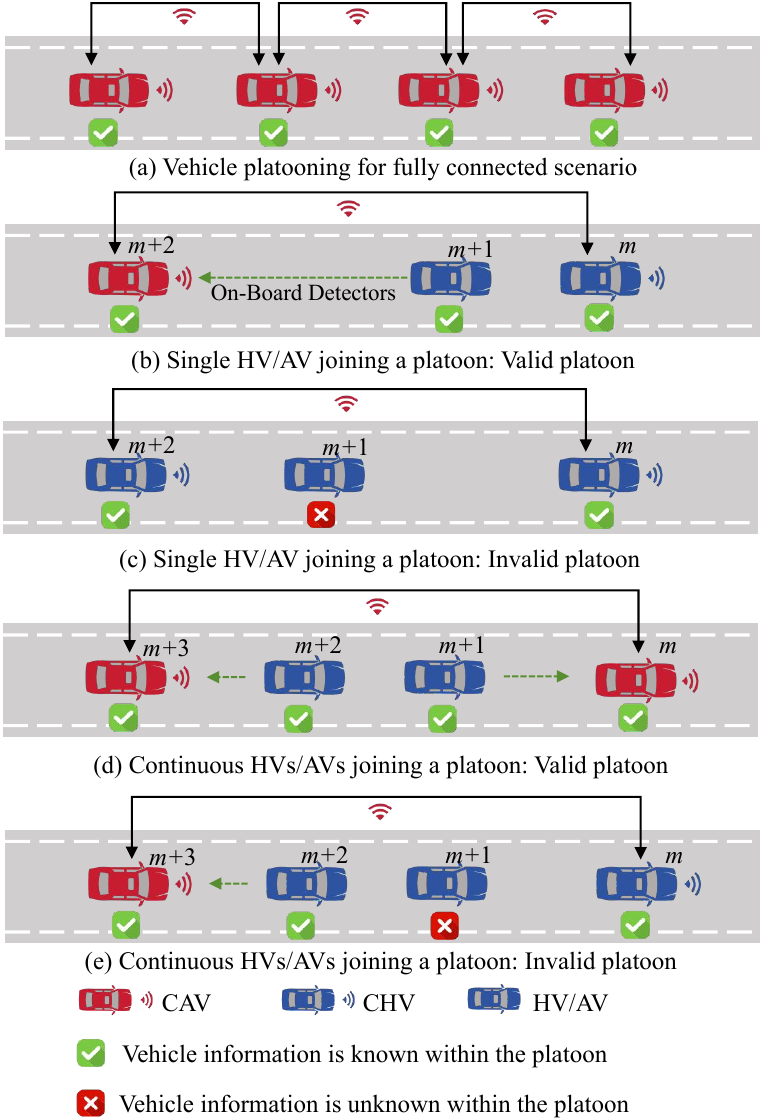}
    \caption{Vehicle platooning scenarios.}
    \label{fig:pla_scenarios}
\end{figure}

As shown in Fig.~\ref{fig:pla_scenarios}, when two adjacent vehicles are both CHVs/CAVs, the follower selects its immediate predecessor as the communication target vehicle (CTV). The follower is allowed to platoon with the preceding vehicle if the following conditions are satisfied: (1) the current platoon size is smaller than the maximum allowable platoon length; and (2) the communication distance constraint is satisfied. The joining probability is given by
\begin{equation}
\begin{aligned}
& P_{\mathrm{pla},Y_\mathrm{ego},Y_\mathrm{front}, Z_m}^{\mathrm{PS2}} \\
& = \begin{cases}
\Pr\!\left(
v\,\Delta T_{\mathrm{ego}}<\min\!\left( d_{Y_\mathrm{ego}}, d_{Y_\mathrm{front}} \right)\right),
& Z_m \le l_{\max}-1,\\
0, & Z_m = l_{\max}.
\end{cases}
\end{aligned}
\label{eq:chap3_pla_strategy_full_con}
\end{equation}
where 
$Z_m $ is the position of vehicle $m\in \{1,2,...,l_\mathrm{max} \}$ in the platoon; and
%$Y_{\text{ego}}$ and $Y_{\text{front}}$ denote the vehicle types of the ego vehicle and its preceding vehicle, respectively; and 
$d_{Y_{\mathrm{ego}}}$ and $d_{Y_{\mathrm{front}}}$ represent the corresponding maximum effective communication ranges.

Figs.~\ref{fig:pla_scenarios} (b) and (c) illustrate the scenario in which a single HVs/AVs vehicle joins a platoon.
As shown in Fig.~\ref{fig:pla_scenarios}(b), the $m$-th vehicle is a CHV, the $(m+1)$-th vehicle is a HV/AV, and the $(m+2)$-th vehicle is a CAV. 
In this case, the HV/AV $(m+1)$ can rely on the CAV to acquire its motion state information via onboard sensing units, which is then relayed to other CHVs/CAVs within the platoon. 
Consequently, the motion states of all vehicles within the platoon become observable to all CHVs/CAVs, enabling the formation of a functional platoon. 
However, under conventional connected platooning strategies, this scenario would prevent effective platoon formation between vehicle $m$ and vehicle $(m+2)$ due to the presence of the HV/AV $(m+1)$.

Furthermore, since HVs/AVs cannot serve as the lead vehicle or the tail vehicle of a platoon, the preceding vehicle $m$ and the following vehicle $(m+2)$ of the HV/AV $(m+1)$ must join the same platoon as vehicle $(m+1)$. 
Therefore, the conditions for vehicle $(m+1)$ to join the platoon are: 
(1) the platoon length constraint remains satisfied after the following vehicle $(m+2)$ joins; and 
(2) the communication distance constraint is satisfied between vehicle $(m+2)$ and its target communication vehicle $m$. 
The probability of vehicle $(m+1)$ joining the platoon can be expressed as:
\begin{equation}
    \begin{aligned}
    & P_{\mathrm{pla},Y_{m},Y_{m+2},Z_m}^\mathrm{PS2}  \\ & =  Pr\{Z_m \leq l_\mathrm{max} - 2 \cap v(\Delta T_{m+1} + \Delta T_{m+2}) \\
    & < \min(d_{Y_{m}}, d_{Y_{m+2}})\}
    \end{aligned}
    \label{eq:chap3_pla_strategy_single_discon}
\end{equation}
where $\Delta T_i$ is the time headway between vehicle $i$ and its preceding vehicle.
%$Z_m $ is the position of vehicle $m\in \{1,2,...,l_{max} \}$ in the platoon.
%$Zm ∈ \{1,2,...,lmax \} 为车辆队列位置状态

Notably, in this scenario, the absence of communication capability in the HV/AV necessitates the presence of a CAV as a communication intermediary among its adjacent vehicles for it to join the platoon. 
As illustrated in Fig.~\ref{fig:pla_scenarios} (c), the lack of a CAV near the HV/AV $(m+1)$ implies that its motion state information cannot be reliably acquired and forwarded to other CHVs/CAVs. 
Consequently, it is unable to form a platoon with its preceding and following vehicles.

%\begin{figure}[!t]
%\centering
%\subfloat[Valid vehicle platoon\label{fig:chap3_valid}]{
%  \includegraphics[width=0.7\linewidth]{fig/3/chap3_pla_strategy_single_discon_success.pdf}
%}
%\hfil
%\subfloat[Invalid vehicle platoon\label{fig:chap3_invalid}]{
%  \includegraphics[width=0.7\linewidth]{fig/3/chap3_pla_strategy_single_discon_fail.pdf}
%}
%\caption{Single non-connected vehicle joining a platoon under different conditions.}
%\label{fig:chap3_pla_strategy_single_discon}
%\end{figure}

As shown in Fig.~\ref{fig:pla_scenarios} (d), vehicle $m$ is a CAV, vehicles $(m+1)$ and $(m+2)$ are HVs/AVs, and vehicle $(m+3)$ is a CHV.
In this configuration, the motion state information of vehicle $(m+1)$ must be acquired and relayed through its preceding vehicle $m$, while the motion state information of vehicle $(m+2)$ requires acquisition and relaying via its following vehicle $(m+3)$. 
Furthermore, since HVs/AVs cannot serve as the lead or tail vehicle of a platoon, vehicles $m$ and $(m+3)$ must form the same platoon encompassing the HVs/AVs. 
Consequently, the conditions for the HVs/AVs $(m+1)$ and $(m+2)$ to join the platoon are:
(1) the platoon length constraint remains satisfied after vehicle $(m+3)$ joins;
(2) the communication distance constraint is met between vehicle $(m+3)$ and its target communication vehicle $m$.
The probability of the HVs/AVs $(m+1)$ and $(m+2)$ joining the platoon can then be expressed as:
\begin{equation}
    \begin{aligned}
    P_{\mathrm{pla},Y_{m},Y_{m+3},Z_m}^\mathrm{PS2} = & Pr\{Z_m \leq l_\mathrm{max} - 3 \cap v \cdot \sum_{i=m+1}^{m+3} \Delta T_i \\
    & < \min(d_{Y_{m}}, d_{Y_{m+3}})\}
    \end{aligned}
    \label{eq:chap3_pla_strategy_2_discon}
\end{equation}

In this scenario, any HVs/AVs must have at least one CAV among its adjacent vehicles. Consequently, the maximum number of consecutive HVs/AVs within a platoon is limited to two.
As illustrated in Fig.~\ref{fig:pla_scenarios}(e), due to the absence of a CAV to probe and relay its motion state, the state information of the HV/AV $(m+1)$ remains unknown to other CHVs/CAVs. Thus, a platoon cannot be formed.

%\begin{figure}[!t]
%\centering
%\subfloat[Valid vehicle platoon]{
%    \includegraphics[width=0.7\linewidth]{fig/4/chap3_pla_strategy_two_discon_success.pdf}
%    \label{fig:chap3_2discon_valid}
%}\\
%\subfloat[Invalid vehicle platoon]{
%    \includegraphics[width=0.7\linewidth]{fig/4/chap3_pla_strategy_two_discon_fail.pdf}
%    \label{fig:chap3_2discon_invalid}
%}
%\caption{Continuous non-connected vehicles joining a platoon under different conditions.}
%\label{fig:chap3_pla_strategy_2_discon}
%\end{figure}

In summary, Table~\ref{tab:chap_3_pla_struct} presents the three distinct platoon structures under the hybrid platooning strategy.

\begin{table}[!t]
\centering
\caption{Platoon structures under hybrid platooning strategy}
\label{tab:chap_3_pla_struct}
\begin{tabularx}{\columnwidth}{
    >{\centering\arraybackslash}m{0.30\columnwidth}
    >{\centering\arraybackslash}X
}
\toprule
\textbf{Platooning structure} & \textbf{Platooning scenario} \\
\midrule
C--C &
Continuous CHV/CAV platooning \\
\midrule
\makecell[c]{CAV--H--CHV\\ CHV--H--CAV\\ CAV--H--CAV} &
Single HV/AV join a platoon \\
\midrule
CAV--H--H--CAV &
Two HVs/AVs join a platoon \\
\bottomrule
\noindent\parbox{\columnwidth}{%
\footnotesize
\textit{Note:} $\mathrm{C}$ denotes CHVs/CAVs and $\mathrm{H}$ represents HVs/AVs.
}
\end{tabularx}
\end{table}

\subsection{Heterogeneous car-following behavior and models}
\label{subsec:cf_behavior}

Based on the vehicle interaction space of Cartesian product $\mathcal{V} \times \mathcal{V}$, car-following behaviors in mixed traffic flow can be classified into two categories according to platoon status:

(1) Fundamental car-following types: Sixteen elementary car-following pairs $s-r \in \mathcal{V}^2$ formed by pairwise combinations of four vehicle types, where $s$ denotes the leading vehicle type and $r$ the following vehicle type. This classification reveals fundamental interaction patterns between heterogeneous vehicles at the microscopic level.

(2) Platoon-based car-following types: Considering the dynamic influence of platooning on control modes, when CAVs, CHVs, AVs, and HVs overcome communication barriers to form hybrid platoons, their intra-platoon and inter-platoon car-following behaviors exhibit significant heterogeneity. Thus, we extend the fundamental pairs into 16 intra-platoon relationships, e.g., $\text{CAV}_{\text{pla}}\text{-CHV}_{\text{pla}}$ and $\text{CHV}_{\text{pla}}\text{-HV}_{\text{pla}}$.

In summary, mixed traffic flow contains 16 fundamental non-platoon car-following types and 16 platoon-based car-following types, totaling 32 car-following configurations.

However, for HVs/AVs, their inability to acquire real-time platoon status via communication implies their time headway remains independent of instantaneous platoon conditions. That is, their car-following behavior is identical both inside and outside platoons. 
Applying the established CHV/CAV demotion criteria, these 32 car-following types can be consolidated into six distinct time headway patterns, called as fundamental autonomous headway ($\Delta T_\mathrm{A}$), manual driving headway ($\Delta T_\mathrm{H}$), connected autonomous headway outside platoons ($\Delta T_\mathrm{CA}$), connected manual headway outside platoons ($\Delta T_\mathrm{CH}$), intra-platoon cooperative autonomous headway ($\Delta T_\mathrm{PA}$), and intra-platoon cooperative manual headway ($\Delta T_\mathrm{PH}$), respectively.  For conciseness, the definitions of the six consolidated time headway patterns and their detailed mappings to the original car-following configurations are provided in Appendix~\ref{app:headway_patterns}.
Generally speaking, autonomous systems are capable of operating at shorter time headways than human drivers. Communication enables cooperative driving that further reduces time headways, while platooning, as an advanced form of cooperative driving, achieves even smaller headways. Consequently, these time headways exhibit the following hierarchical relationship:

\begin{equation}
\Delta T_\mathrm{H} > \Delta T_\mathrm{A} > \Delta T_\mathrm{CH} > \Delta T_\mathrm{CA} > \Delta T_\mathrm{PH} > \Delta T_\mathrm{PA}
\label{eq:chap3_delta_T_relation}
\end{equation}

Based on the above classification of heterogeneous car-following behaviors, the car-following process can be further represented through acceleration functions. Given the headway, velocity difference, and velocity of the ego vehicle, the car-following dynamics can be expressed by the following ordinary differential equations:
\begin{equation}
\dot{x}_{n}(t) = v_{n}(t)
\label{eq:chap2_stable_analysis_general_cf1}
\end{equation}
\begin{equation}
\dot{v}_{n}(t) = f(s_n(t), v_n(t), \Delta v_n(t))
\label{eq:chap2_stable_analysis_general_cf2}
\end{equation}
where $x_n(t)$ and $v_n(t)$ denote the position and velocity of vehicle $n$ at time $t$; $s_n(t) = x_{n-1}(t) - x_n(t)$ represents the headway between vehicle $n$ and the preceding vehicle $n-1$; and $\Delta v_n(t) = v_{n-1}(t) - v_n(t)$ is the velocity difference.
To characterize the heterogeneous driving behaviors associated with different vehicle types and connectivity conditions, three car-following models are adopted in this study. The detailed model formulations, parameter definitions, and parameter settings are provided in Appendix~\ref{app:cf_models}.

\section{Platooning strategy using deep reinforcement learning}
\label{sec:PS_DRL}

%This study proposes a hybrid platooning strategy for  mixed traffic environments using DRL, building upon the proposed platooning frameworks. 
%The strategy controls the platooning behavior of CHVs/CAVs to jointly optimize traffic volume and traffic flow stability.

\begin{figure}[!htbp]
   \centering
   \includegraphics[width=\textwidth]{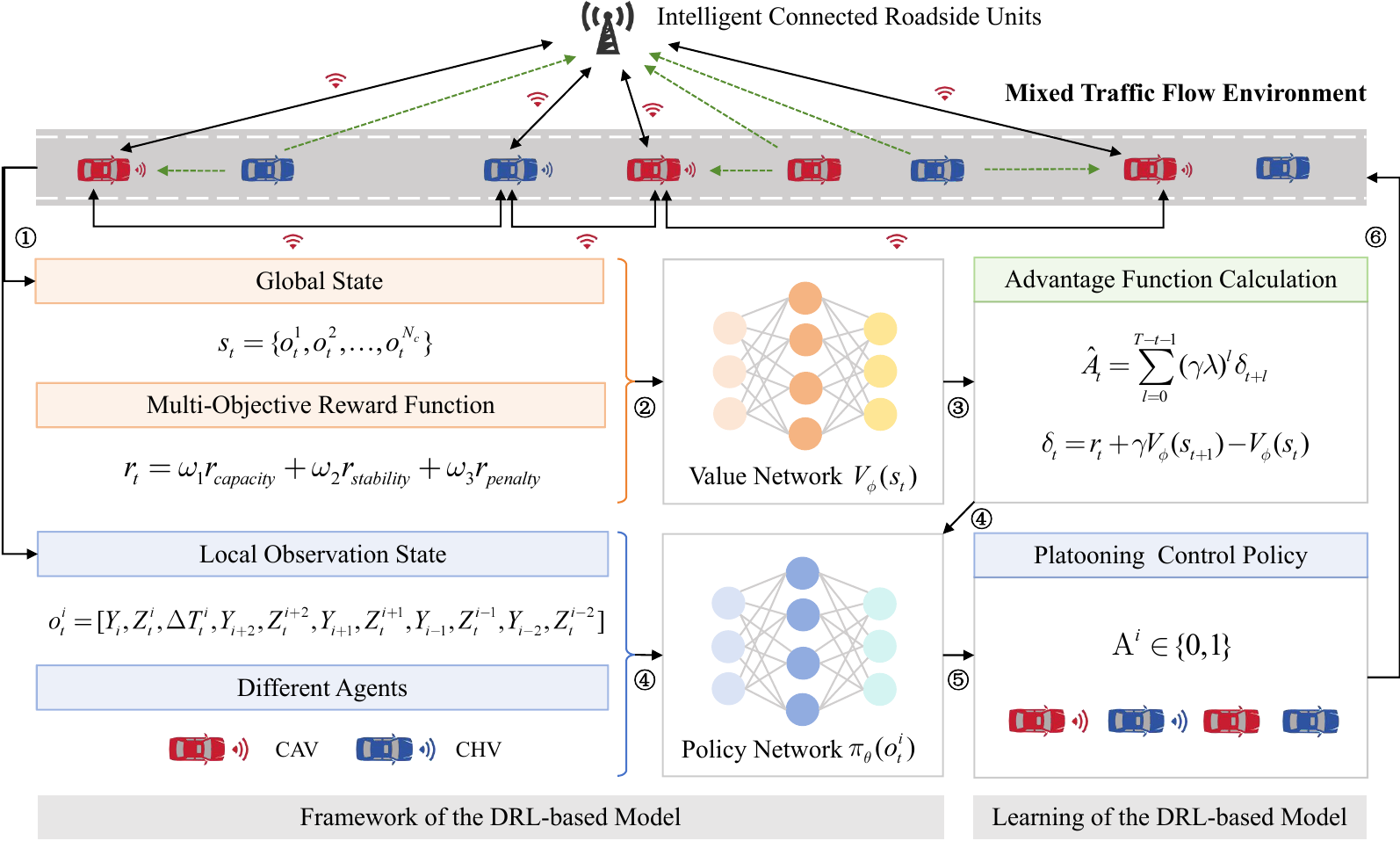}
   \caption{Structure of the DRL-based hybrid platooning model considering multi-type vehicles.}
   \label{fig:chap4_struct}
\end{figure}

The design of the hybrid platooning strategy should comprehensively integrate multiple factors including heterogeneous vehicle dynamics, communication topology constraints, and traffic flow stability. 
Formulated within a Partially Observable Markov Decision Process (POMDP) framework, we develop a distributed platooning strategy for hybrid platooning in mixed traffic, that
%. As illustrated in Fig. \ref{fig:chap4_struct}, 
comprises state space, action space, multi-objective reward functions, and training architecture.
The primary structure of the DRL-based platooning strategy is illustrated in Fig.\ref{fig:chap4_struct}.

\subsection{State space}
%\subsection 状态空间

Connected vehicles (CHVs/CAVs) possess communication capabilities that enable them to acquire partial state information of surrounding vehicles from vehicle-to-everything (V2X) links. 
To enable multi-dimensional environmental perception and coordinated platooning decisions in mixed traffic flows, the state space integrates multi-modal data, including vehicle dynamics parameters, platoon topological relationships, and traffic environment characteristics. This integration ensures real-time capture of dynamic traffic conditions, instantaneous platoon structure control, and optimized traffic efficiency/stability.
This integration lays the groundwork for establishing spatiotemporal coupling relationships between microscopic vehicle states and macroscopic traffic flow. 
The state-space vector $o_t^i \in \mathbb{R}^{11}$, where $i$ denotes the index of the ego vehicle, and $t$ denotes the discrete time step, is constructed by concatenating ego-vehicle states with ambient environment states derived from multi-source heterogeneous perception data.
\begin{equation}
    o_t^i = \{s_t^\mathrm{ego}, s_t^\mathrm{env}\}
\end{equation}
where $s_t^\mathrm{ego}$ denotes the ego-vehicle state; and $s_t^\mathrm{env}$ represents the ambient environment state.

The ego-vehicle state $s_t^\mathrm{ego}$ comprises three elements, including vehicle type, position within the platoon, and time headway.
\begin{equation}
    s_t^\mathrm{ego} = \{Y_{i}, Z^{i}_t, \Delta T^{i}_t\}
\end{equation}

Vehicle type and platoon position serve as critical inputs for platooning decisions. The introduction of HVs/AVs significantly increases platoon length. 
To capture essential queue state information while avoiding the curse of dimensionality from expanding input features, which would escalate model training complexity, this study selects vehicle types and positions from the two immediately preceding and following vehicles, respectively, to construct the state $s_t^\mathrm{env}$.
\begin{equation}
    s_t^\mathrm{env} = \{Y_{i+2}, Z^{i+2}_t,
    Y_{i+1}, Z^{i+1}_t,
    Y_{i-1}, Z^{i-1}_t,
    Y_{i-2}, Z^{i-2}_t\}
\end{equation}

Leveraging local observations, including ego-vehicle platoon position, time headway, and surrounding vehicle states, CHVs/CAVs dynamically generate platoon control decisions through policy networks. 
These decisions enable real-time structural adjustments to platoon formations, achieving communication efficiency and traffic flow stability in mixed traffic scenarios.

\subsection{Action space}

The platooning decision states for vehicles in mixed traffic flow comprise three categories: join platoon, leave platoon, maintain independent driving.
%The leave platoon and maintain independent driving states can be consolidated into a platoon-disbanding action for vehicles. 
The platooning decision action space $\mathcal{A}^i$ can be formulated as,
\begin{equation}
\label{eq:action}
\mathcal{A}^i = \{ a_t^i \in \{0,1\}  \}
\end{equation}
where
$a_t^i = 1$ denotes vehicle $i$ executing a joining-platoon action at timestep $t$; while
$a_t^i = 0$ denotes vehicle $i$ executing either a leaving-platoon action or maintaining independent driving at $t$.

\subsection{Multi-objective reward function}

Guided by multi-objective optimization principles, the reward function 
$r_t$ is formulated to achieve a dynamic trade-off between traffic efficiency and stability, and is defined as:
\begin{equation}
r_t = \omega_1 r_\mathrm{capacity} + \omega_2 r_\mathrm{stability} + \omega_3 r_\mathrm{penalty}
\end{equation}
where 
$r_\mathrm{capacity}$, $r_\mathrm{stability}$ and $r_\mathrm{penalty}$  denote the traffic efficiency reward, traffic flow stability reward, and capacity constraint penalty, respectively; 
$\omega_i,\ i \in \{1,2,3\}$ represent their corresponding weight coefficients.

(1) The traffic efficiency reward is,
\begin{equation}
r_\mathrm{capacity} = \frac{C_{RL} - C_\mathrm{base}}{C_\mathrm{base}} \times 100\%
\end{equation}
where $C_{RL}$ is the lane capacity under the DRL-based platooning strategy. 
Without loss of generality, the baseline capacity  $C_\mathrm{base}$ is defined as the lane capacity under the $\text{PS}2$ hybrid platooning scheme.

(2) The traffic flow stability reward is,
\begin{equation}
r_\mathrm{stability} = \frac{G_h^{RL} - G_h^\mathrm{base}}{G_h^\mathrm{base}} \times 100\%
\end{equation}
where $G_h^{RL}$ is the hybrid traffic flow stability discriminant value under the DRL-based platooning strategy; and $G_h^\mathrm{base}$ is the corresponding value under the $\text{PS}2$ hybrid platooning scheme.

Following the mixed traffic stability criterion established by \cite{Ngoduy_2013_analytical_studies_nonlinear_sci} and \cite{Xie2019MixedTraffic}, 
$G_h$ is evaluated under the DRL-based and $\text{PS}2$ platooning strategies to compare their impacts on mixed traffic flow stability: 
\begin{equation}
G_h = \sum_{j=1}^{J} \left[P_{H_j} \frac{(f_v^j)^2 - 2 f_s^j - 2 f_v^j f_{\Delta v}^j}{(f_s^j)^2}\right]
\label{eq:mixed_stability_judge}
\end{equation}
%其中，$j\in J=\{1,2,...,6 \}$对应于\ref{subsec:cf_behavior} 的6类车辆跟驰类型，$P_{H_j}$是对应的比例。$f_v^j$, $f_s^j$, and $f_{\Delta v}^j$分别是跟驰类别$j$对应车辆跟驰模型对车辆速度、车间距和速度差的微分项。
where $j\in J=\{1,2,...,6 \}$ corresponds to the six car-following behavior types in Section \ref{subsec:cf_behavior}, and 
$P_{H_j}$ is the corresponding proportion. 
The terms $f_v^j$, $f_s^j$, and $f_{\Delta v}^j$ represent the partial derivatives of the car-following model for type $j$ with respect to velocity, spacing, and velocity difference, respectively.

(3) The capacity constraint penalty term is,
\begin{equation}
\begin{aligned}
r_\mathrm{penalty} =
\left\{
\begin{array}{ll}
-1  ,& \frac{C_{RL} - C_\mathrm{base}}{C_\mathrm{base}} < 0.05 \\
0  ,else \\
\end{array}
\right.
\end{aligned}
\end{equation}
This reward term penalizes significant lane capacity losses.

To prevent single-objective dominance in policy updates and ensure the model improves traffic flow stability while maintaining traffic efficiency, this study assigns equal weights to the traffic efficiency and stability reward terms. 
Furthermore, since the impact of platoon structure changes on traffic efficiency and flow stability exhibits coupling characteristics, identical weights compel the agents (CHVs/CAVs) to dynamically trade off between these objectives and avoid extreme strategies. 
To prioritize preventing significant capacity degradation and avoid sacrificing traffic efficiency for excessive stability optimization, a fixed efficiency loss threshold is established as a boundary condition. 
This is reinforced by assigning a larger weight coefficient to amplify the impact of excessive efficiency loss.

\begin{figure}[!t]
    \centering
    \subfloat[Reward]{
        \includegraphics[width=0.32\linewidth]{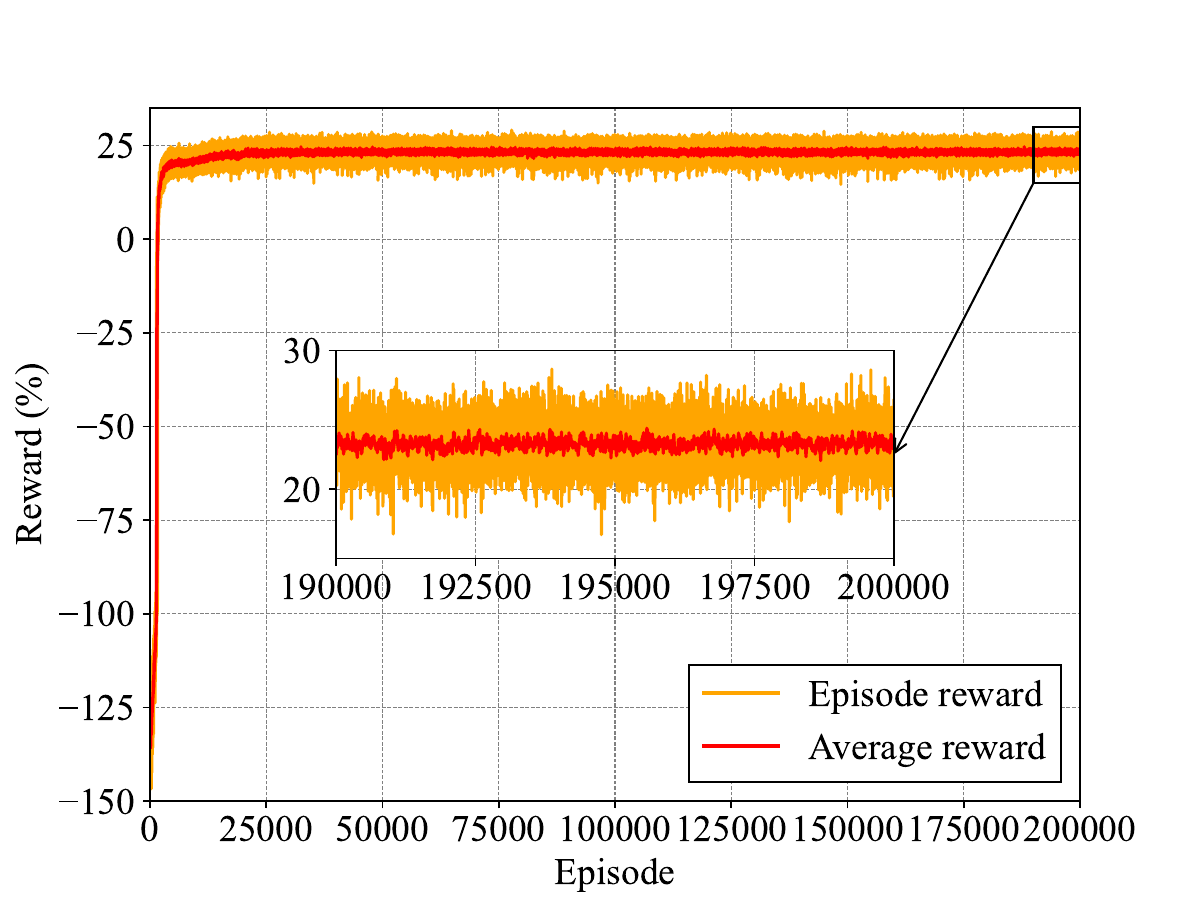}
    }
    \subfloat[Lane capacity]{
        \includegraphics[width=0.32\linewidth]{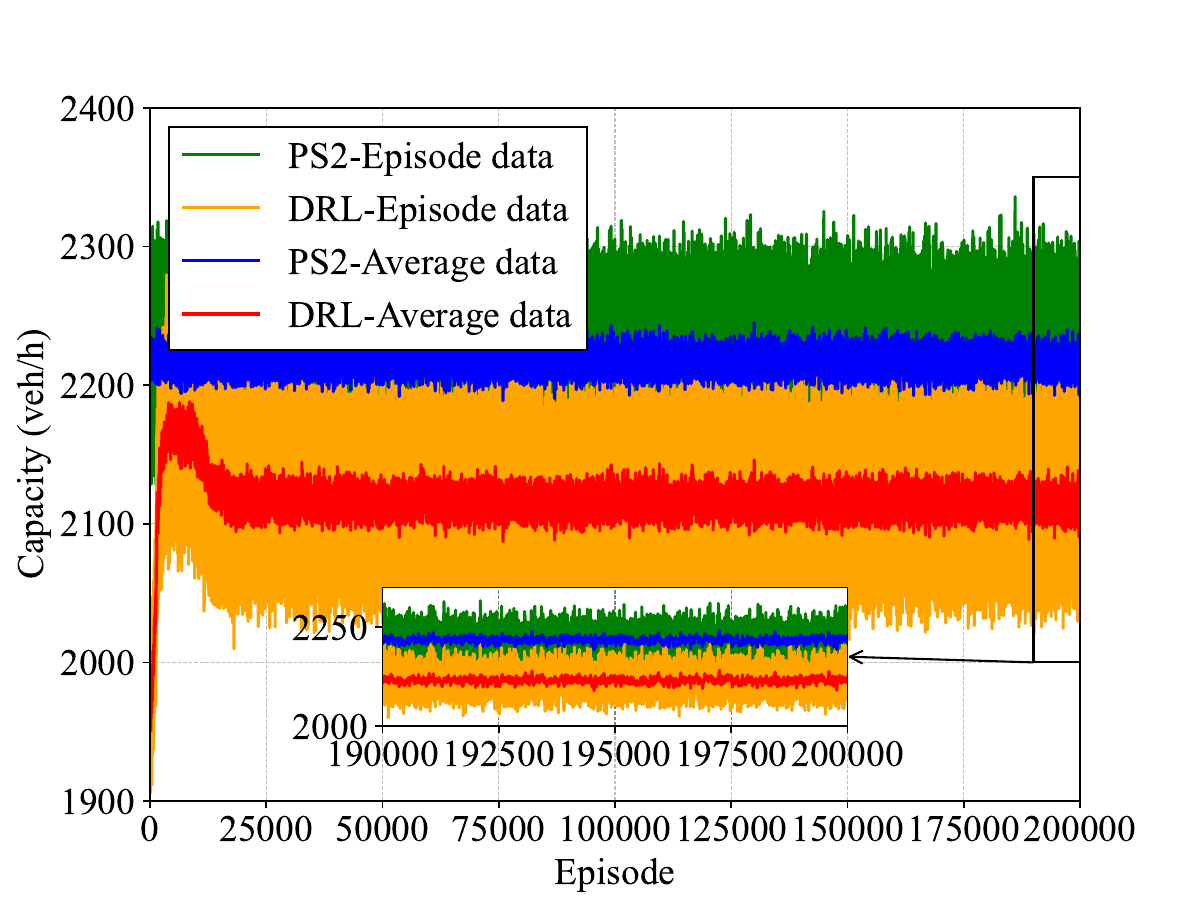}
    }
    \subfloat[$G_{h}$]{
        \includegraphics[width=0.32\linewidth]{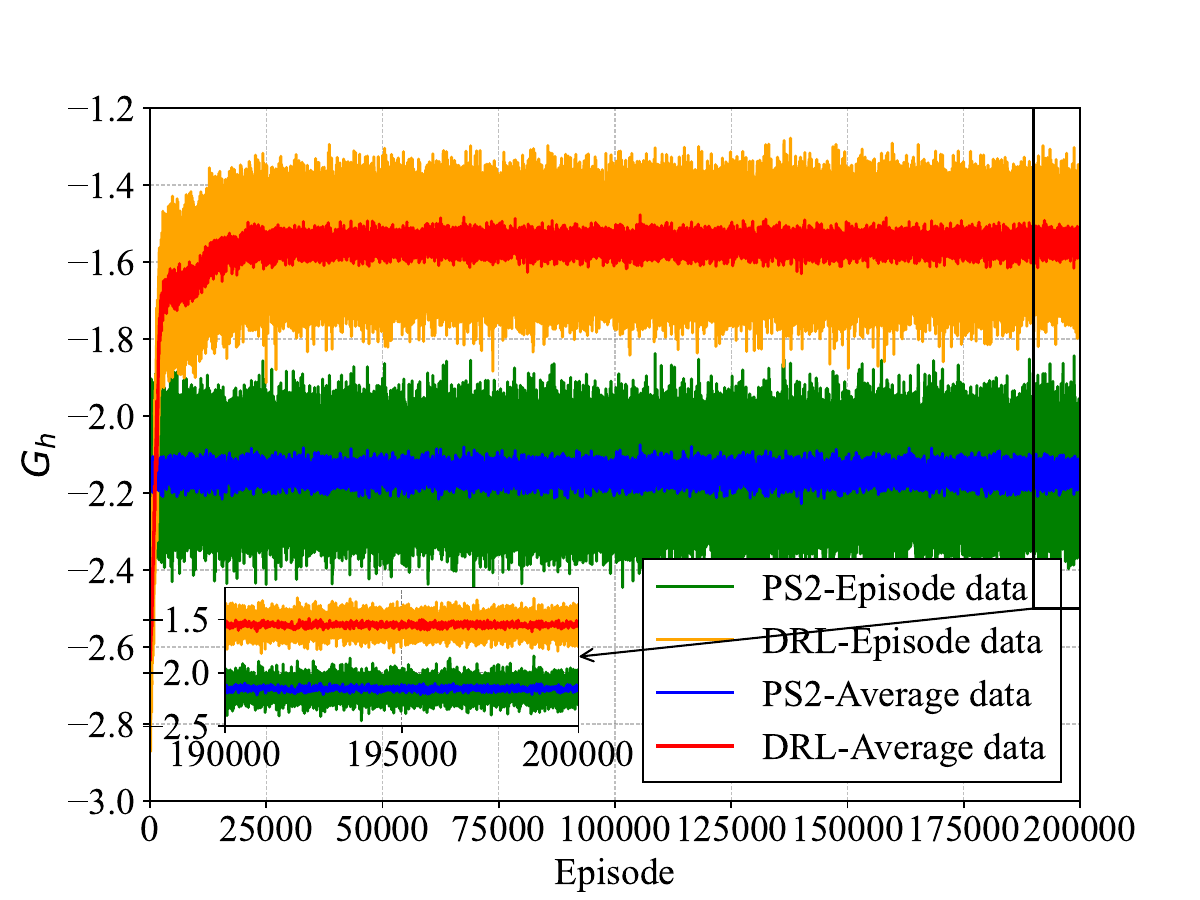}
    }
    \caption{Evolution of reward, lane capacity, and $G_{h}$ during the training process.}
    \label{fig:chap4_train_reward_capacity_gh}
\end{figure}

\begin{figure}[!htbp]
    \centering
    \subfloat[Average platoon size]{
        \includegraphics[width=0.32\linewidth]{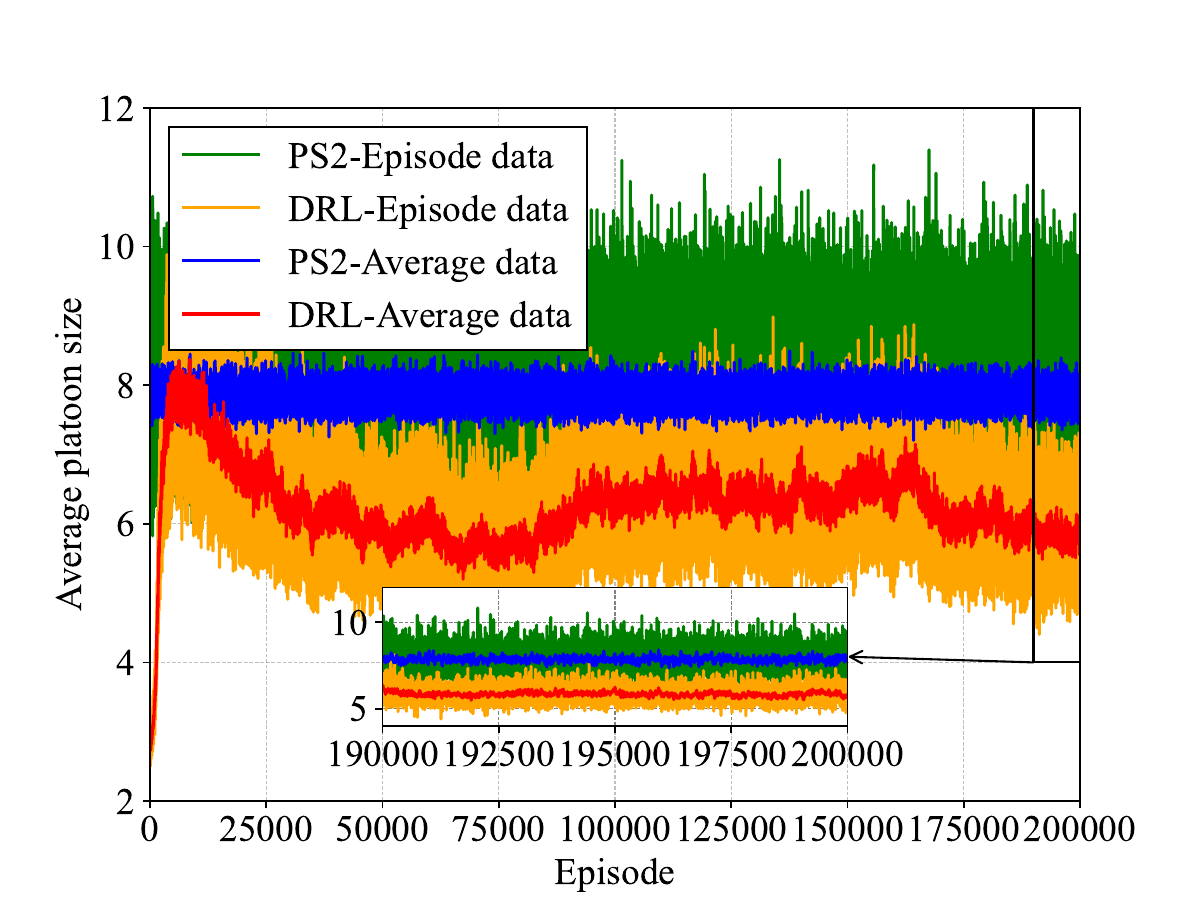}
    }
    \subfloat[Maximum platoon size]{
        \includegraphics[width=0.32\linewidth]{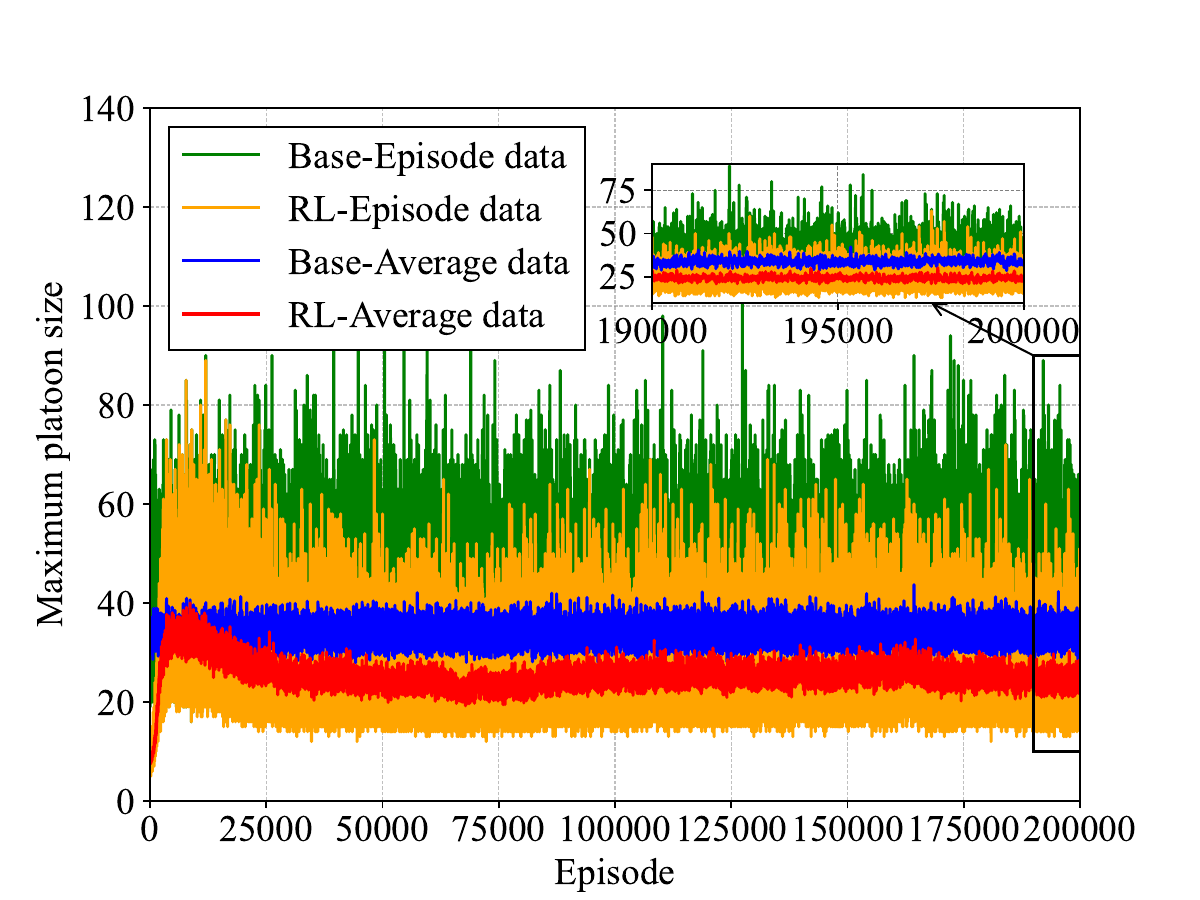}
    }
    \subfloat[Total number of platoons]{
        \includegraphics[width=0.32\linewidth]{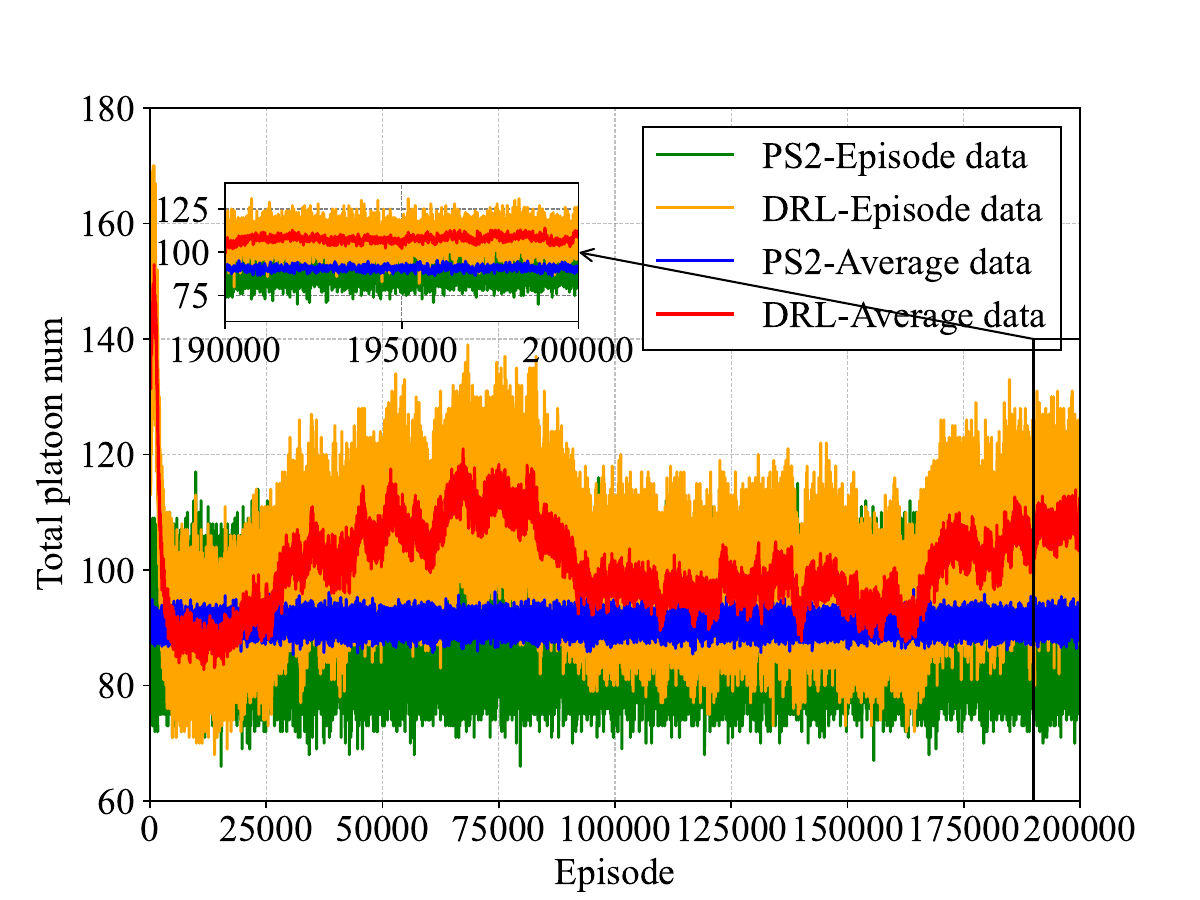}
    }
    \caption{Evolution of platoon characteristics during the training process.}
    \label{fig:chap4_train_pla_info}
\end{figure}

\subsection{Training framework}

To address platooning behavior decision-making in large-scale mixed traffic flows, this study adopts the Centralized Training with Decentralized Execution (CTDE) framework \citep{Bernstein_2000_ctde}. 
This architecture obtains globally optimal policies through centralized training, while during decentralized execution, each vehicle makes real-time decisions relying solely on local observations, effectively mitigating the curse of dimensionality in large-scale traffic flows. 

(1) Policy network $\pi_{\theta}$:
The policy network employs a dual-network structure. The feature extraction layer consists of bilinear fully-connected layers and ReLU activation functions to capture nonlinear vehicle features. 
The final layer outputs action probability distributions through a Sigmoid activation function.

(2) Value network $V_{\phi}$:
The value network serves as a global state value estimator. It fuses multi-agent state information via mean pooling, with linear layers outputting value estimates, thereby establishing a mapping between global states and values.

(3) PPO algorithm:
The PPO algorithm ensures policy update stability through a clipped objective function and experience replay \citep{Lin_1991_data_memory}.

(4) Generalized advantage estimation (GAE) algorithm \citep{Schulman_2015_gae}:
The GAE algorithm combines temporal difference errors and multi-step prediction concepts, computing advantage estimates via exponentially weighted averaging to enhance the stability and efficiency of policy gradient methods.

The detailed training procedure is outlined in Algorithm \ref{alg:chap4_ctde_ppo}.

\begin{algorithm}[t]
\caption{PPO-based platooning algorithm with CTDE architecture}
\label{alg:chap4_ctde_ppo}
\renewcommand{\algorithmicrequire}{\textbf{Inputs:}}
\renewcommand{\algorithmicensure}{\textbf{Outputs:}}
\begin{algorithmic}[1]

\REQUIRE Vehicle number $N$, observation dimension $d_s$, training epochs $K$, discount factor $\gamma$, 
GAE parameter $\lambda$, learning rate $\alpha_{\mathrm{ppo}}$, clipping parameter $\epsilon$, 
loss weights $c_1$, $c_2$, reward weights $\omega_1$, $\omega_2$, $\omega_3$
\ENSURE Optimal policy $\pi^{*}_{\theta}$

\STATE Initialize policy network $\pi_{\theta}$ and value network $V_{\phi}$
\STATE Initialize experience buffer $\mathcal{B}$

\FOR{$k = 1$ \TO $K$}
    \FOR{$t = 0$ \TO $T-1$}
        \STATE Observe global state $\mathbf{s}_t = \{o_t^i\}_{i=1}^{N}$
        \FOR{$i = 1$ \TO $N$}
            \STATE Sample action $a_t^i \sim \pi_{\theta}(o_t^i)$
        \ENDFOR
        \STATE Execute joint action $\mathbf{a}_t$, observe $(\mathbf{s}_{t+1}, r_t)$
        \STATE Store $(\mathbf{s}_t, \mathbf{a}_t, r_t, V_{\phi}(\mathbf{s}_t))$ in $\mathcal{B}$
    \ENDFOR

    \STATE Compute GAE advantage estimates $\hat{A}_t$
    
    \FOR{$m = 1$ \TO $M$}
        \STATE Sample minibatch from $\mathcal{B}$
        \STATE Compute importance ratio $\rho_t$
        \STATE Update policy $\pi_{\theta}$ using clipped PPO objective
        \STATE Update value function $V_{\phi}$ with squared error loss
    \ENDFOR
\ENDFOR

\end{algorithmic}
\end{algorithm}

\subsection{Training of the DRL-based platooning model}

This subsection systematically analyzes the training process of the DRL-based hybrid platooning strategy, examining policy optimization dynamics through the convergence of the reward function, lane capacity, traffic flow stability, and platoon state evolution. The parameters for this training procedure are provided in Table \ref{tab:chap4_train_params}.

\begin{table}[!t]
    \centering
    \caption{Parameters for training the DRL-based platooning model}
    \label{tab:chap4_train_params}
    \begin{tabularx}{\columnwidth}{
        >{\raggedright\arraybackslash}m{0.7\columnwidth}
        >{\centering\arraybackslash}m{0.15\columnwidth}
    }
        \toprule
        \textbf{Parameter} & \textbf{Value} \\
        \midrule
        Total vehicle number, $N$ & 1000 \\
        Local observation state dimension, $d_s$ & 11 \\
        Total training epochs, $K$ & $2\times10^{5}$ \\
        Discount factor, $\gamma$ & 0.99 \\
        GAE parameter, $\lambda$ & 0.95 \\
        PPO learning rate, $\alpha_{\mathrm{ppo}}$ & $3\times10^{-4}$ \\
        Policy loss clipping hyperparameter, $\epsilon$ & 0.2 \\
        Value loss weight coefficient, $c_1$ & 0.5 \\
        Entropy regularization weight coefficient, $c_2$ & 0.01 \\
        Traffic efficiency reward weight, $\omega_1$ & 0.5 \\
        Traffic flow stability reward weight, $\omega_2$ & 0.5 \\
        Traffic capacity constraint penalty weight, $\omega_3$ & 100 \\
        \bottomrule
    \end{tabularx}
\end{table}

Fig. \ref{fig:chap4_train_reward_capacity_gh}(a) illustrates the reward progression during training, with a moving average (window size: 20 episodes) computed to mitigate stochasticity from vehicle sequencing. During initial training ($<$5000 episodes), the reward exhibits significant volatility with negative moving averages. This originates from the DRL-based platooning strategy's initial inability to effectively coordinate platoon formation with traffic efficiency. As evidenced in Figs. \ref{fig:chap4_train_pla_info}(a)-(b), both average and maximum platoon lengths remain suboptimal during this phase, resulting in insufficient vehicles maintaining small time headways for platooning. Consequently, lane capacity experiences significant degradation (Fig. \ref{fig:chap4_train_reward_capacity_gh}(b)).

As training progresses, the value network progressively learns the mapping between platoon formation and lane capacity through experience replay. Concurrently, the policy network increasingly favors multi-vehicle cooperative platooning actions. The average reward subsequently demonstrates rapid improvement, ultimately stabilizing within a narrow band centered at $23.30\%$, confirming the stability of the DRL-based platooning strategy.

Fig. \ref{fig:chap4_train_reward_capacity_gh} illustrates the mechanistic impact of DRL-based platooning strategy on macroscopic traffic flow characteristics. During the training phase with episodes $\geq 25000$, the single-lane capacity under the DRL-based platooning strategy converges to $2115 \text{veh/h}$, representing a $4.6\%$ reduction compared to the hybrid platooning strategy $(2217 \text{veh/h}$). This capacity compromise, however, achieves a substantial stability enhancement: the hybrid traffic flow stability discriminant value increases from $-2.15$ (hybrid strategy $\text{PS}2$) to $-1.55$, marking a $27.9\%$ improvement.
This performance trade-off empirically validates the inherent efficiency-stability contradiction in mixed traffic flows. The DRL strategy effectively balances these competing objectives through persistent exploratory training, demonstrating its capability to navigate complex traffic optimization landscapes.

Fig. \ref{fig:chap4_train_pla_info} further validates the microscopic evolutionary mechanisms of DRL-based platooning strategy optimization through hybrid traffic flow platoon state evolution. The DRL strategy regulates the average platoon length within a steady-state band centered at 5.84, representing a $25.5\%$ reduction compared to the hybrid platooning scheme $\text{PS}2$. The maximum platoon length is reduced by $28.3\%$ to $24.3$, effectively mitigating stability risks from large-scale platoon aggregation. Notably, the number of platoons increases by $19.3\%$ to $108.3$ under the DRL-based platooning strategy, indicating optimized platoon structure through enhanced distribution density of medium-scale platoons while maintaining stable total platooned vehicles.
This platooning approach simultaneously mitigates stability degradation from oversized platoons via optimized length parameters and alleviates capacity loss from reduced platoon sizes through increased platoon count.

\begin{figure}[!htbp]
    \centering
    \subfloat[Reward]{
        \includegraphics[width=0.32\linewidth]{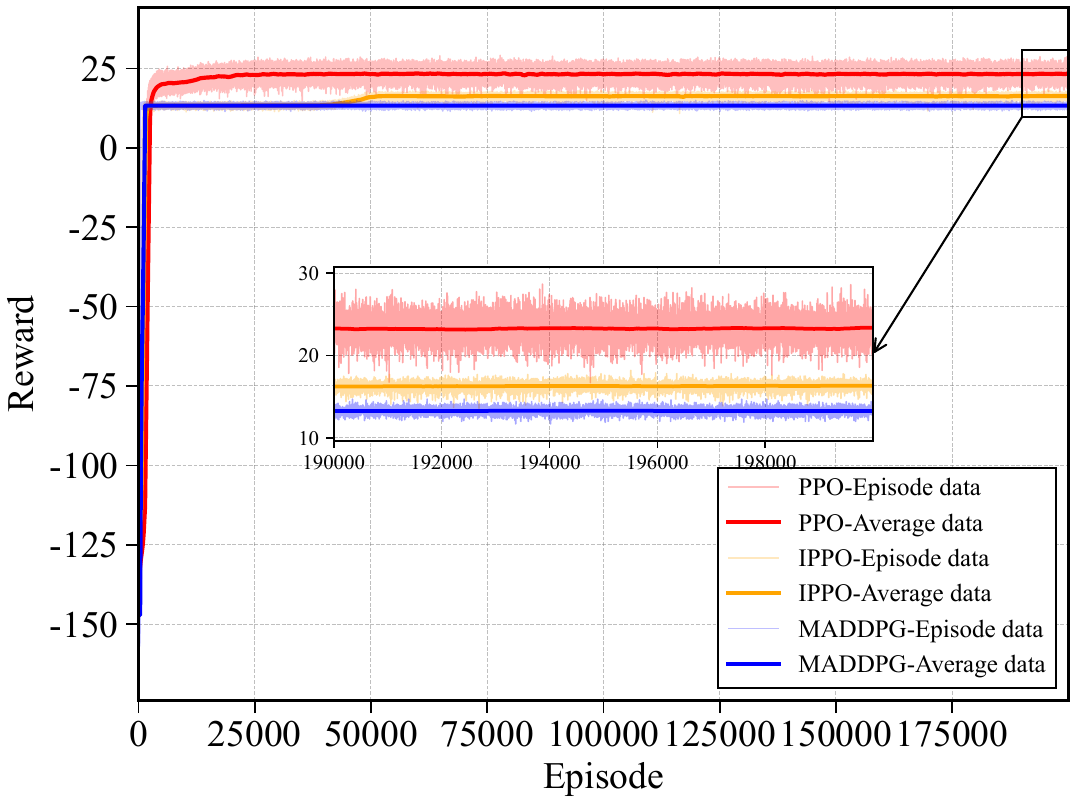}
    }
    \subfloat[Lane capacity]{
        \includegraphics[width=0.32\linewidth]{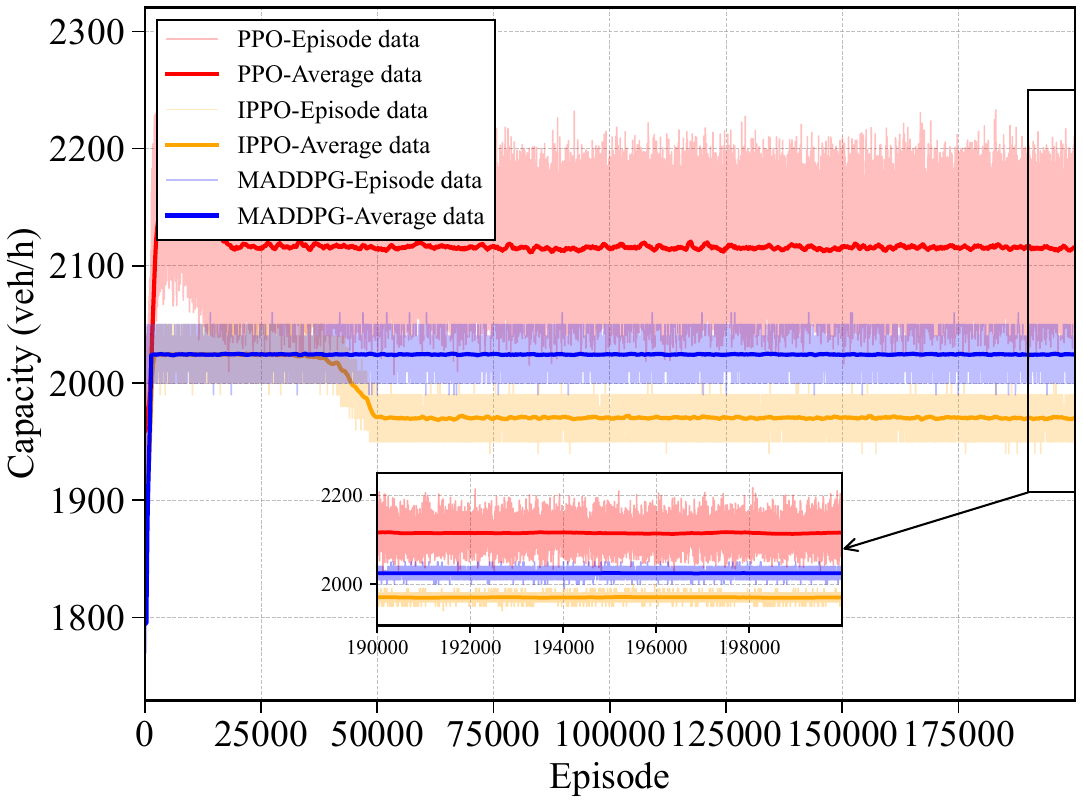}
    }
    \subfloat[$G_h$]{
        \includegraphics[width=0.32\linewidth]{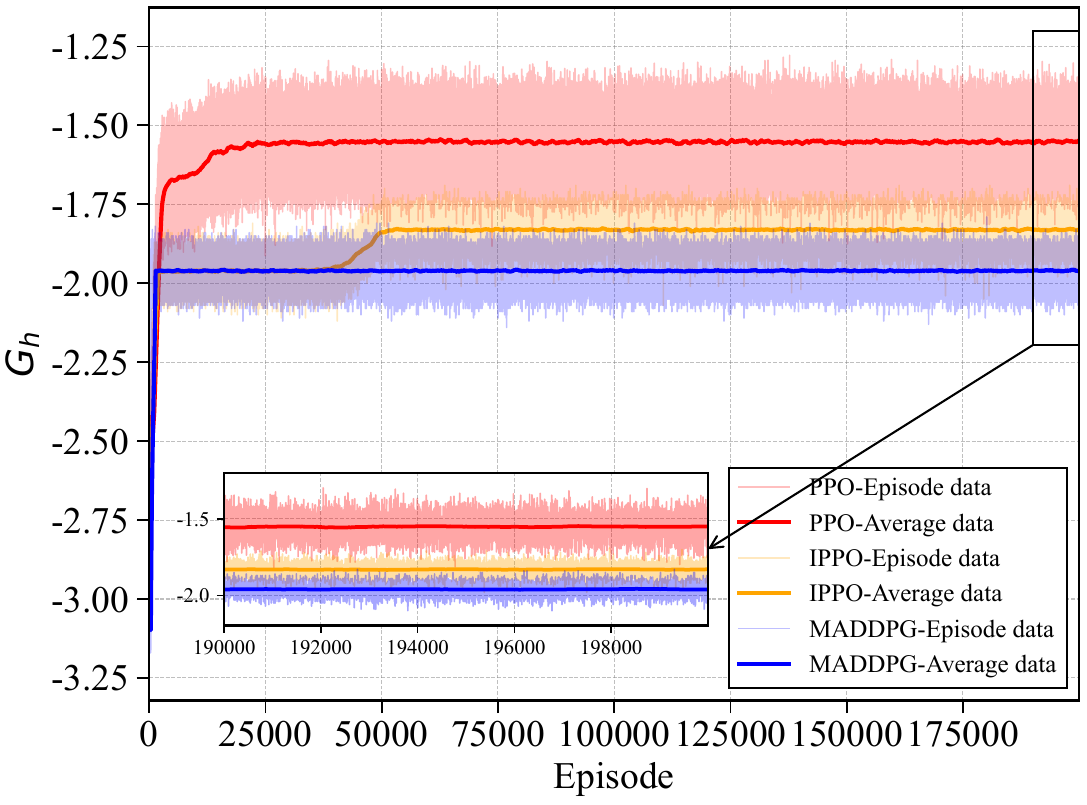}
    }
    \hfill
    \subfloat[Average platoon size]{
        \includegraphics[width=0.32\linewidth]{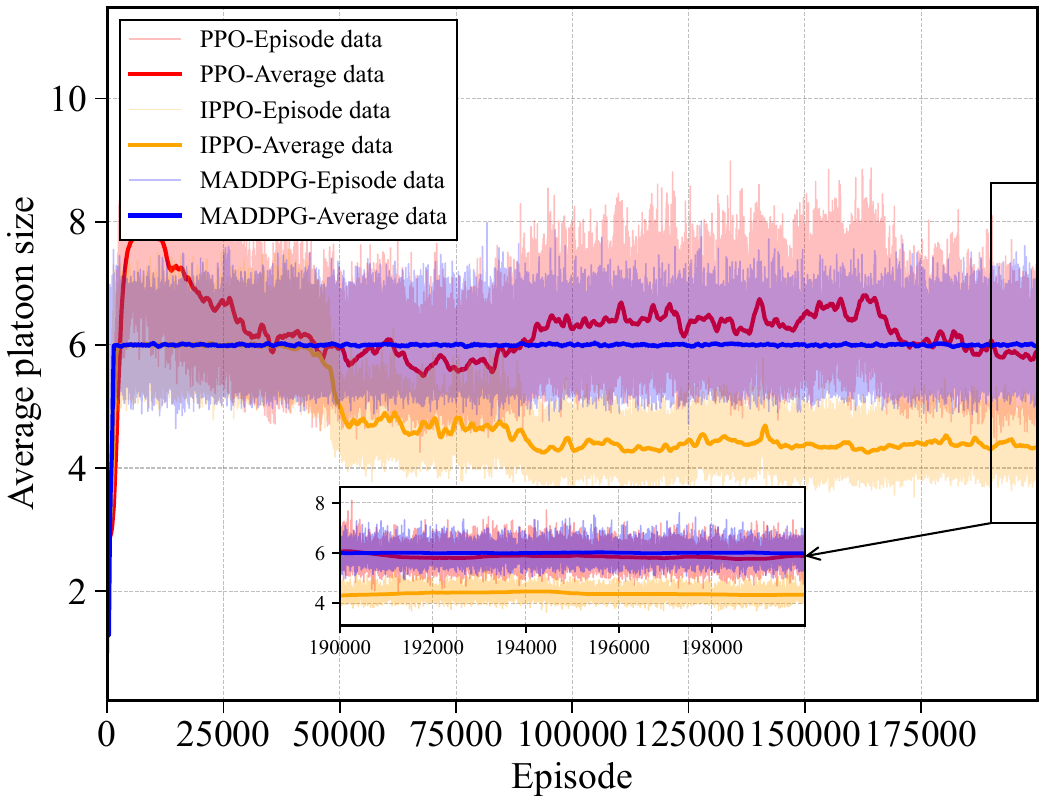}
    }
    \subfloat[Maximum platoon size]{
        \includegraphics[width=0.32\linewidth]{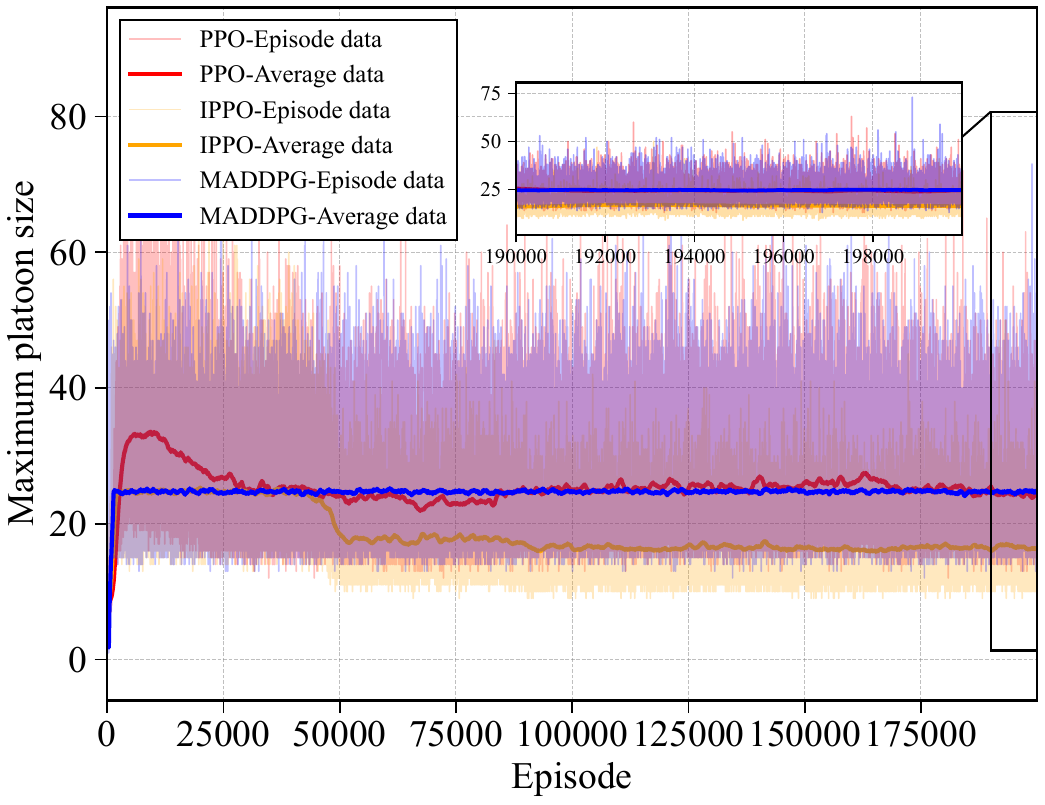}
    }
    \subfloat[Total number of platoons]{
        \includegraphics[width=0.32\linewidth]{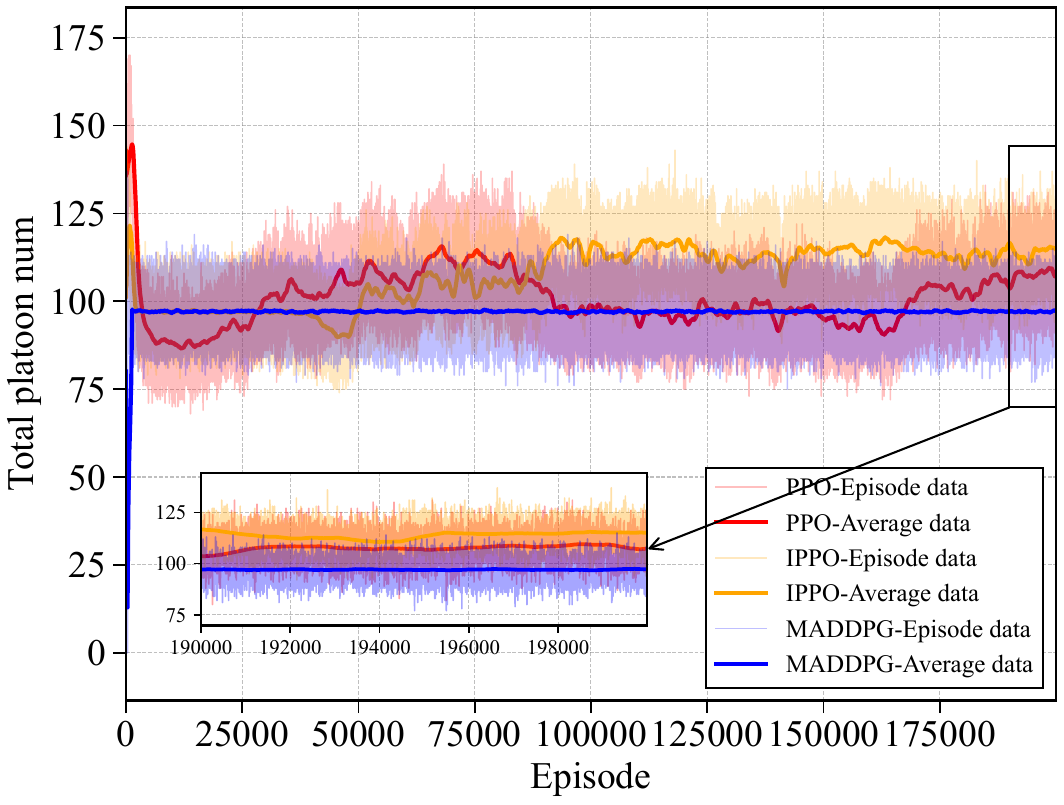}
    }
    \caption{Comparison of training performance among PPO, IPPO, and MADDPG.}
    \label{fig:ppo_ippo_maddpg_training}
\end{figure}

To further examine the suitability of the adopted PPO-based training framework, we compare it with two representative alternatives, namely IPPO \citep{Shi2025MixedVehiclePlatoonForming} and MADDPG \citep{Shi2023DRLDistributedControl}. 
The comparison includes: reward, lane capacity, the traffic-flow stability criterion $G_h$, average platoon size, maximum platoon size, and the total number of platoons.
As shown in Fig. ~\ref{fig:ppo_ippo_maddpg_training} (a)--(c), PPO achieves the highest and most stable reward and converges to better macroscopic traffic performance, with higher lane capacity and a more favorable traffic-flow stability criterion $G_h$. 
This improvement can be attributed to the clipped policy update in PPO, which constrains abrupt policy changes and thus helps maintain a more consistent platoon-joining strategy during training. 
The platoon-level results in Fig.~\ref{fig:ppo_ippo_maddpg_training} (d)--(f) further show that PPO forms platoons with sufficient scale while avoiding excessive platoon expansion. 
In contrast, IPPO relies on more independent agent-wise learning and tends to produce smaller and more fragmented platoons, which benefits stability but limits capacity improvement. 
MADDPG forms platoons with sizes comparable to those of PPO, but its lower final reward and $G_h$ suggest less effective coordination in the proposed hybrid platooning task.

In summary, experimental results from the training process validate the effectiveness of the DRL-based platooning strategy in addressing multi-objective optimization challenges in mixed traffic flows. 
The policy network achieves a nonlinear equilibrium between lane capacity and traffic flow stability. 
This capability fundamentally stems from optimizing hybrid traffic flow platoon structures through the policy network, whereby regulating platoon scale distributions enables nonlinear balancing of traffic efficiency and flow stability. 
Training data quantitatively demonstrate that compared to traditional fixed hybrid-platooning schemes, the DRL approach significantly enhances traffic flow stability within acceptable traffic efficiency margins.

\section{Performance evaluation of the DRL-based platooning strategy}
\label{sec:Performance_evaluation}

To validate the effectiveness of the DRL-based platooning strategy, this section conducts numerical simulations across four critical dimensions: model generalization capability, traffic flow stability, traffic safety, and energy consumption/emissions. 
Performance of the DRL-based strategy is comparatively evaluated against hybrid platooning scheme $\text{PS}2$ and traditional connected vehicle platooning scheme $\text{PS}1$.

\subsection{Generalization capability of the strategy}

Through rigorous numerical experiments, this subsection validates the generalization capability of the proposed DRL-based platooning strategy in terms of penetration rates of CHVs/CAVs. 
Employing a controlled-variable approach, the experimental framework incorporates 10000 independent randomized trials to mitigate stochastic variations.
%, establishing a comprehensive parameter sensitivity analysis architecture.

\begin{figure}[!htbp]
\centering
\subfloat[Reward value]{
    \includegraphics[width=0.48\linewidth]{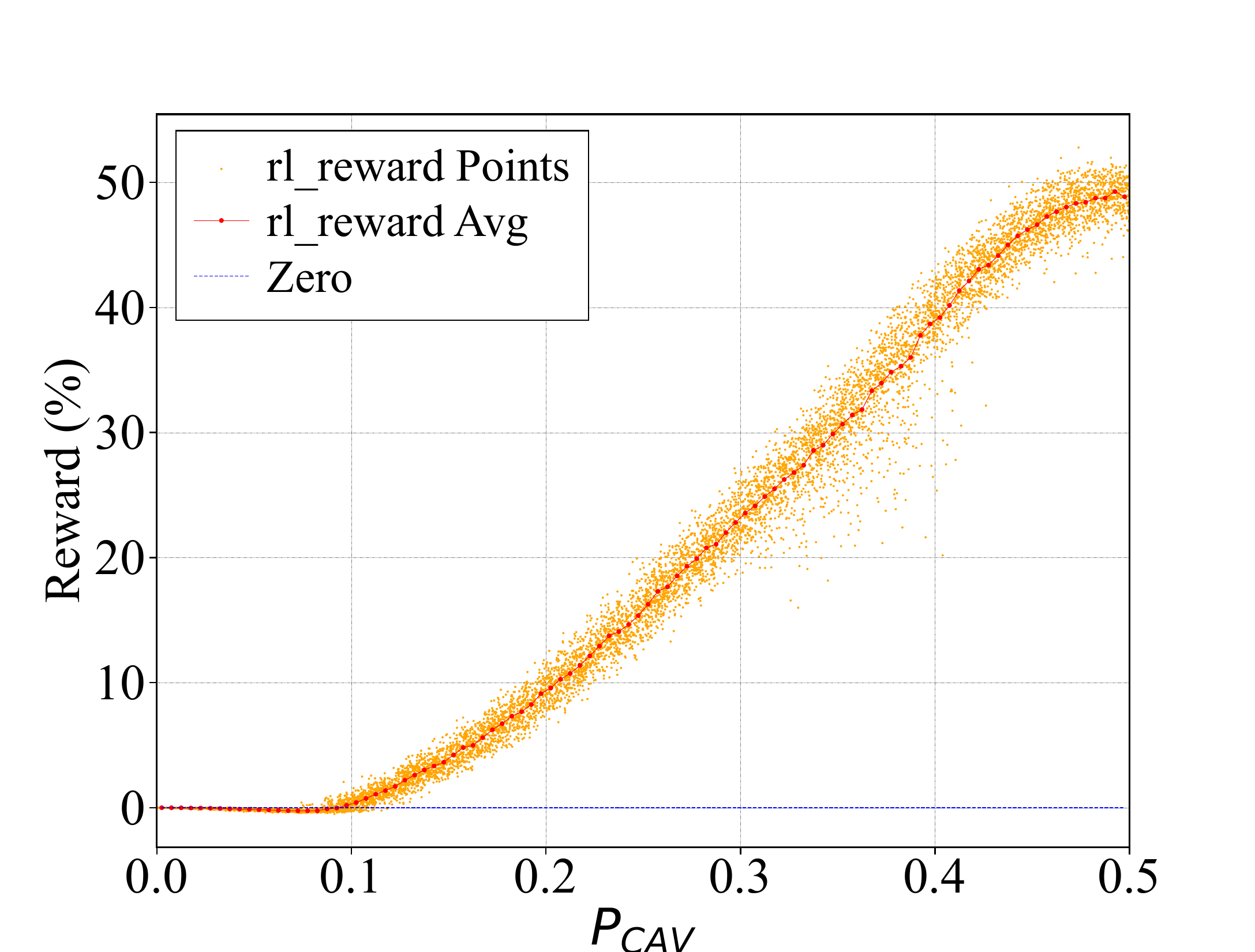}
    \label{fig:gen_reward}
}
\subfloat[Lane capacity]{
    \includegraphics[width=0.48\linewidth]{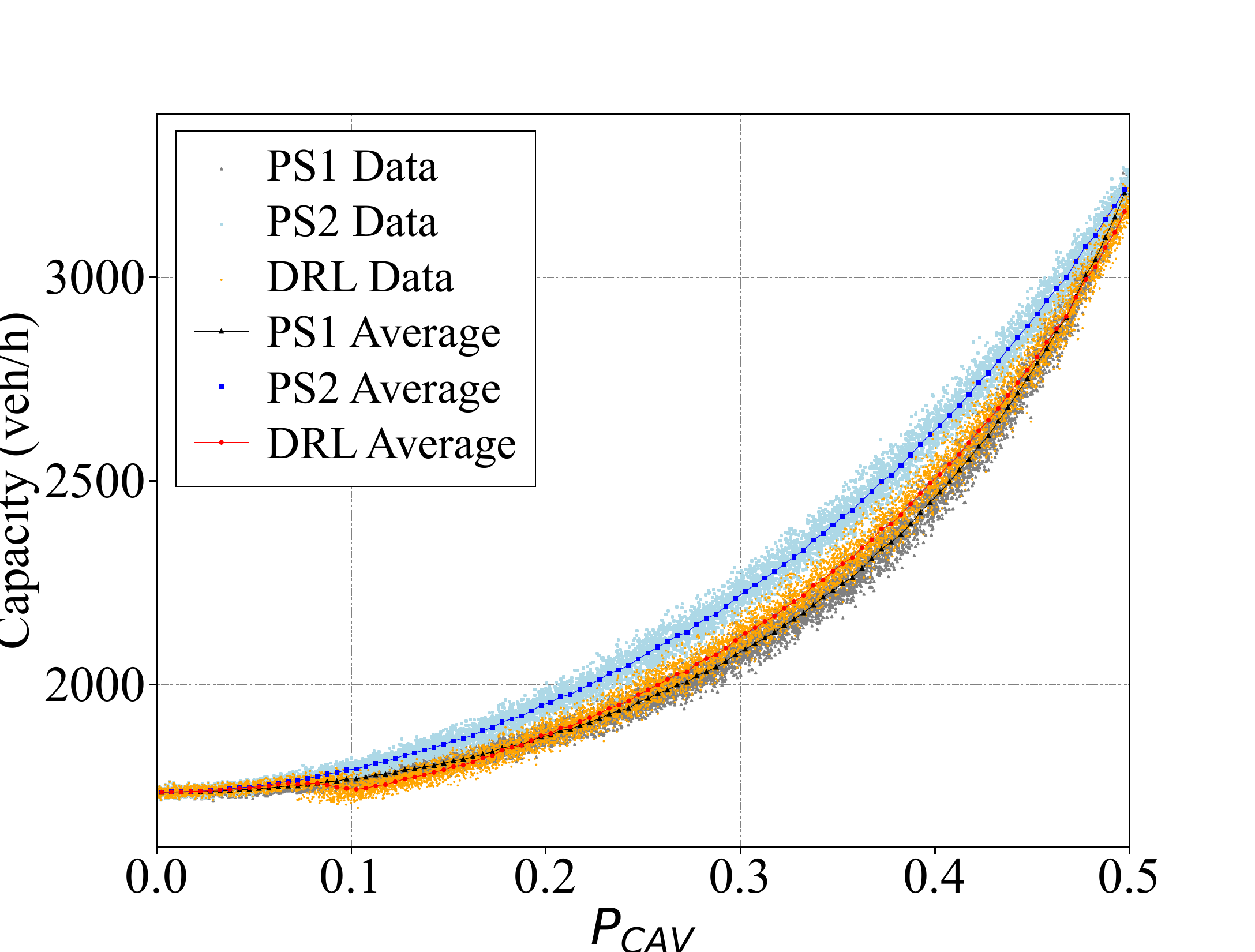}
    \label{fig:gen_capacity}
}
\\
\subfloat[$G_{h}$]{
    \includegraphics[width=0.48\linewidth]{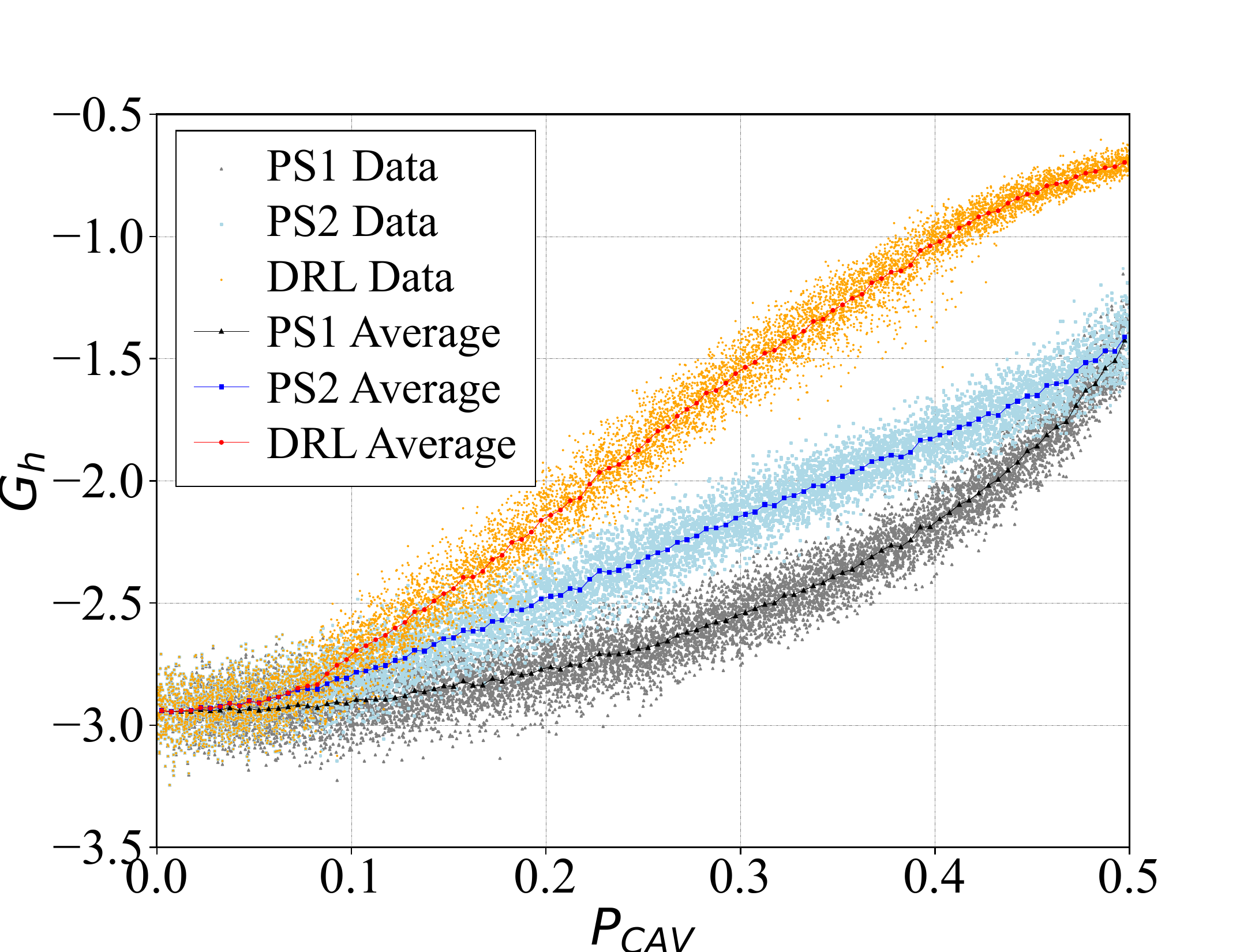}
    \label{fig:gen_gh}
}
\subfloat[Average platoon size]{
    \includegraphics[width=0.48\linewidth]{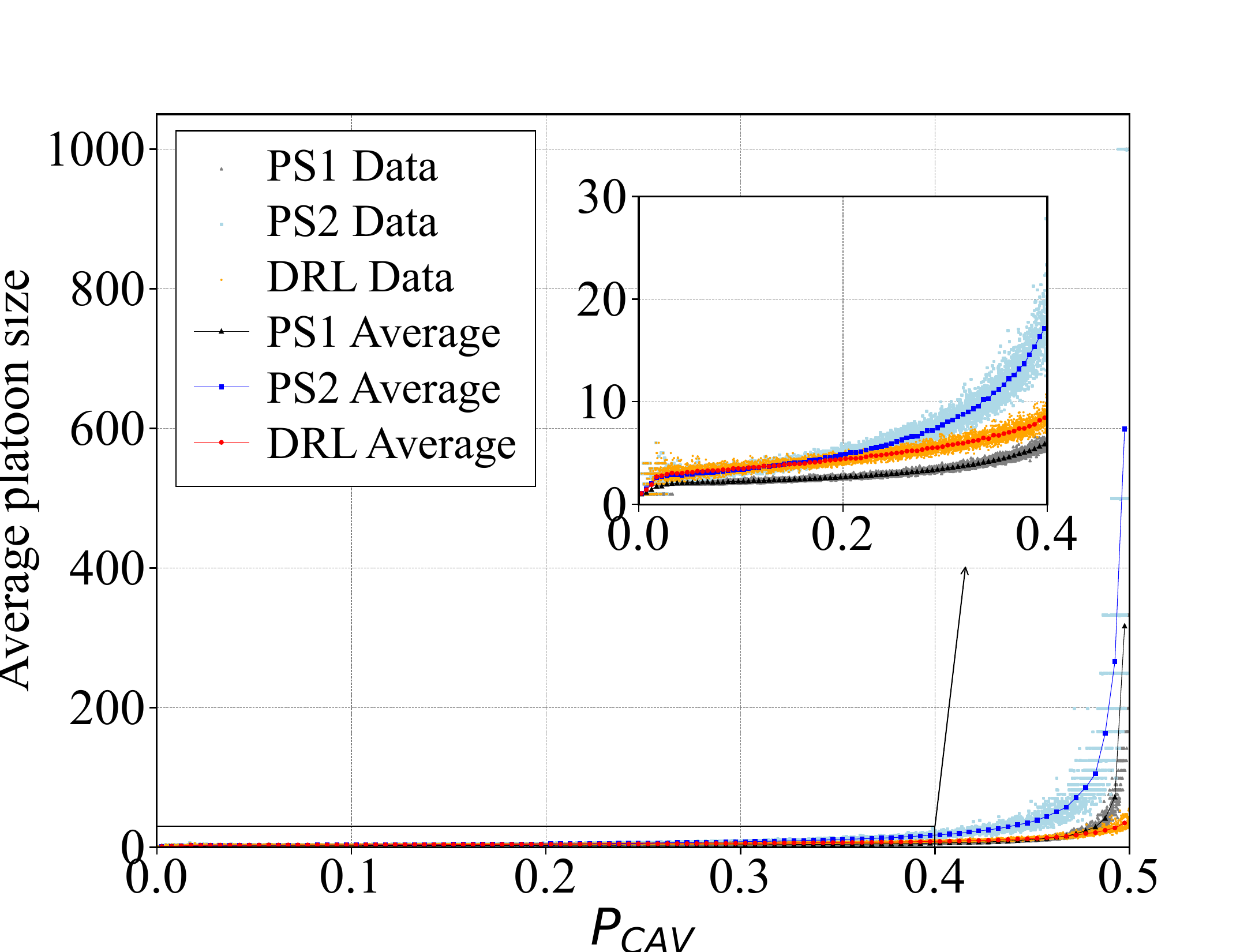}
    \label{fig:gen_avg_pla}
}
\\
\subfloat[Maximum platoon size]{
    \includegraphics[width=0.48\linewidth]{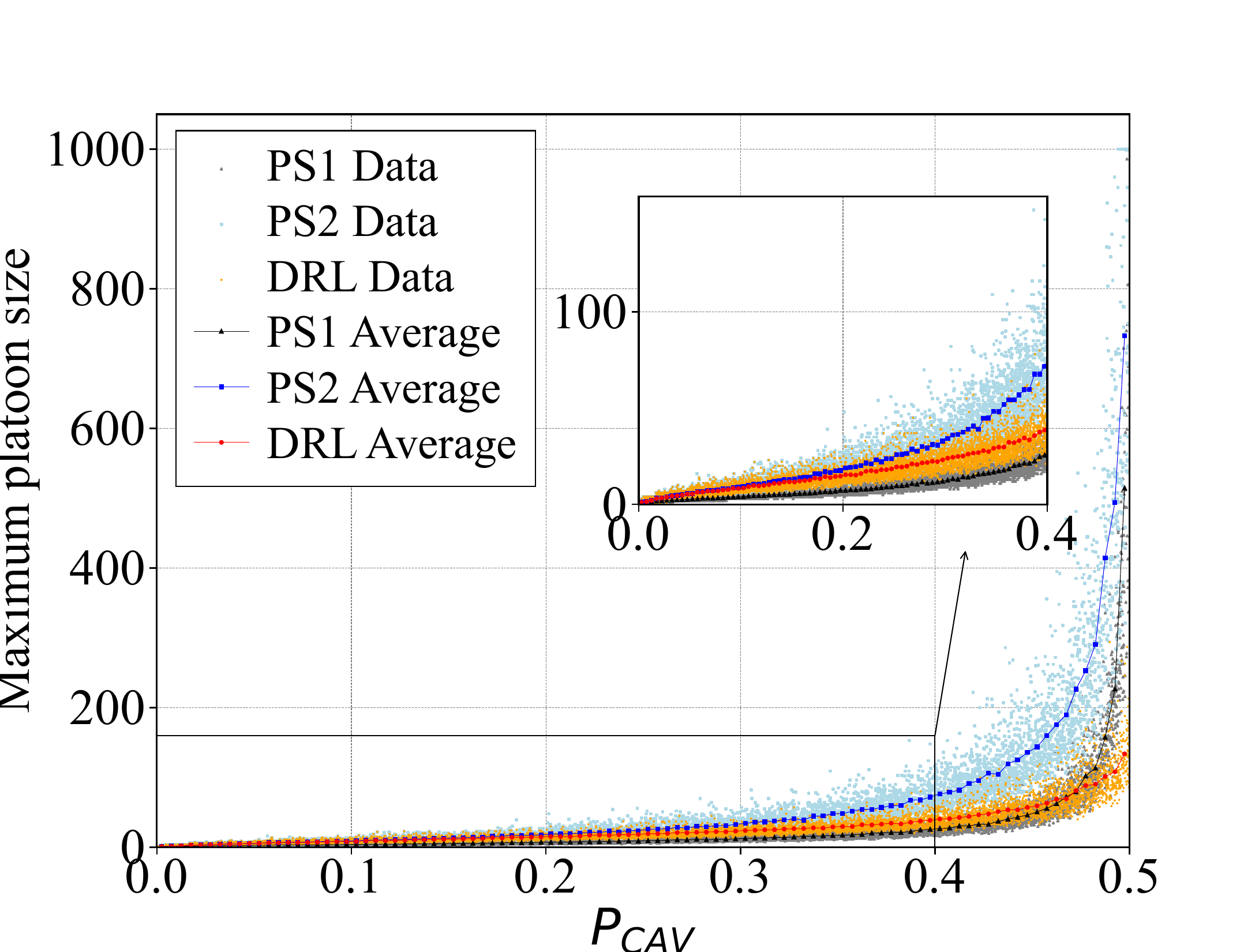}
    \label{fig:gen_max_pla}
}
\subfloat[Total platoon number]{
    \includegraphics[width=0.48\linewidth]{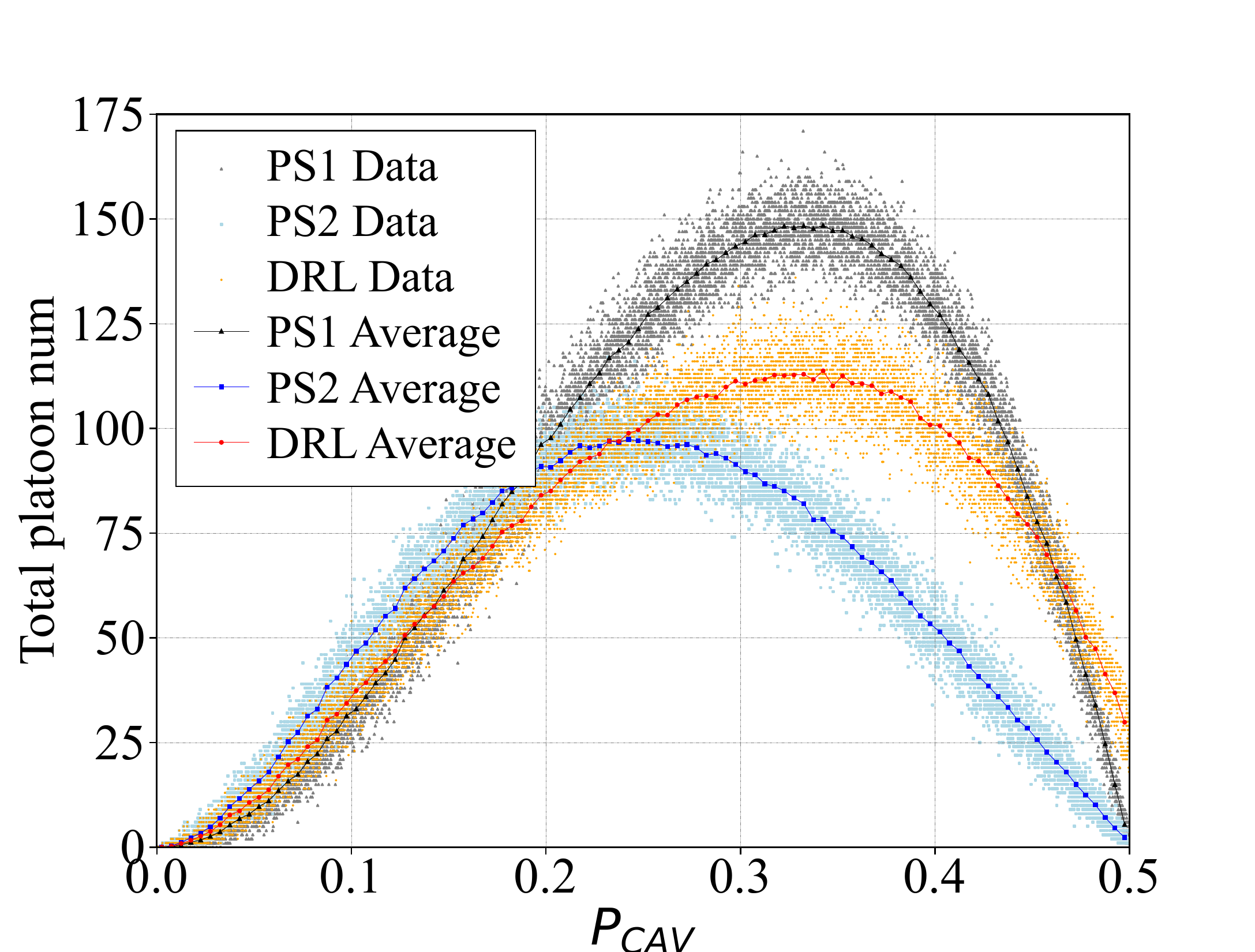}
    \label{fig:gen_pla_num}
}
\caption{Generalization performance of the proposed model under different $P_{\text{CAV}}$ conditions.}
\label{fig:chap4_generalization_ability_pcav}
\end{figure}

%To evaluate the impact of CVs/CAVs penetration rate on the generalization performance of the DRL-based platooning strategy, this study employs 10000 independent randomized experiments. 
The experimental framework configures vehicle penetration parameters as $P_{\mathrm{CAV}} = P_\mathrm{CHV}$, $P_\mathrm{AV} = P_\mathrm{HV}$, with  $P_{\mathrm{CAV}}$ following uniform distribution over $[0.0, 0.5]$. Resultant performance metrics are illustrated in Fig. \ref{fig:chap4_generalization_ability_pcav}.

Fig. \ref{fig:chap4_generalization_ability_pcav}(a) illustrates the evolutionary trend of the reward function, indicating that the reward value approaches $0$ when $P_{\mathrm{CAV}} < 0.1$. Under this condition, the sparse distribution of CHVs and CAVs in mixed traffic flow impedes platoon formation, resulting in negligible performance differences among the three platooning strategies. 
As the CAV penetration rate increases within the range $P_{\mathrm{CAV}} \in (0.1, 0.5)$, the DRL-based strategy demonstrates progressively enhanced optimization of platoon structure, yielding an approximately linear increase in reward value.

Fig. \ref{fig:chap4_generalization_ability_pcav}(b) reveals that lane capacity exhibits a distinct quadratic growth pattern with increasing CAV penetration. 
This phenomenon stems from CAVs' facilitation of platoon formation in mixed traffic flows. Notably, lane capacity shows insignificant variation across the three strategies when $P_{\mathrm{CAV}} < 0.1$ or $P_{\mathrm{CAV}} \rightarrow 0.5$. Within the interval $P_{\mathrm{CAV}} \in (0.2, 0.45)$, the hybrid platooning strategy maximizes collaborative gains through CAVs' information relay and topology reconfiguration capabilities, achieving superior lane capacity compared to both DRL-based and conventional connected platooning strategies. However, the performance advantage comes at the cost of reduced traffic flow stability.

Fig. \ref{fig:chap4_generalization_ability_pcav}(c) demonstrates that the DRL-based platooning strategy delivers a $51.08\%$ improvement in stability discriminant value over both hybrid and conventional connected strategies when $P_{\mathrm{CAV}} > 0.1$. This enhancement is attributed to the strategy's online learning optimization of platoon structures in mixed traffic flows. 
In contrast, hybrid and conventional schemes ($\text{PS}2$ and $\text{PS}1$), constrained by fixed formation patterns, struggle to balance traffic efficiency with flow stability. Crucially, the DRL-based strategy maintains optimal traffic stability across all CAV penetration rates, fundamentally enabled by its integration of centralized optimization's global perspective and distributed control's local adaptability.

Figs. \ref{fig:chap4_generalization_ability_pcav}(d)-(f) further illustrate the evolutionary mechanisms of platoon characteristics in mixed traffic flows in terms of platoon size and number. 
Fig. \ref{fig:chap4_generalization_ability_pcav}(d) demonstrates that under both hybrid and conventional platooning schemes ($\text{PS}2$ and $\text{PS}1$), increased CAV penetration at higher rates induces exponential growth in average platoon length. In contrast, the DRL-based platooning strategy maintains this metric at 35 vehicles through platoon optimization, effectively mitigating length inflation caused by rigid formation patterns in hybrid strategies. 
As shown in Fig. \ref{fig:chap4_generalization_ability_pcav}(e), maximum platoon length exhibits unbounded exponential growth under schemes $\text{PS}2$ and $\text{PS}1$, posing significant stability challenges to mixed traffic flows. The DRL-based platooning strategy, however, prevents irrational expansion of maximum length by incorporating stability reward constraints. 
Fig. \ref{fig:chap4_generalization_ability_pcav}(f) reveals a parabolic trend in platoon quantity, initially increasing then decreasing with rising $P_{\mathrm{CAV}}$. Peak values occur at $P_{\mathrm{CAV}}=0.24$ for both DRL-based strategy and $\text{PS}1$ scheme (97 platoons), while the hybrid scheme $\text{PS}2$ peaks later at $P_{\mathrm{CAV}}=0.34$ (148 platoons).

Collectively, repeated independent experiments verify the generalizability and effectiveness of the DRL-based platooning strategy across diverse penetration scenarios. 
By optimizing platoon structures to reduce both average and maximum lengths, the DRL-based platooning strategy significantly enhances mixed traffic flow stability and achieves robust strategy transferability irrespective of vehicle penetration rates.

\subsection{Stability of traffic flow}
\label{subsec:chap4_traffic_stability}

This study conducts traffic flow simulations under open boundary conditions to quantitatively analyze the impact of three platooning strategies on disturbance propagation characteristics in mixed traffic flows. The experimental setup involves 100 heterogeneous vehicles randomly distributed on a single-lane road, with all vehicles strictly adhering to their respective vehicle-type-specific car-following dynamics and platoon behavioral models.

Under fixed proportions $P_{\mathrm{CAV}} = P_\mathrm{CHV}$, $P_\mathrm{AV} = P_\mathrm{HV}$, four penetration scenarios with $P_{\mathrm{CAV}} = {0.1, 0.2, 0.3, 0.4}$ were tested, respectively. 
Following system initialization, traffic flow operates at steady-state velocity $v_e = 20\ \mathrm{m/s}$ for $600\mathrm{s}$ to achieve dynamic equilibrium. 
Subsequently, a disturbance is applied to the leading vehicle: a constant deceleration of $-1\ \mathrm{m/s^2}$ sustained for $3\mathrm{s}$, followed by acceleration at $1\ \mathrm{m/s^2}$ until $v_e$ is recovered. This disturbance velocity profile is mathematically described as a piecewise linear function:
\begin{equation}
v_{1}(t) =
\left\{
\begin{array}{ll}
v_e, & 0 < t \leq 600 \\
v_e - (t - 600), & 600 < t \leq 603 \\
v_e + (t - 606), & 603 < t \leq 606 \\
v_e, & t > 606 \\
\end{array}
\right.
\end{equation}

\begin{figure}[!htbp]
\centering
% P_CAV = 0.1
\subfloat[DRL, $P_{\text{CAV}} = 0.1$]{
    \includegraphics[width=0.3\linewidth]{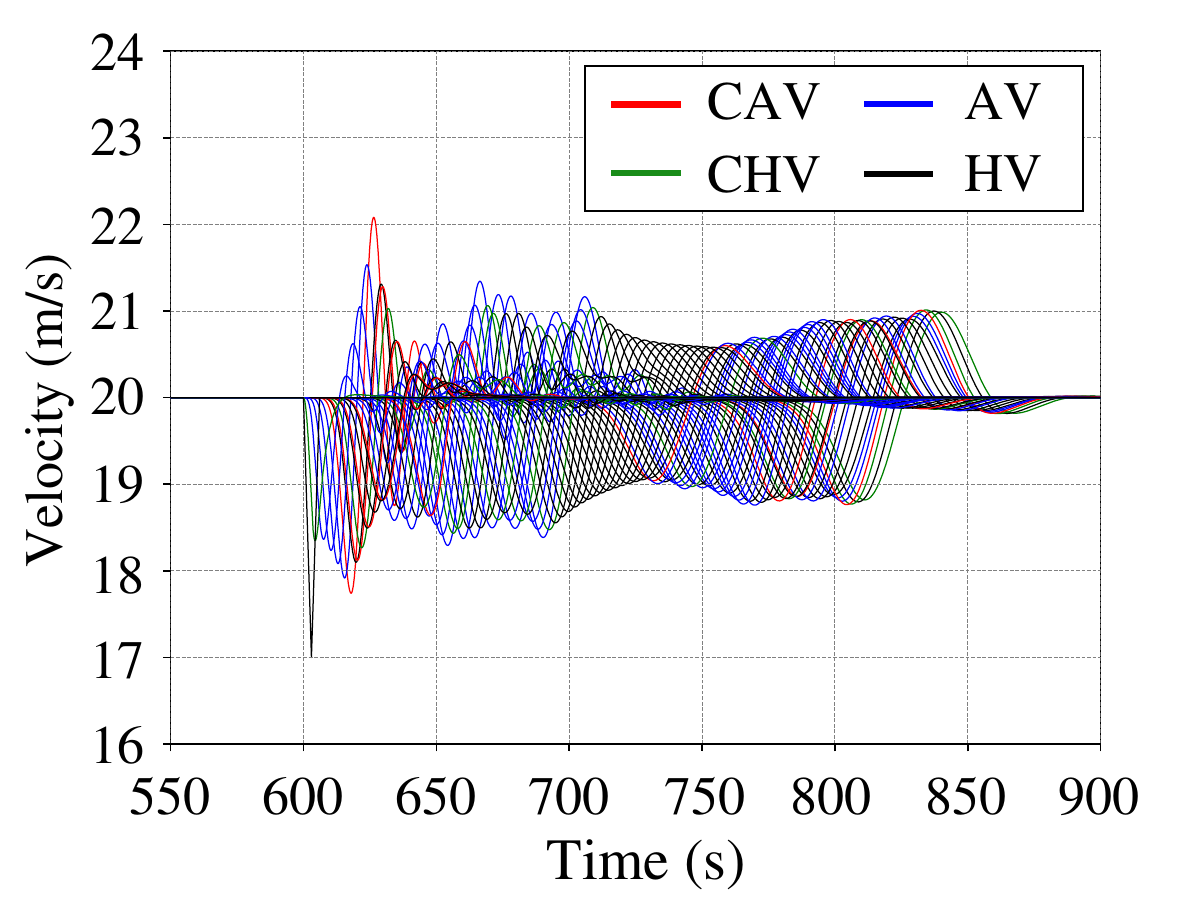}
}
\subfloat[PS2, $P_{\text{CAV}} = 0.1$]{
    \includegraphics[width=0.3\linewidth]{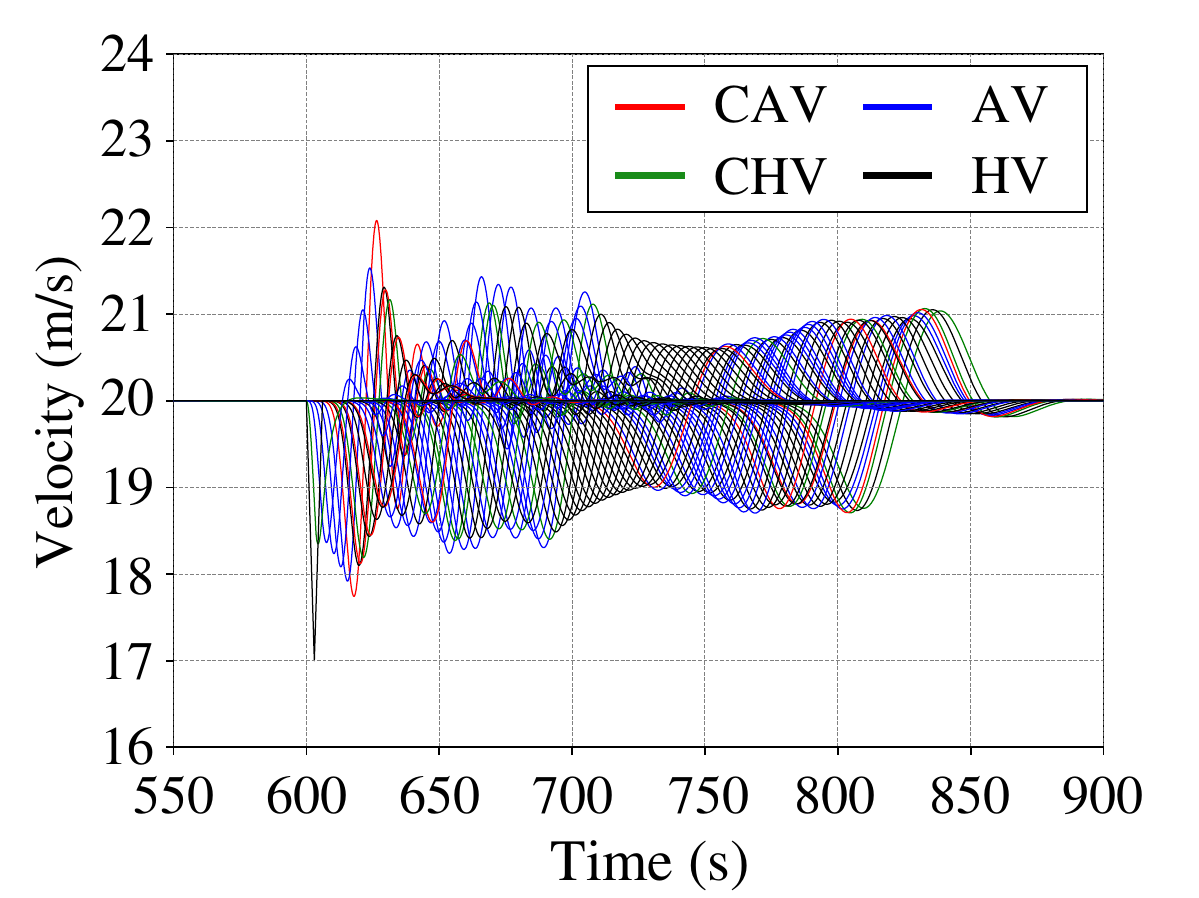}
}
\subfloat[PS1, $P_{\text{CAV}} = 0.1$]{
    \includegraphics[width=0.3\linewidth]{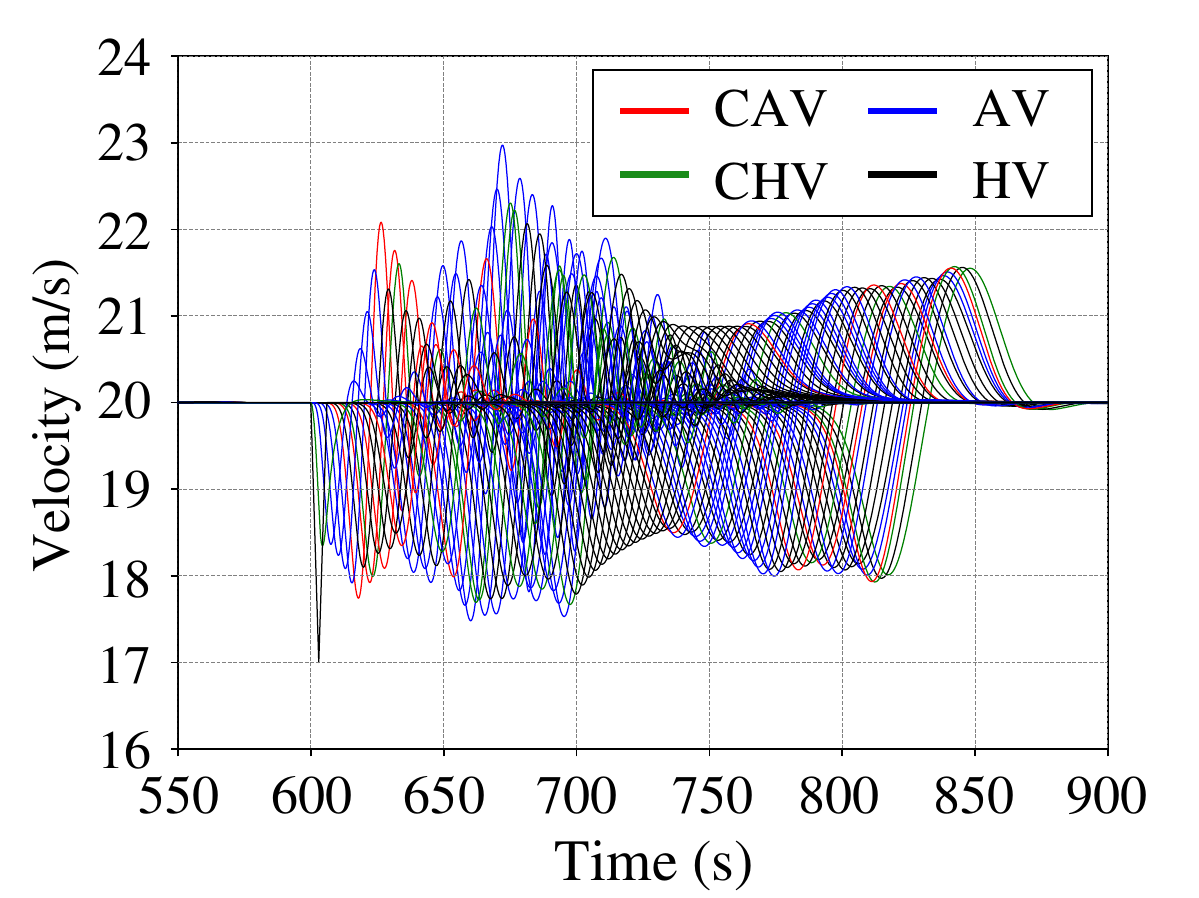}
}\\

% P_CAV = 0.2
\subfloat[DRL, $P_{\text{CAV}} = 0.2$]{
    \includegraphics[width=0.3\linewidth]{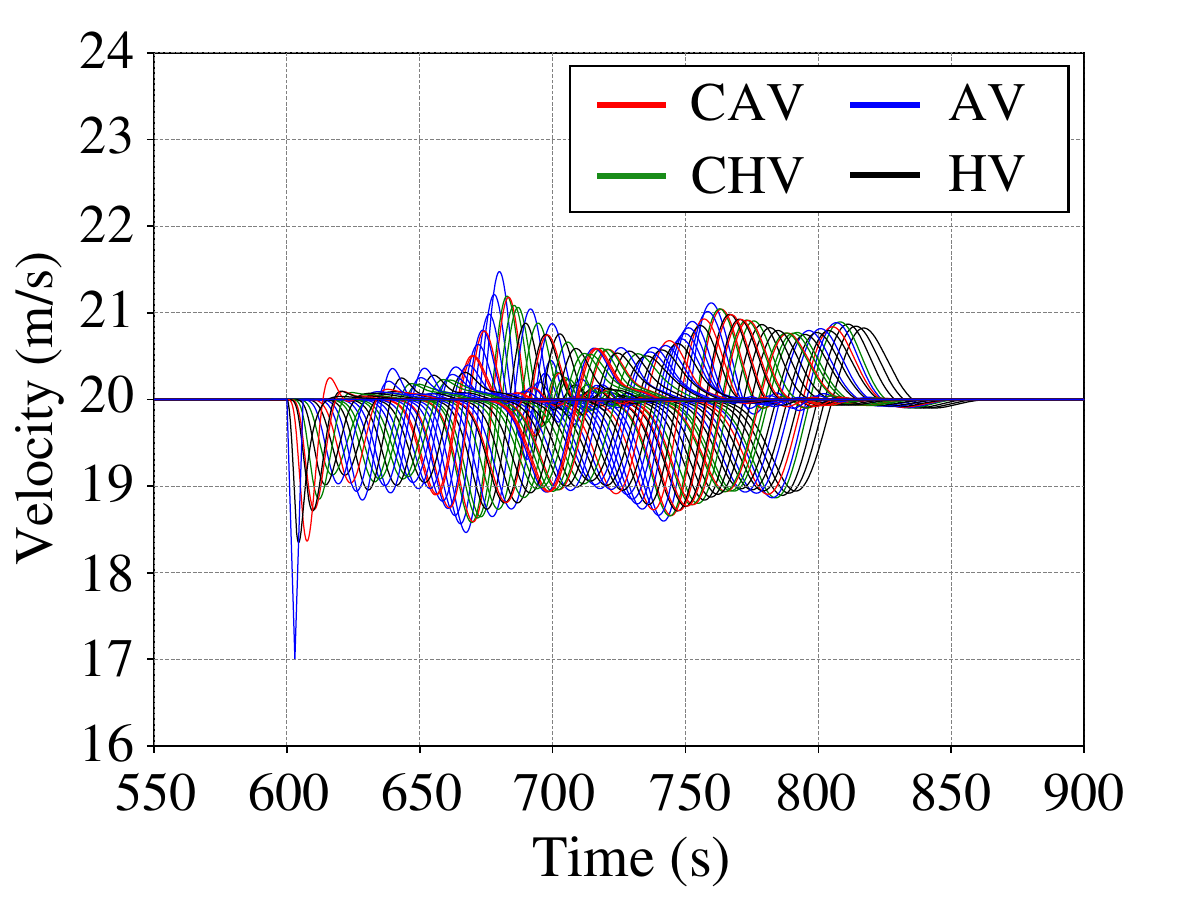}
}
\subfloat[PS2, $P_{\text{CAV}} = 0.2$]{
    \includegraphics[width=0.3\linewidth]{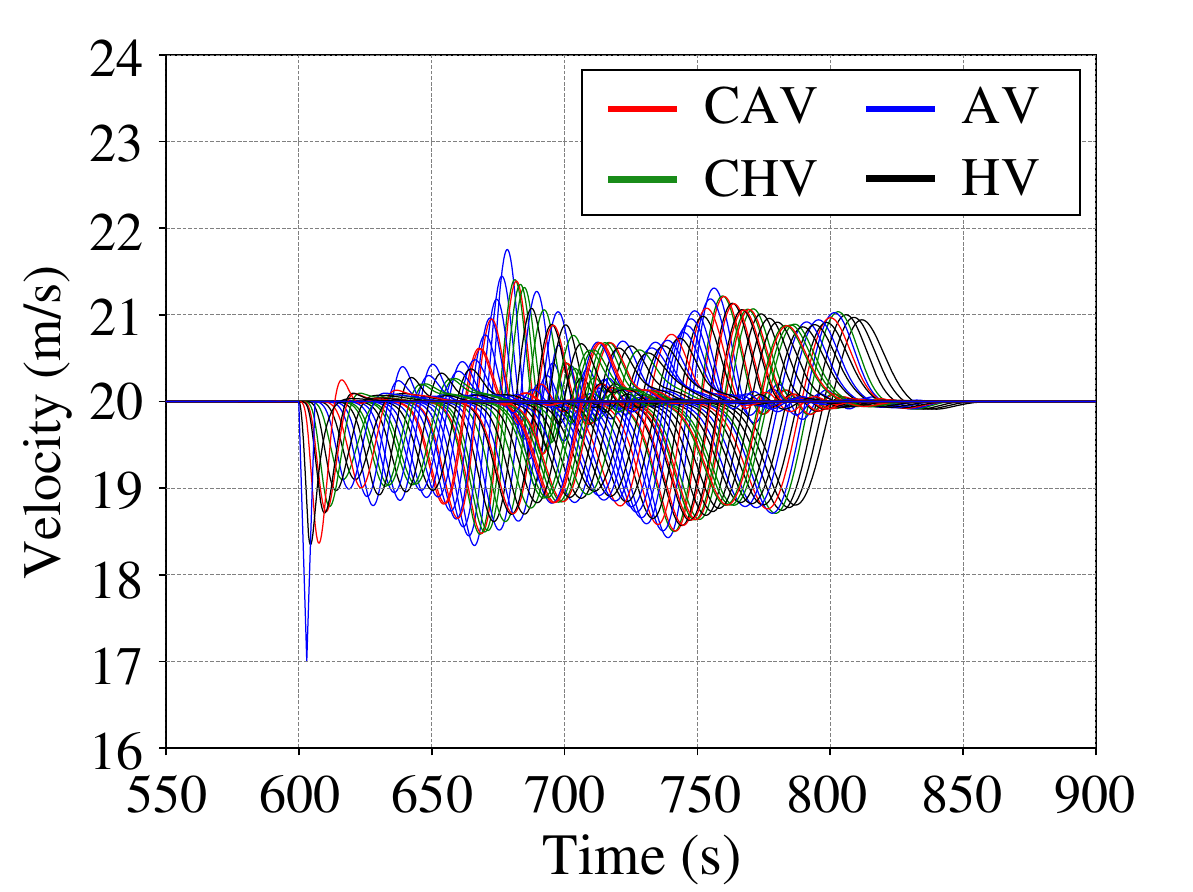}
}
\subfloat[PS1, $P_{\text{CAV}} = 0.2$]{
    \includegraphics[width=0.3\linewidth]{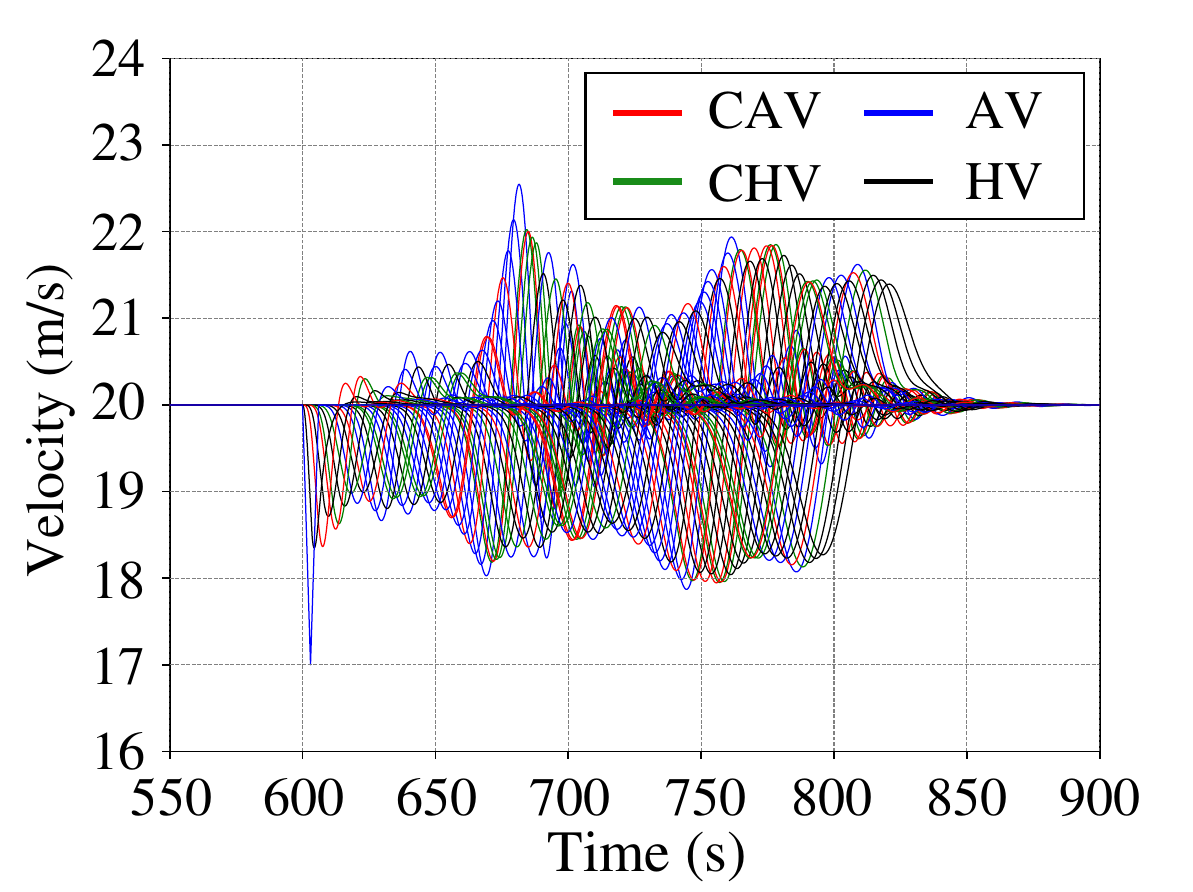}
}\\

% P_CAV = 0.3
\subfloat[DRL, $P_{\text{CAV}} = 0.3$]{
    \includegraphics[width=0.3\linewidth]{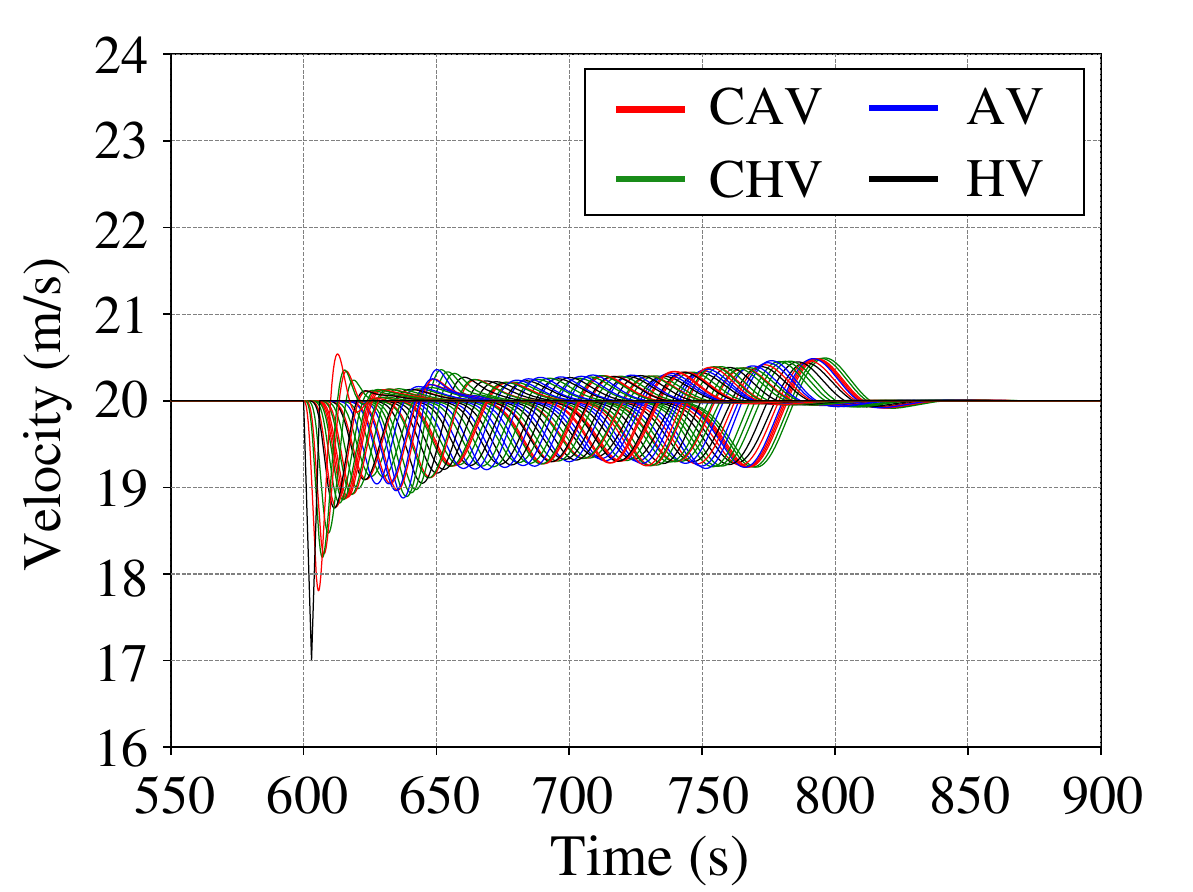}
}
\subfloat[PS2, $P_{\text{CAV}} = 0.3$]{
    \includegraphics[width=0.3\linewidth]{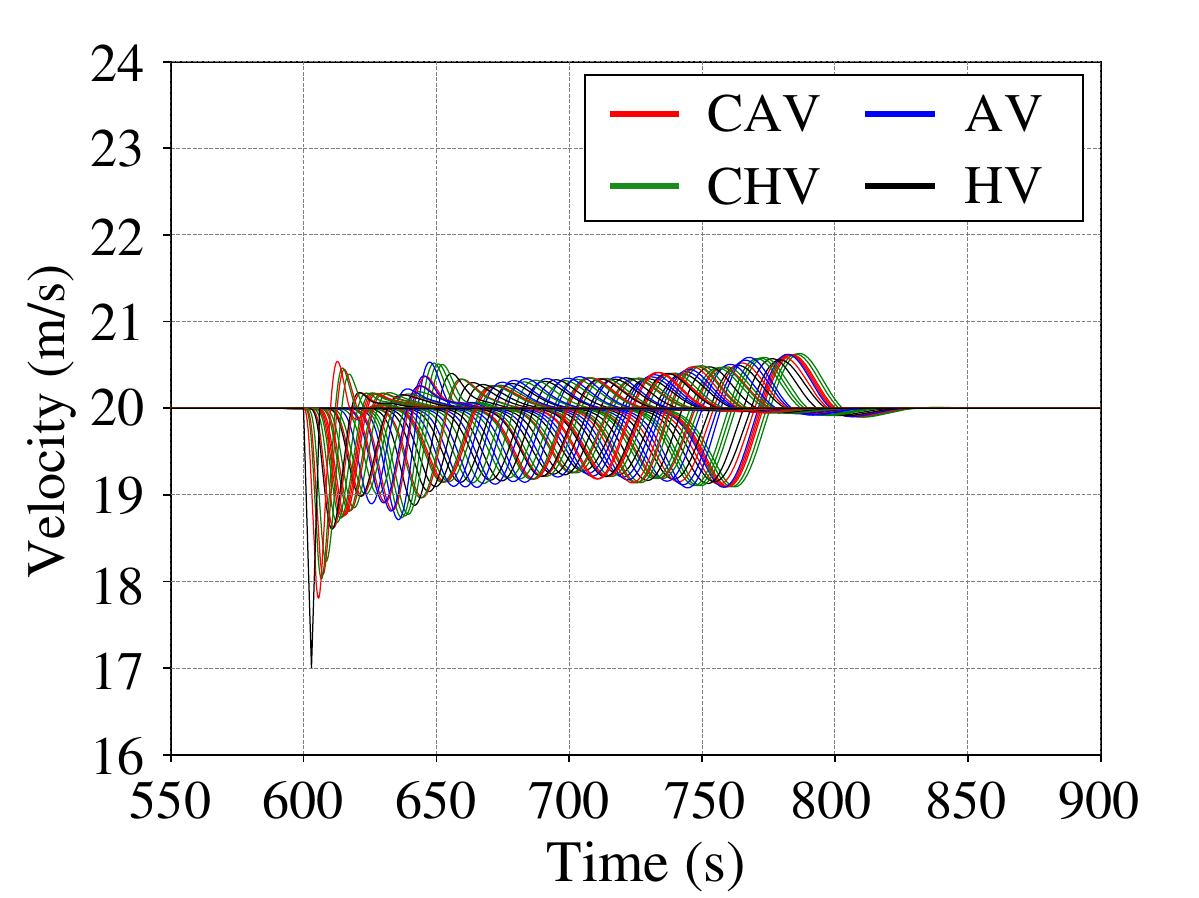}
}
\subfloat[PS1, $P_{\text{CAV}} = 0.3$]{
    \includegraphics[width=0.3\linewidth]{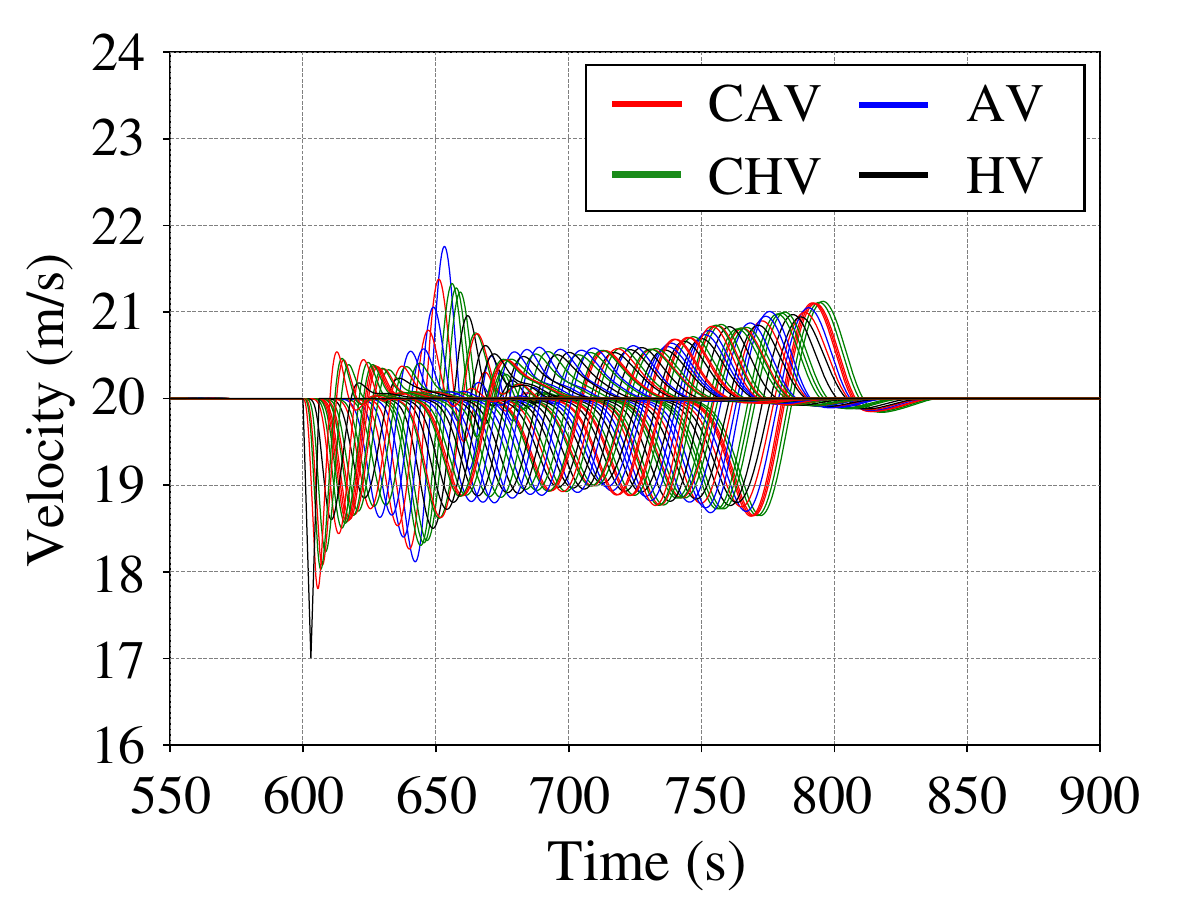}
}\\

% P_CAV = 0.4
\subfloat[DRL, $P_{\text{CAV}} = 0.4$]{
    \includegraphics[width=0.3\linewidth]{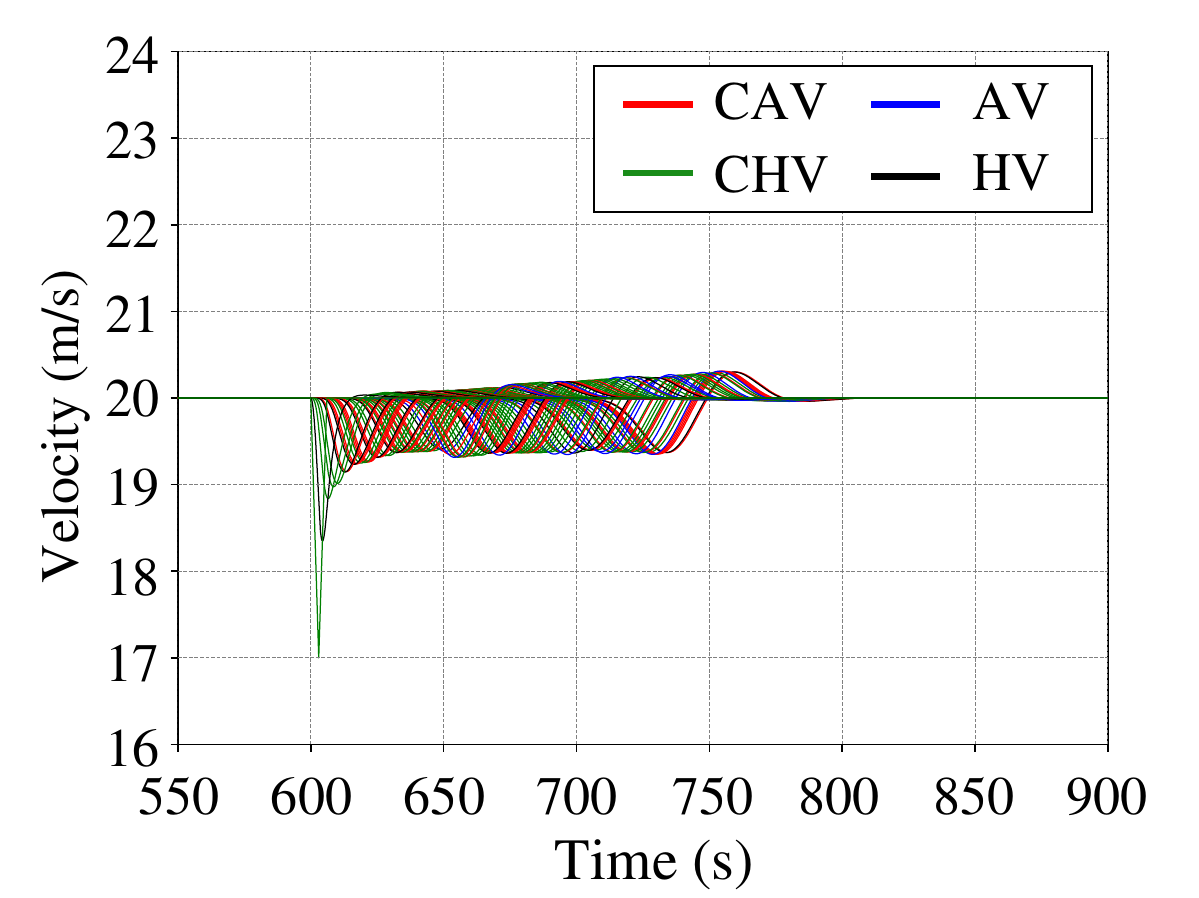}
}
\subfloat[PS2, $P_{\text{CAV}} = 0.4$]{
    \includegraphics[width=0.3\linewidth]{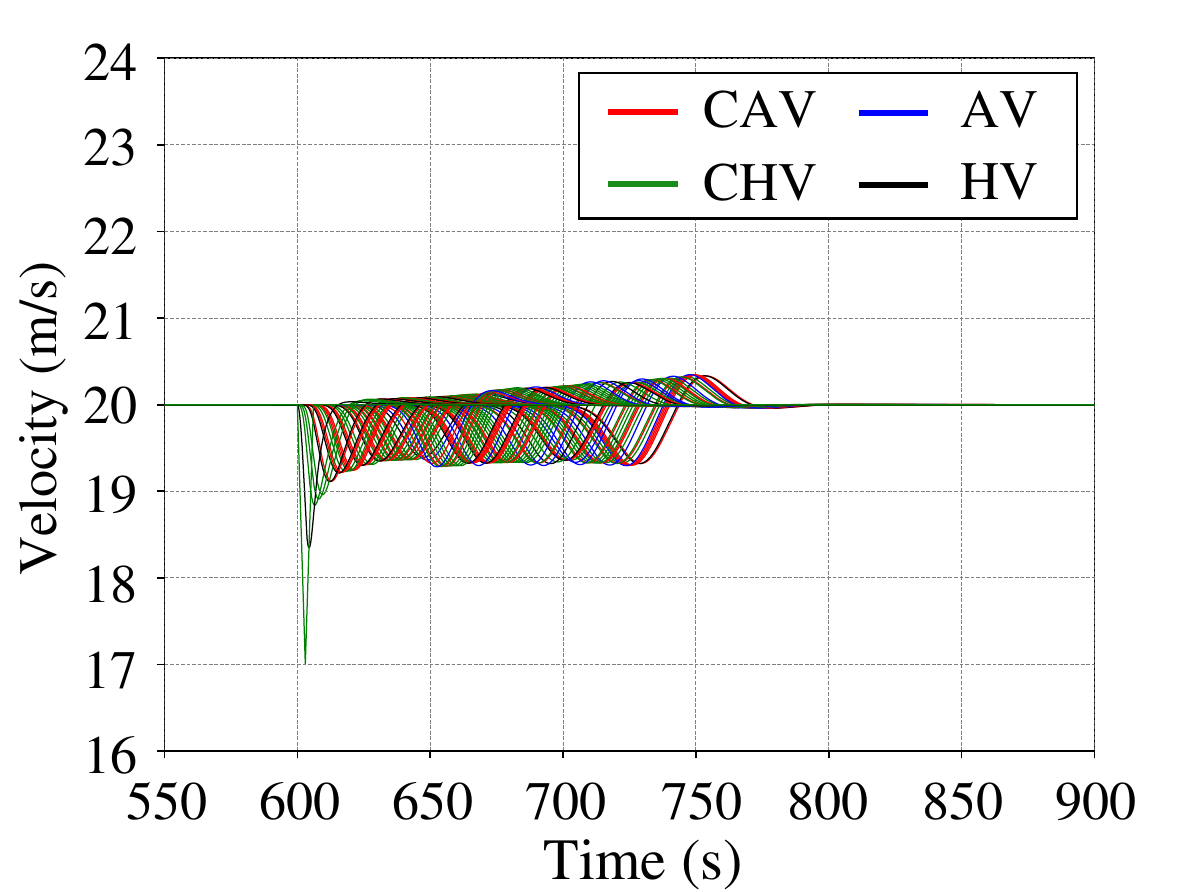}
}
\subfloat[PS1, $P_{\text{CAV}} = 0.4$]{
    \includegraphics[width=0.3\linewidth]{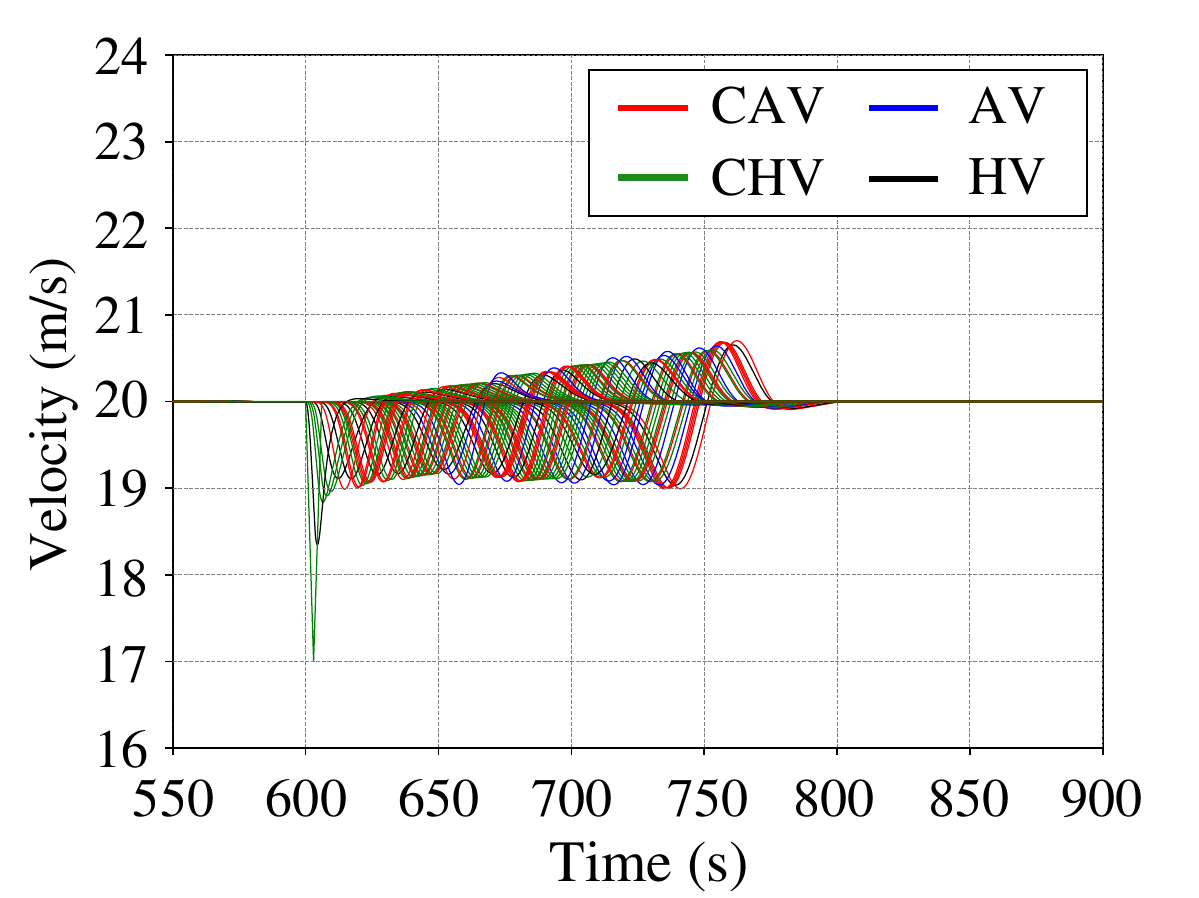}
}

\caption{Velocity responses to head-vehicle perturbations under different penetration rates $P_{\mathrm{CAV}}$.
For $P_{\mathrm{CAV}} = 0.1$, $0.2$, $0.3$, and $0.4$, the three subfigures from left to right correspond to the DRL-based strategy, hybrid strategy (PS2), and conventional strategy (PS1), respectively.
%From top to bottom, $P_{\mathrm{CAV}} = 0.1$, $0.2$, $0.3$, and $0.4$.
}
\label{fig:chap4_stability_velocity}
\end{figure}

Fig. \ref{fig:chap4_stability_velocity}  illustrates the velocity propagation characteristics  of each platooning strategy across four penetration scenarios. 
Experimental data reveal that under conventional connected platooning $\text{PS}1$, steady-state recovery times corresponding to $P_{\mathrm{CAV}}$ values of $0.1$, $0.2$, $0.3$, and $0.4$ measure $300 \mathrm{s}$, $265 \mathrm{s}$, $240 \mathrm{s}$, and $190 \mathrm{s}$ respectively. Reduced recovery times at higher CAV penetration confirm a positive stability-penetration correlation.

As depicted in Figs. \ref{fig:chap4_stability_velocity}(c), (f), and (i), conventional platooning $\text{PS}1$ exhibits significant velocity oscillation amplification in low-penetration scenarios ($P_{\mathrm{CAV}} \leq 0.3$). 
In contrast, both DRL-based strategy (Figs. \ref{fig:chap4_stability_velocity}(a), (d), and (g)) and hybrid scheme $\text{PS}2$ (Figs. \ref{fig:chap4_stability_velocity}(b), (e), and (h)) reduce average velocity fluctuations by approximately 1 m/s. Furthermore, the DRL-based platooning strategy achieves lower velocity  fluctuation compared to the other strategies.

When $P_{\mathrm{CAV}}$ reaches 0.4 (Figs. \ref{fig:chap4_stability_velocity}(j), (k), and (l)), the DRL-based strategy demonstrates notable platoon synergy effects, maintaining smaller velocity and acceleration fluctuations while enabling faster recovery to pre-disturbance steady-state operation.

Collectively, these experiments validate the effectiveness of DRL-based platooning in controlling disturbance propagation within mixed traffic environments through dual perspectives: velocity oscillation mitigation and acceleration response optimization.

\subsection{Traffic safety}

To evaluate the impact of platooning strategies on the safety of mixed traffic, this study employs the inverse time to collision (TTCI) and velocity standard deviation (SD) as safety metrics \citep{yao2023analysis,LI_2024_safety_Enhancing,HUA_2023_sd_Impact,LUO_2024_safety_Modeling}. The TTCI dynamically quantifies longitudinal collision risk for vehicle $i$ at time $t$, which is:
\begin{equation}
\mathrm{TTCI}_i(t) =
\left\{
\begin{array}{ll}
\displaystyle\frac{v_{i}(t) - v_{i-1}(t)}{x_{i-1}(t) - x_{i}(t)}, & v_{i}(t) > v_{i-1}(t) \\
0, & \text{Otherwise}
\end{array}
\right.
\end{equation}
where $x_{i}(t)$ and $v_{i}(t)$ denote the position and velocity of vehicle $i$ at time $t$, respectively. This dimensionally consistent metric assesses collision risk by calculating the velocity differential relative to inter-vehicle spacing, with higher values indicating elevated collision probability.

The velocity SD is mathematically defined as:
\begin{equation}
\mathrm{SD}(t) = \sqrt{\frac{1}{N}\sum_{i=1}^{N}\left(v_{i}(t) - \overline{v}(t)\right)^2}
\end{equation}
where $\overline{v}(t)$ denotes the space-mean velocity of traffic flow at time $t$. This metric quantifies velocity dispersion within the traffic stream, with its temporal evolution capturing spatiotemporal patterns of velocity oscillations during disturbance propagation.

\begin{figure}[!htbp]
\centering
\subfloat[$P_{\text{CAV}} = 0.1$]{
    \includegraphics[width=0.45\linewidth]{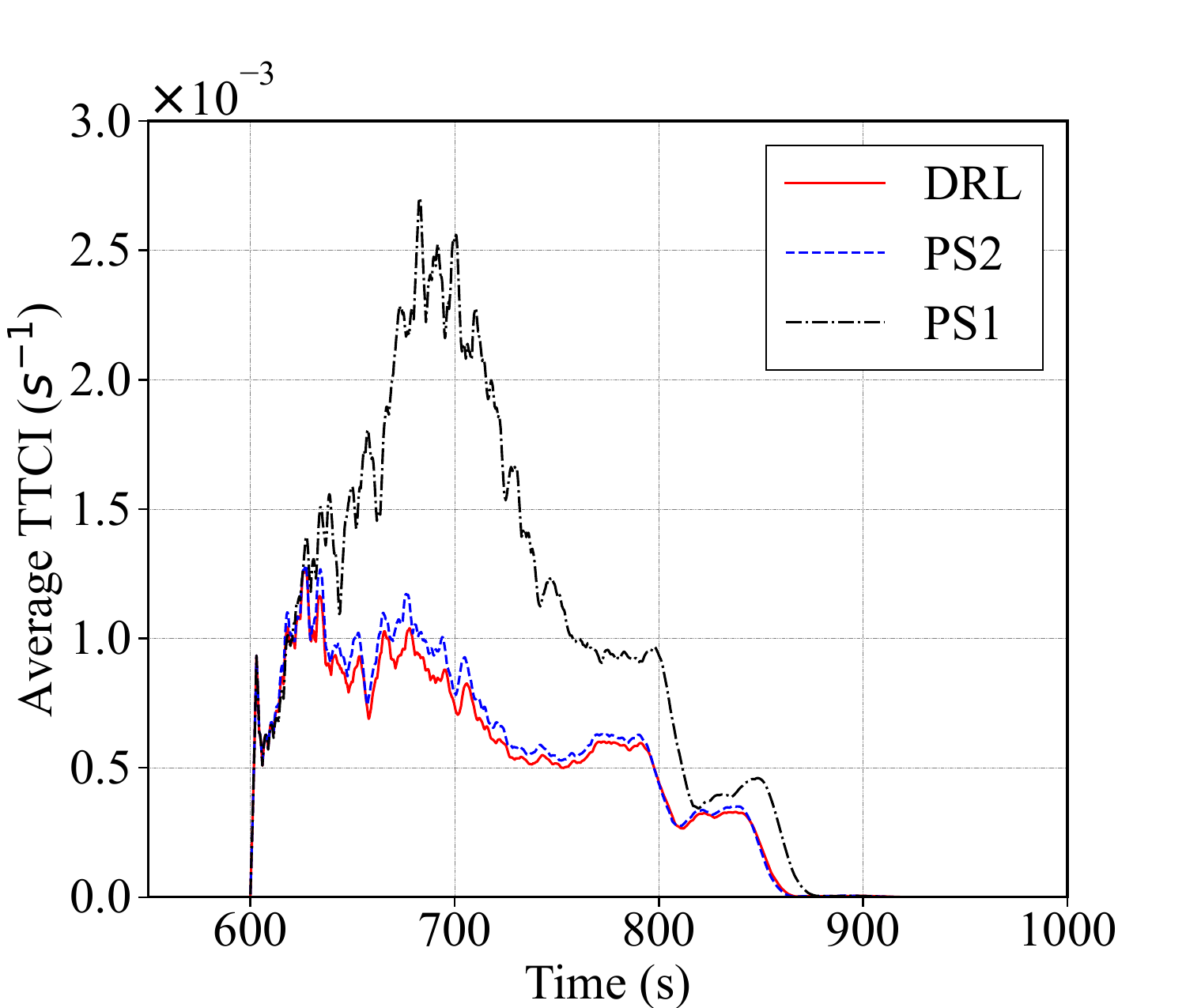}
}
\subfloat[$P_{\text{CAV}} = 0.2$]{
    \includegraphics[width=0.45\linewidth]{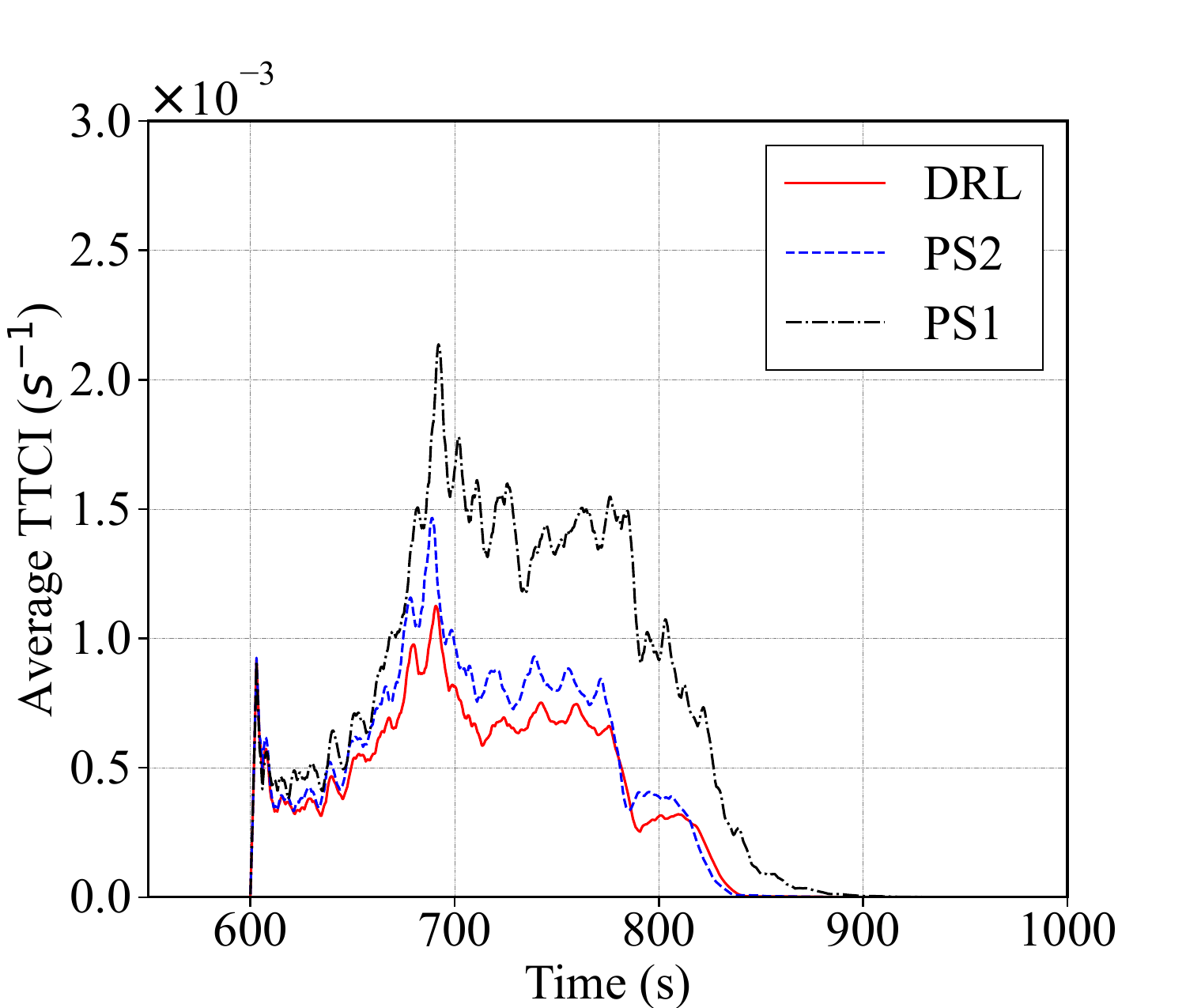}
}\\
\subfloat[$P_{\text{CAV}} = 0.3$]{
    \includegraphics[width=0.45\linewidth]{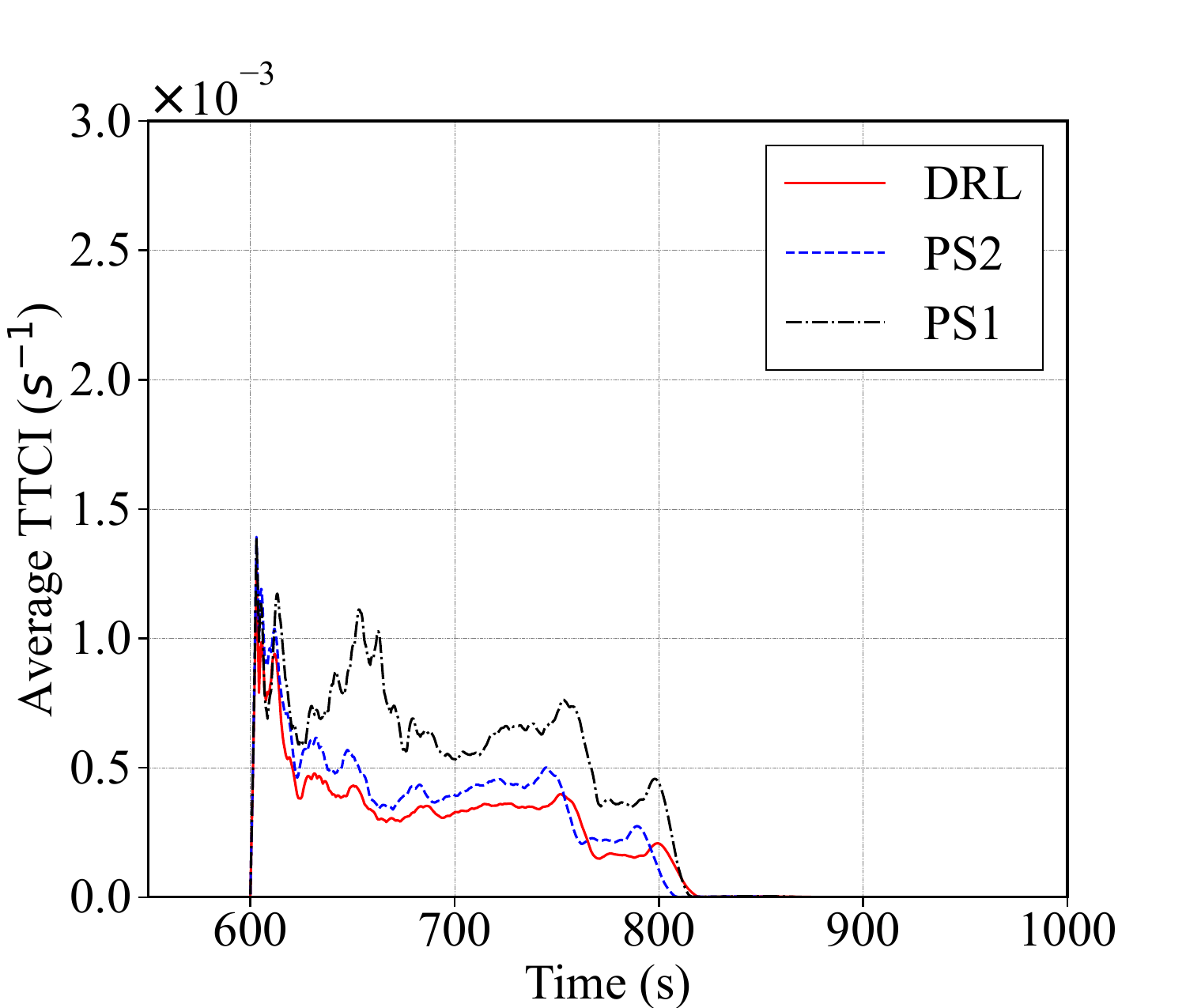}
}
\subfloat[$P_{\text{CAV}} = 0.4$]{
    \includegraphics[width=0.45\linewidth]{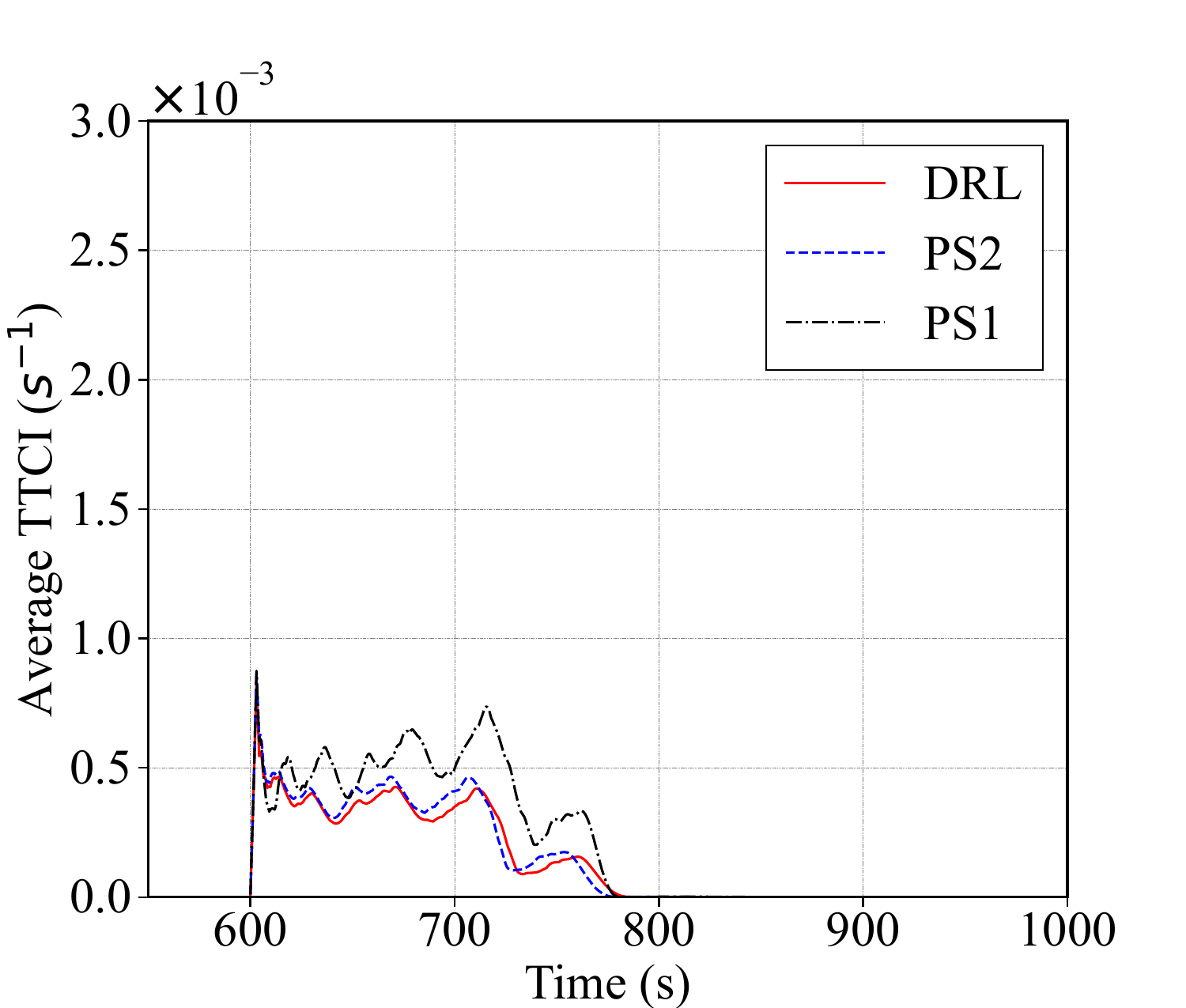}
}
\caption{Average inverse time-to-collision (TTCI) under different penetration rates $P_{\text{CAV}}$. 
%From top left to bottom right, $P_{\text{CAV}} = 0.1$, $0.2$, $0.3$, and $0.4$, respectively.
}
\label{fig:chap4_avg_ttci}
\end{figure}

\begin{figure}[!t]
\centering
\subfloat[$P_{\text{CAV}} = 0.1$]{
    \includegraphics[width=0.45\linewidth]{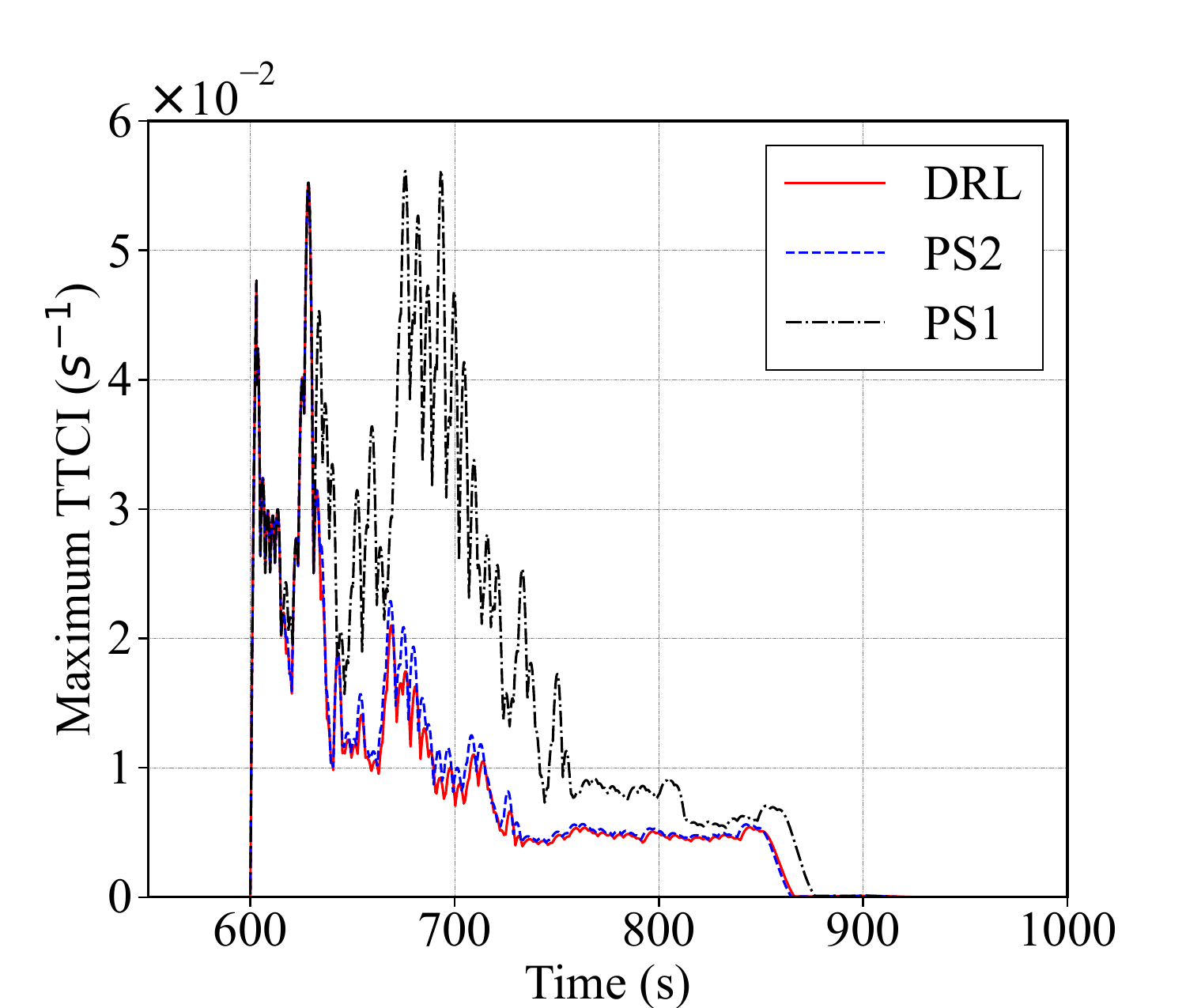}
}
\subfloat[$P_{\text{CAV}} = 0.2$]{
    \includegraphics[width=0.45\linewidth]{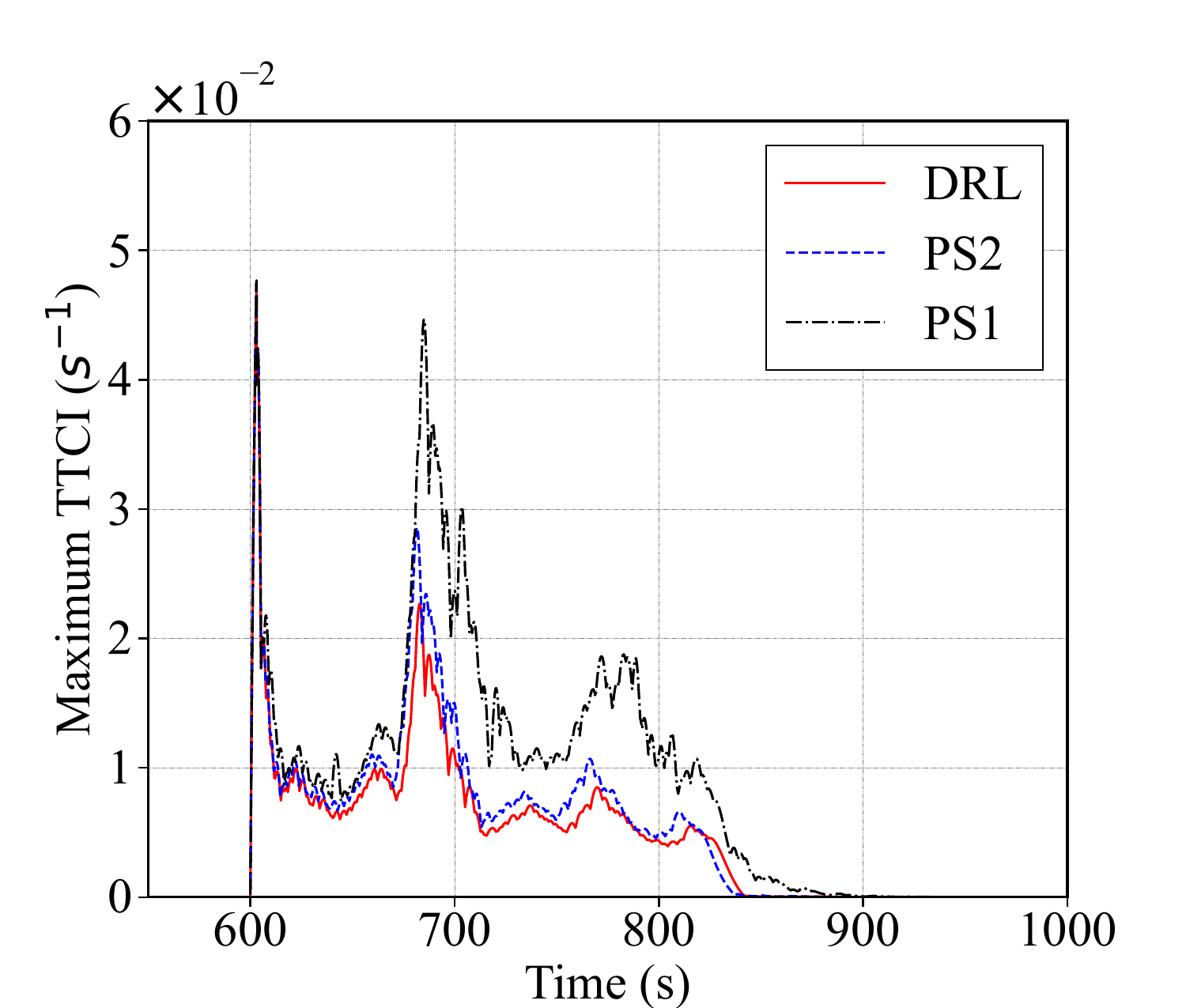}
}\\
\subfloat[$P_{\text{CAV}} = 0.3$]{
    \includegraphics[width=0.45\linewidth]{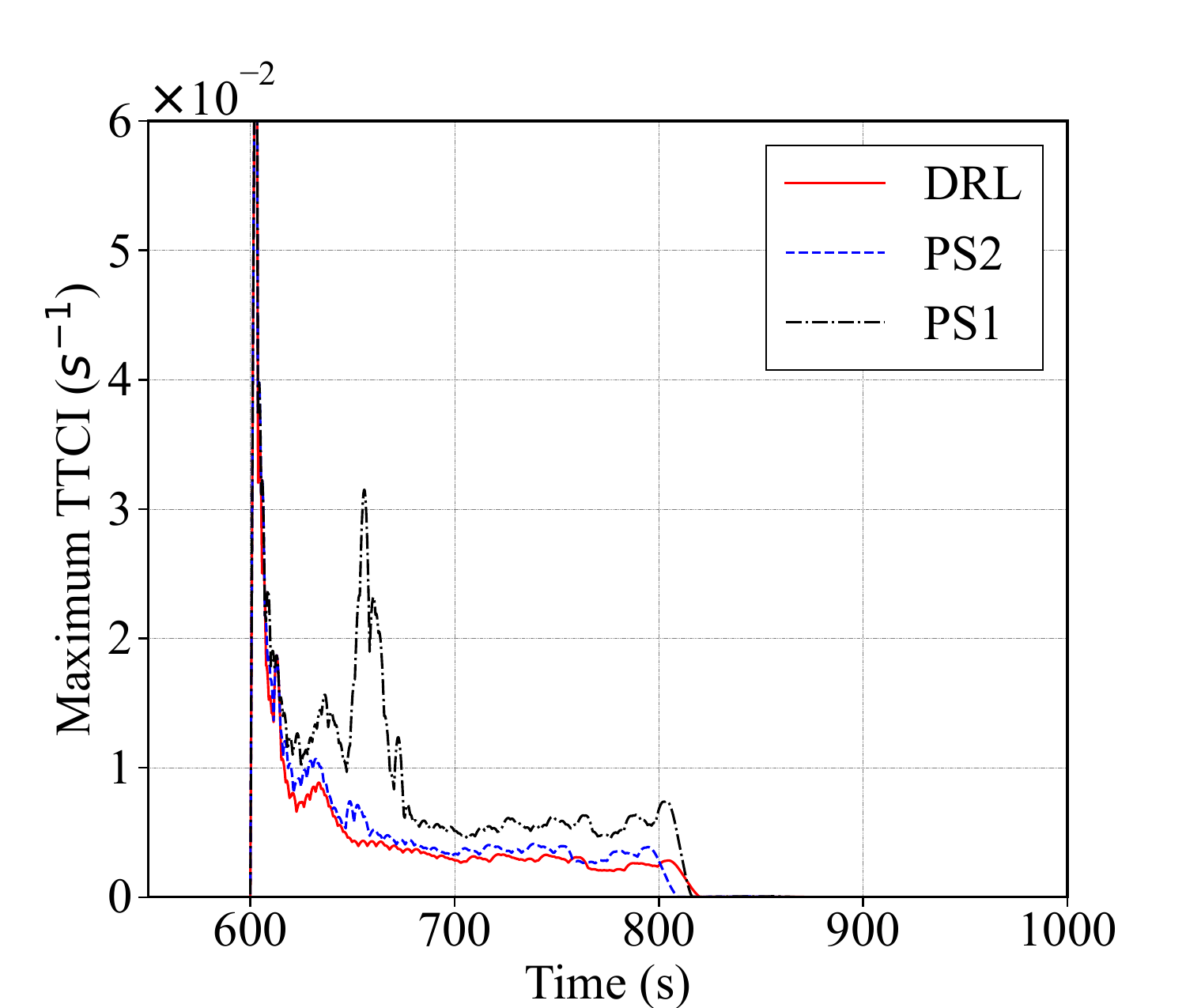}
}
\subfloat[$P_{\text{CAV}} = 0.4$]{
    \includegraphics[width=0.45\linewidth]{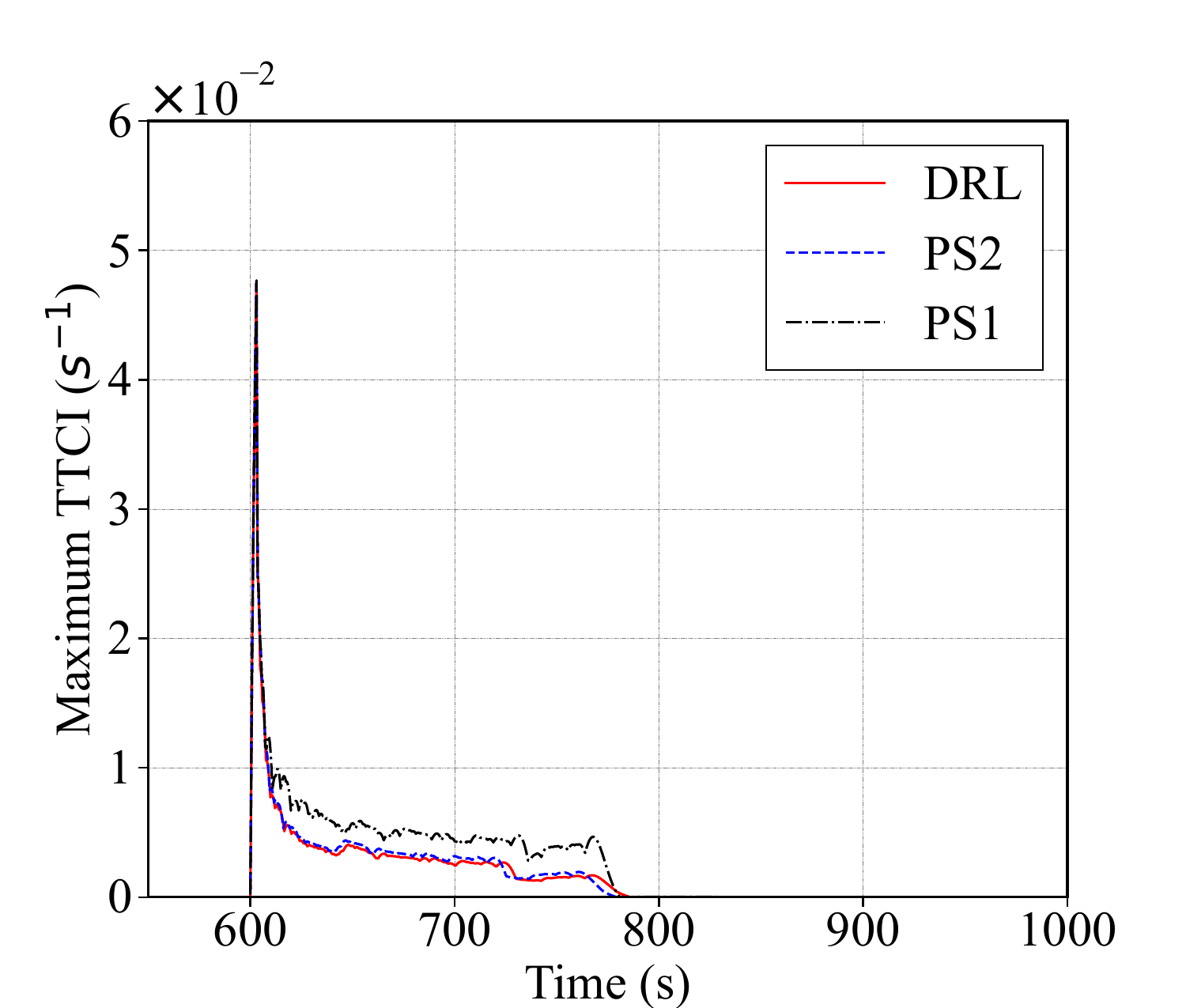}
}
\caption{Peak inverse time-to-collision (TTCI) under different penetration rates $P_{\text{CAV}}$. 
%From top left to bottom right, $P_{\text{CAV}} = 0.1$, $0.2$, $0.3$, and $0.4$, respectively.
}
\label{fig:chap4_max_ttci}
\end{figure}

Similar to the simulation scenarios in Section \ref{subsec:chap4_traffic_stability}, Figs. \ref{fig:chap4_avg_ttci} and \ref{fig:chap4_max_ttci} present the dynamic responses of average ($\overline{\mathrm{TTCI}}$) and maximum ($\mathrm{TTCI}_\mathrm{max}$) with various penetration rates. Experimental data indicate that under high CAV penetration ($P_{\mathrm{CAV}}=0.4$), all three platooning strategies maintain $\overline{\mathrm{TTCI}}$ below $0.001\ \mathrm{s^{-1}}$, confirming CAV cooperative control's effectiveness in suppressing risk propagation. 
Notably, the DRL-based strategy demonstrates significant advantages at low penetration ($P_{\mathrm{CAV}}=0.1$), reducing peak $\overline{\mathrm{TTCI}}$ by $52.95\%$ compared to conventional platooning $\text{PS}1$ and by $23.21\%$ versus the hybrid strategy $\text{PS}2$.

Regarding extreme collision risks, Fig. \ref{fig:chap4_max_ttci}(a) reveals violent $\mathrm{TTCI}_\mathrm{max}$ oscillations under conventional platooning $\text{PS}1$ at $P_{\mathrm{CAV}}=0.1$, indicating collision risk accumulation with transient peaks reaching $0.0562\ \mathrm{s^{-1}}$. In contrast, the DRL-based strategy maintains stable extreme risk values, demonstrating superior control over critical collision scenarios.

\begin{figure}[!htbp]
\centering
\subfloat[$P_{\text{CAV}} = 0.1$]{
    \includegraphics[width=0.45\linewidth]{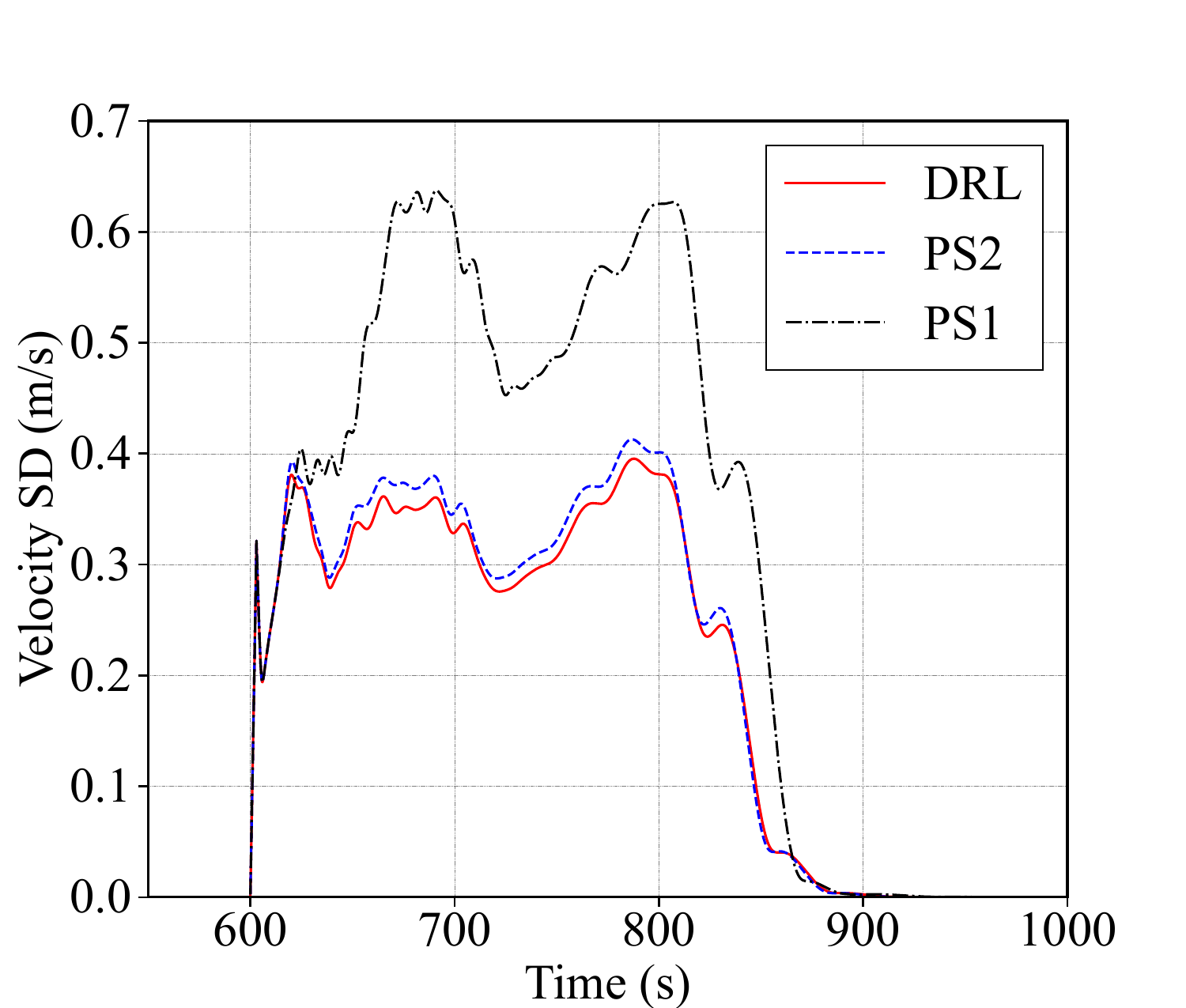}
}
\subfloat[$P_{\text{CAV}} = 0.2$]{
    \includegraphics[width=0.45\linewidth]{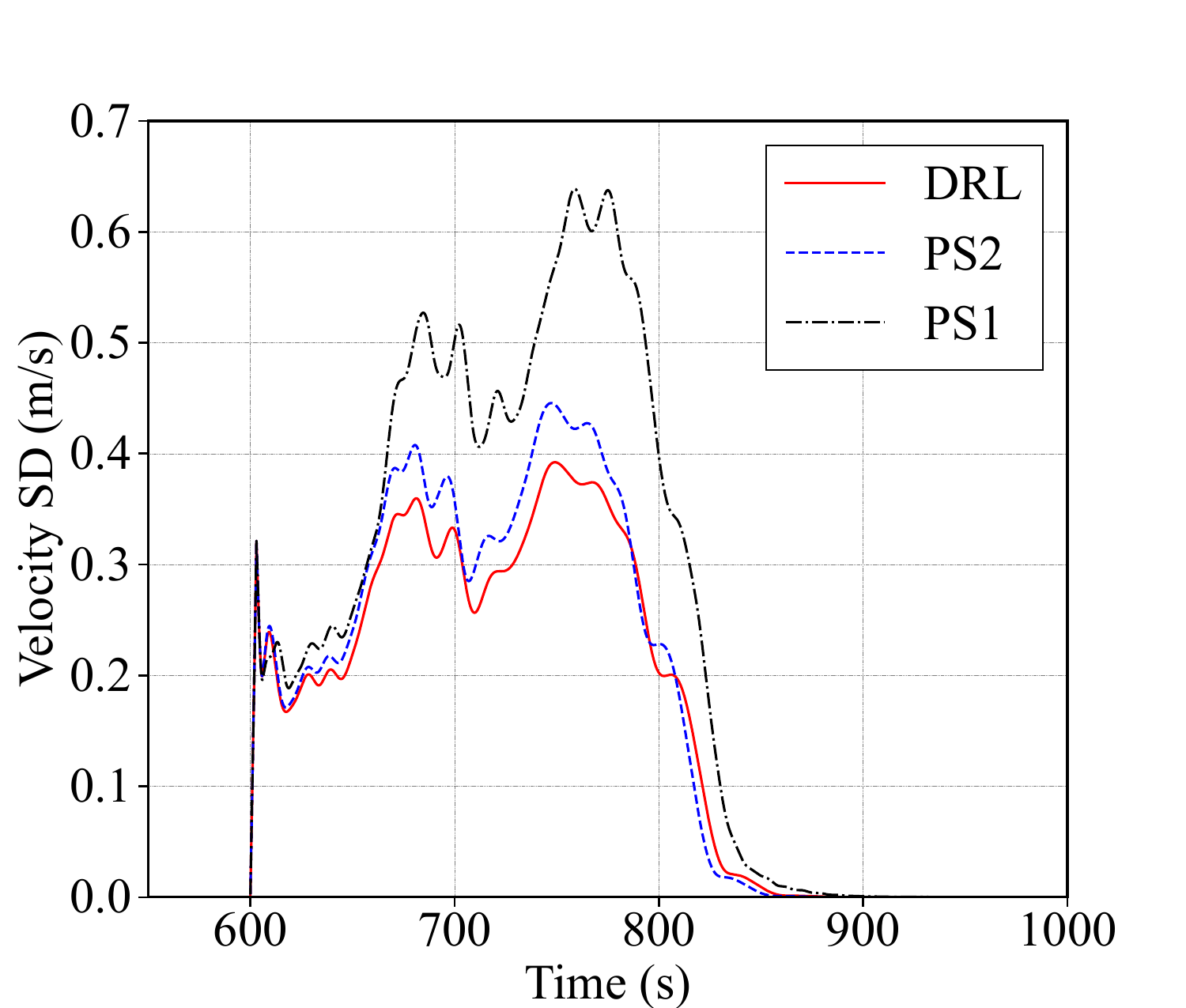}
}\\
\subfloat[$P_{\text{CAV}} = 0.3$]{
    \includegraphics[width=0.45\linewidth]{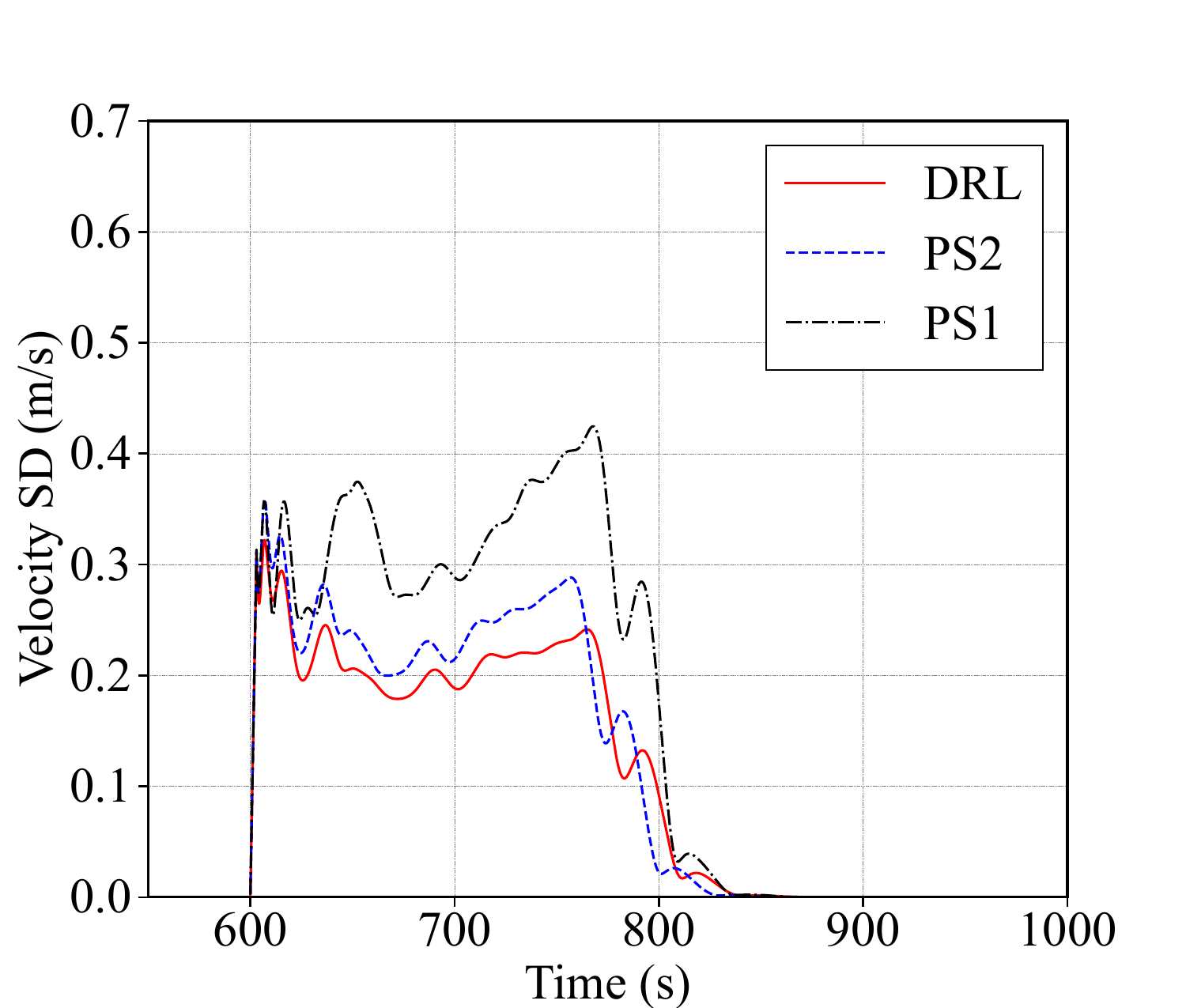}
}
\subfloat[$P_{\text{CAV}} = 0.4$]{
    \includegraphics[width=0.45\linewidth]{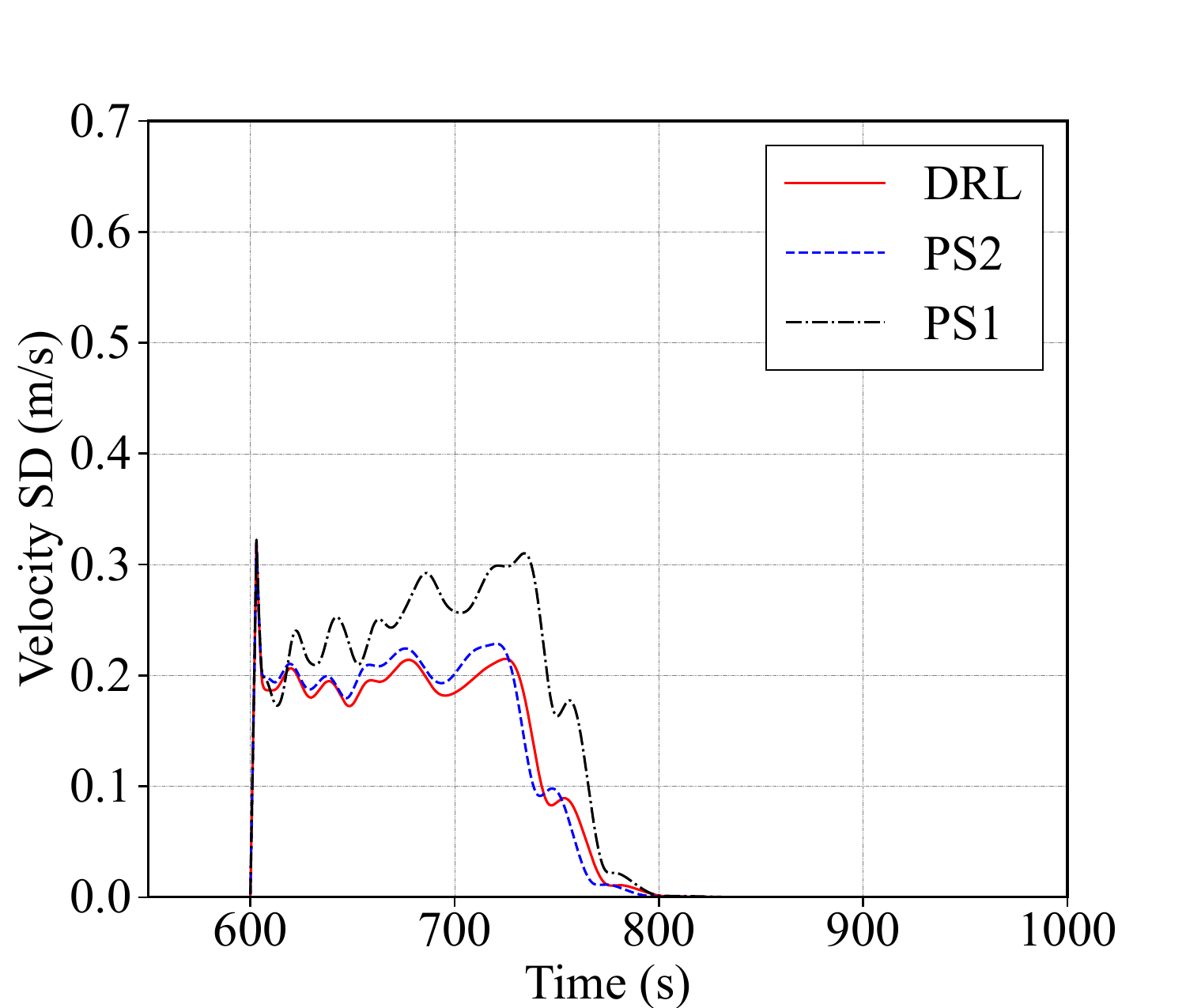}
}
\caption{Velocity standard deviation (SD) under different penetration rates $P_{\text{CAV}}$. 
%From top left to bottom right, $P_{\text{CAV}} = 0.1$, $0.2$, $0.3$, and $0.4$, respectively.
}
\label{fig:chap4_velocity_sd}
\end{figure}

Fig. \ref{fig:chap4_velocity_sd} reveals significant stability differences among platooning strategies through velocity SD analysis. 
The DRL-based scheme reduces peak velocity SD by up to $38.60\%$ compared to conventional platooning and $11.97\%$ versus the hybrid scheme $\text{PS}2$, demonstrating its effectiveness in suppressing velocity oscillation propagation within mixed traffic flows under disturbance scenarios.

The experiments demonstrate platooning strategy effects on traffic safety from various perspectives. 
The results confirm that the DRL-based strategy, through synergistic platoon structure optimization, effectively mitigates longitudinal collision risks, suppresses velocity oscillation propagation, and ensures operational safety in mixed traffic environments.

It should be noted that the above safety evaluation is conducted under ideal communication and perception conditions. In practice, communication failures, sensor faults, or observation noise may impair vehicle-state acquisition, especially for HVs/AVs, thereby weakening platoon integrity and safety. Future work will therefore examine fault-induced robustness, fail-safe control strategies, and more comprehensive traffic-risk assessment methods, such as causal graph-based conflict risk prediction \citep{Lyu2025ConflictRisk}.

\subsection{Energy consumption and emissions}

Recent years have witnessed extensive research by scholars and institutions on vehicle fuel consumption and emissions using the Measure of Effectiveness (MOE) framework. 
As a quantitative metric, MOE quantifies fuel consumption and emissions per unit distance traveled. 
Studies establish that instantaneous MOE represents a mathematical function of a vehicle's instantaneous velocity and acceleration, yielding unique values when these parameters are specified.

This study employs the VT-Micro model, originally developed by  
\cite{Ahn1998}. 
and validated through extensive experimental observations, to compute fuel consumption and emissions for vehicles in mixed traffic flows. Within this model, MOE is calculated as:
\begin{equation}
\mathrm{MOE}_{e}(t) = e^{P(v(t), a(t))}
\end{equation}
where $\mathrm{MOE}_{e}(t)$ denotes instantaneous fuel consumption or emission values at time $t$; $e \approx 2.71828$ represents the natural logarithm base, and $P(v(t), a(t))$ constitutes a function of instantaneous velocity and acceleration:
\begin{equation}
P(v(t), a(t)) = \sum_{i=0}^{3} \sum_{j=0}^{3} K_{i,j} v(t)^i a(t)^j
\end{equation}
where $K_{i,j}$ represents regression coefficients with power exponent $i$ for velocity and $j$ for acceleration.

Fuel consumption and emission test data across diverse vehicle types have revealed distinct energy and emission characteristics during acceleration and deceleration phases \citep{LI_2014_Stop_and_go}.
Consequently, three independent sets of regression coefficients $K_{i,j}$ were derived through VT-Micro modeling for fuel consumption, $\mathrm{CO_{2}}$, and $\mathrm{NO_{x}}$ emissions, respectively.

\begin{figure}[!t]
\centering
\subfloat[$P_{\text{CAV}} = 0.1$]{
    \includegraphics[width=0.45\linewidth]{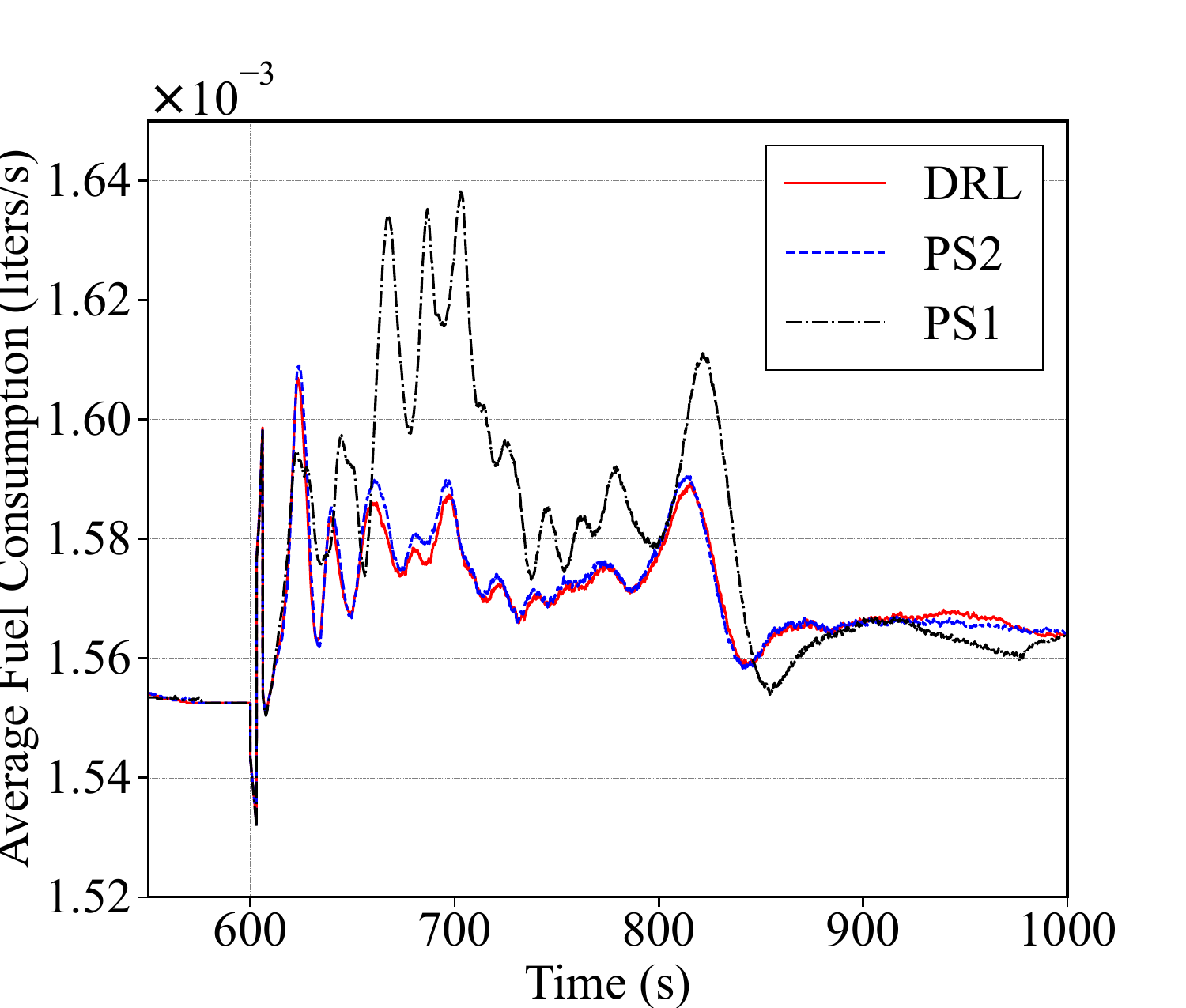}
}
\subfloat[$P_{\text{CAV}} = 0.2$]{
    \includegraphics[width=0.45\linewidth]{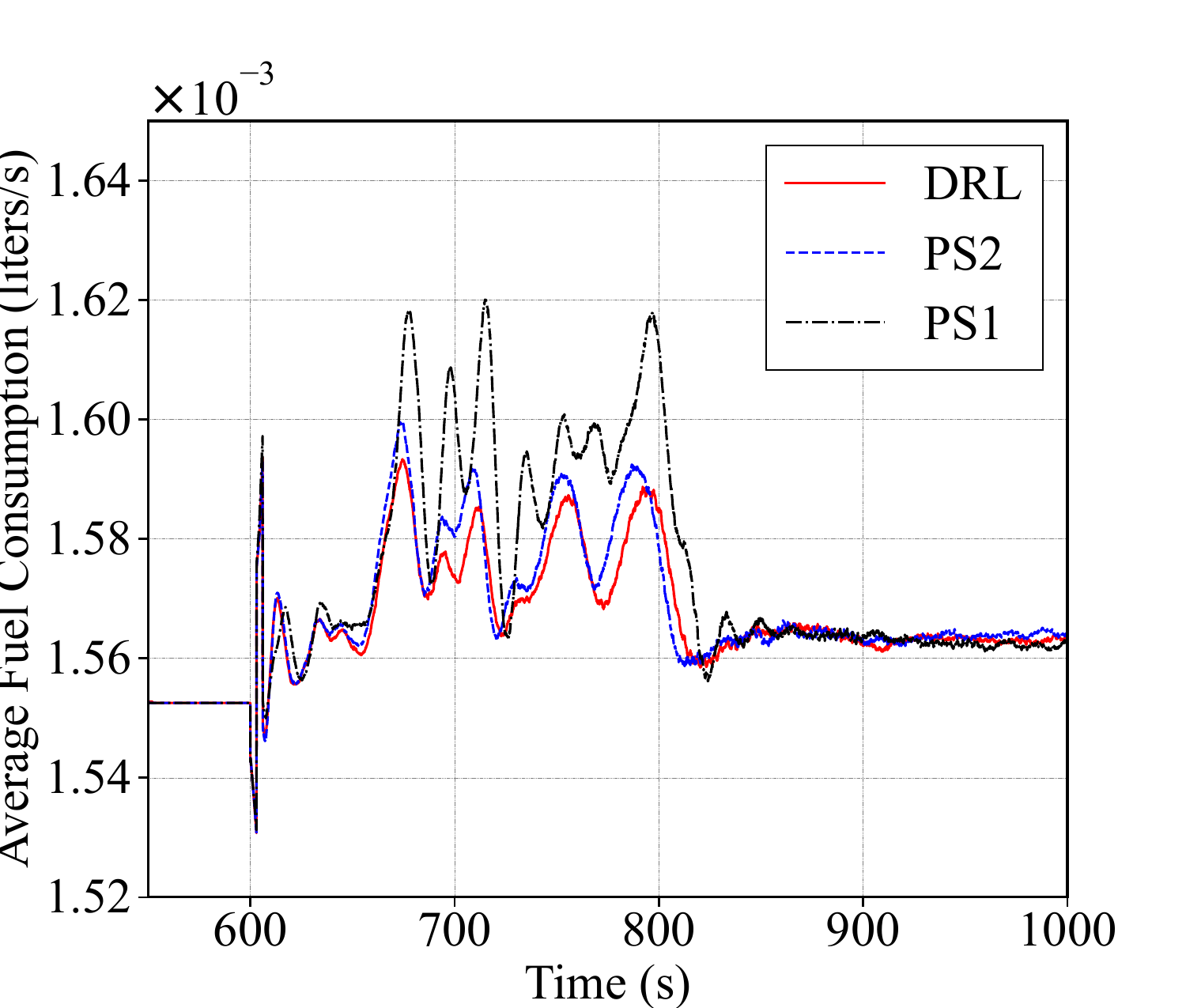}
}\\
\subfloat[$P_{\text{CAV}} = 0.3$]{
    \includegraphics[width=0.45\linewidth]{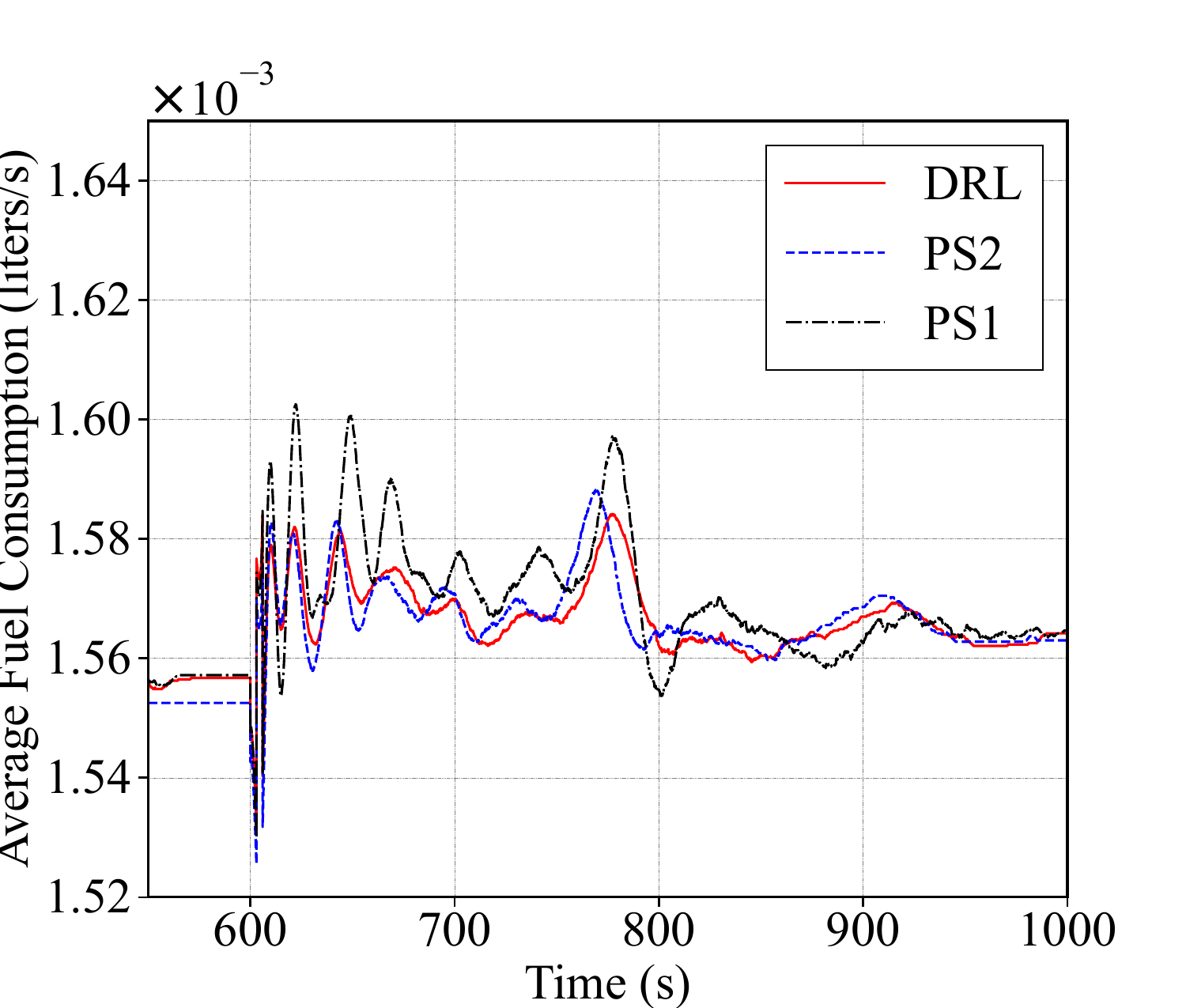}
}
\subfloat[$P_{\text{CAV}} = 0.4$]{
    \includegraphics[width=0.45\linewidth]{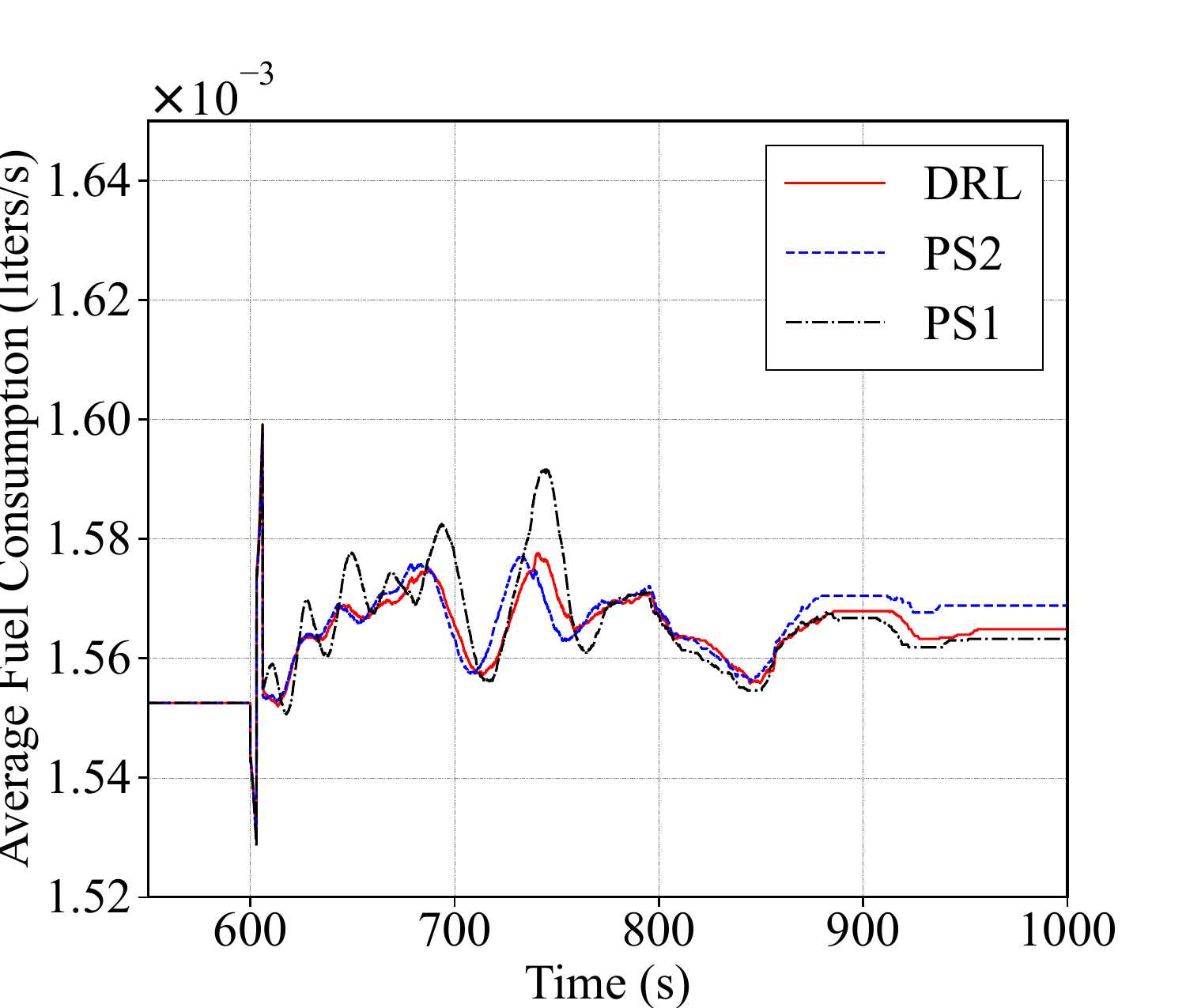}
}
\caption{Average fuel consumption under different penetration rates $P_{\text{CAV}}$. 
%From top left to bottom right, $P_{\text{CAV}} = 0.1$, $0.2$, $0.3$, and $0.4$, respectively.
}
\label{fig:chap4_avg_fuel}
\end{figure}

\begin{figure}[!htbp]
\centering
\subfloat[$P_{\text{CAV}} = 0.1$]{
    \includegraphics[width=0.45\linewidth]{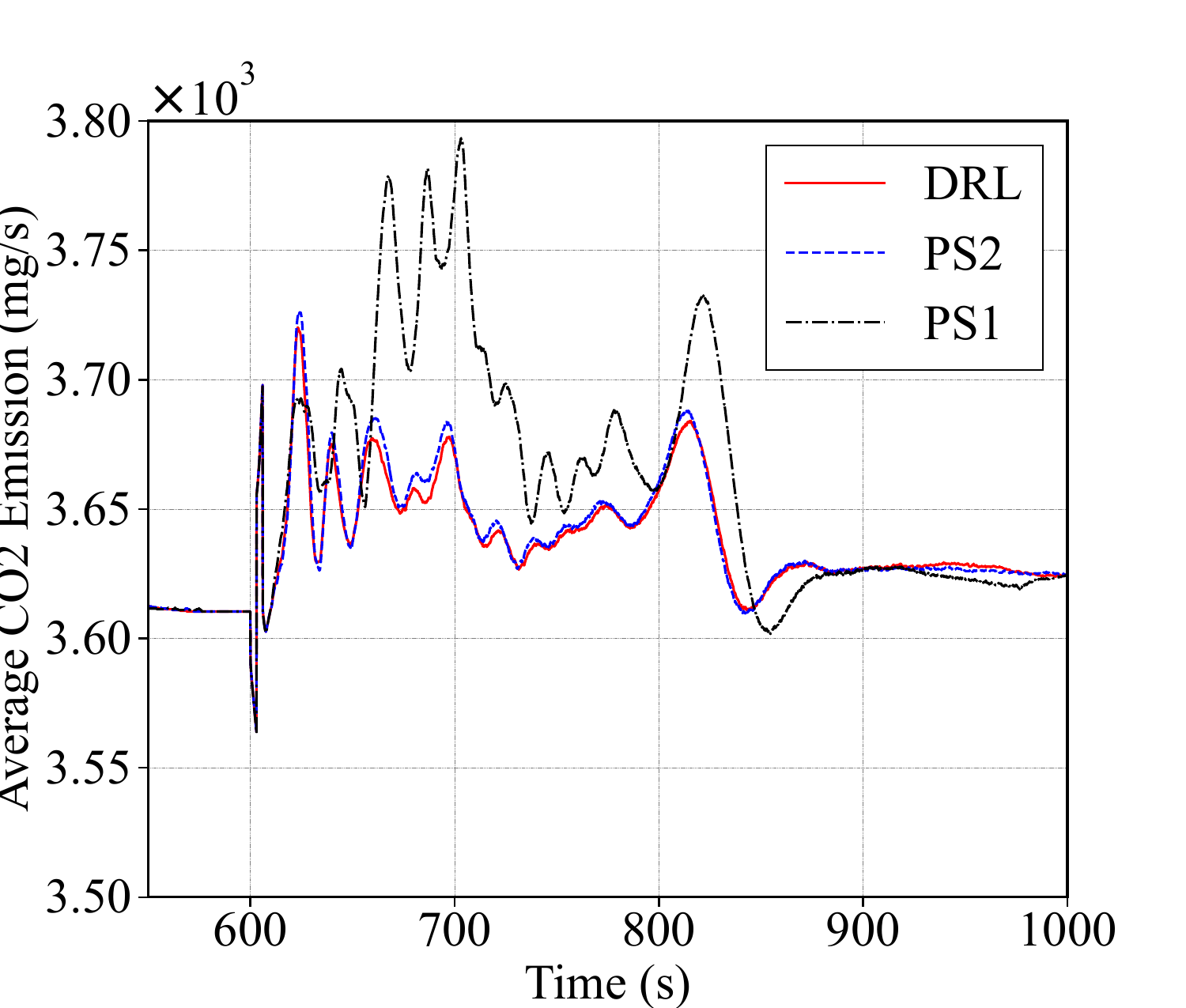}
}
\subfloat[$P_{\text{CAV}} = 0.2$]{
    \includegraphics[width=0.45\linewidth]{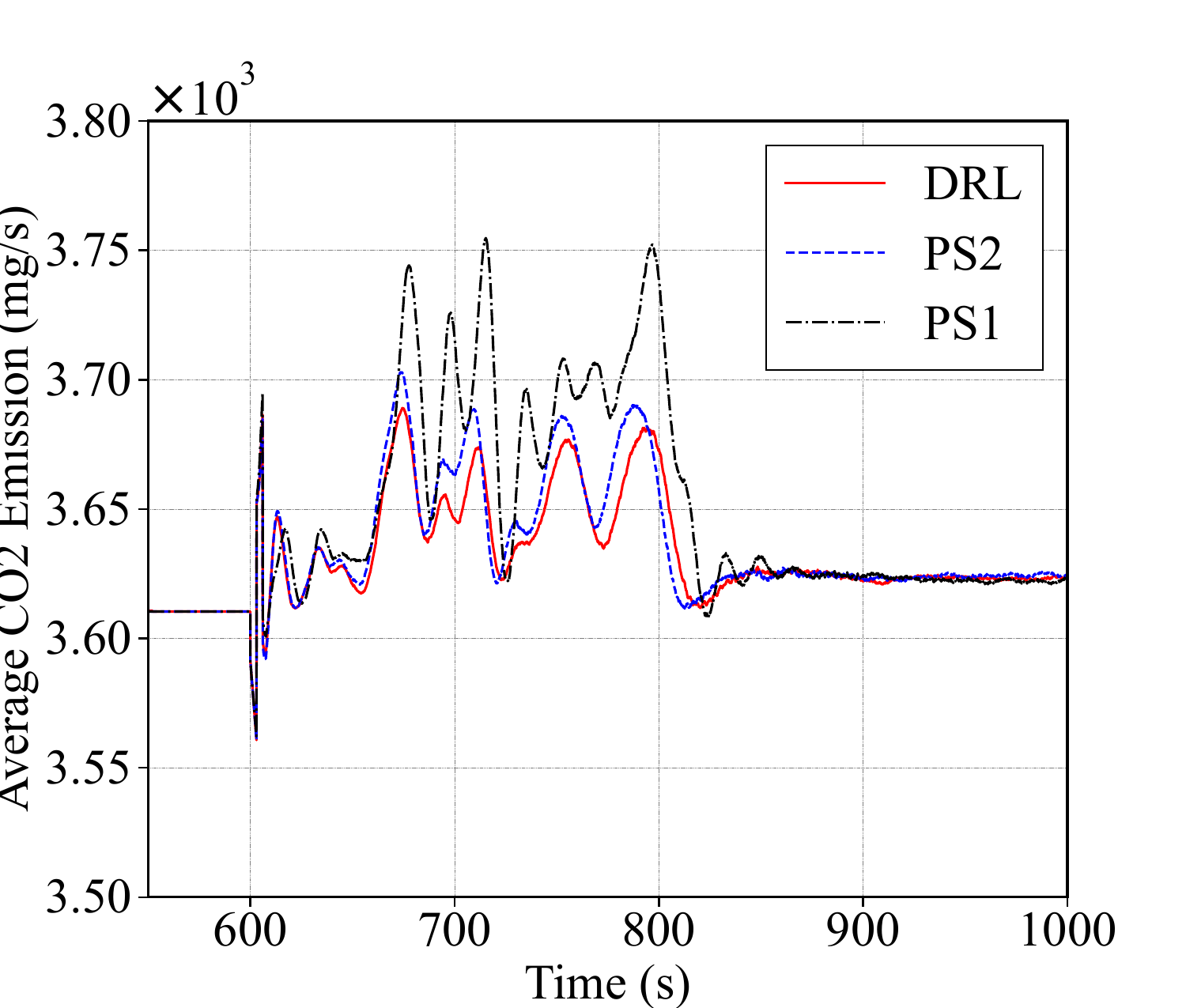}
}\\
\subfloat[$P_{\text{CAV}} = 0.3$]{
    \includegraphics[width=0.45\linewidth]{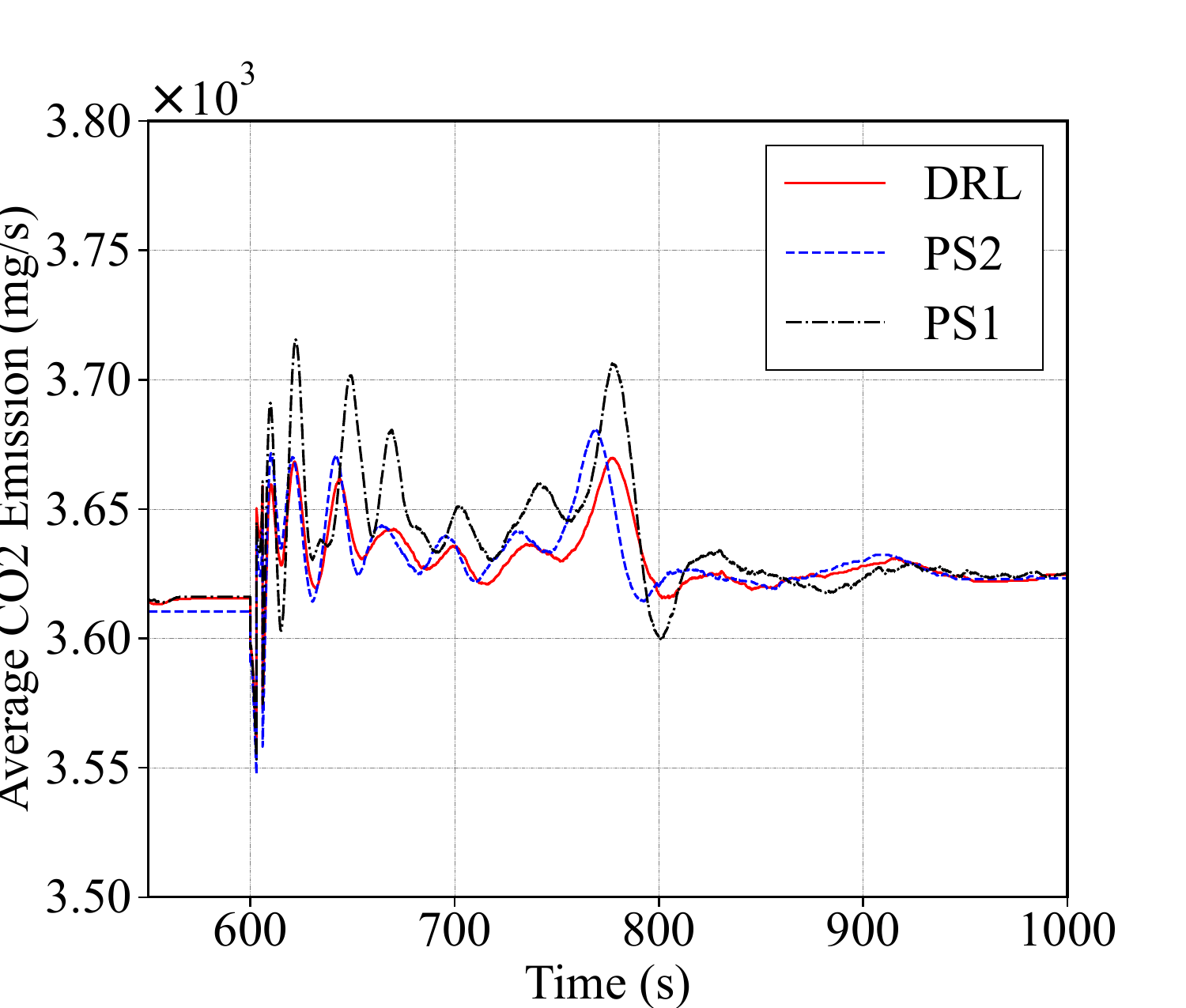}
}
\subfloat[$P_{\text{CAV}} = 0.4$]{
    \includegraphics[width=0.45\linewidth]{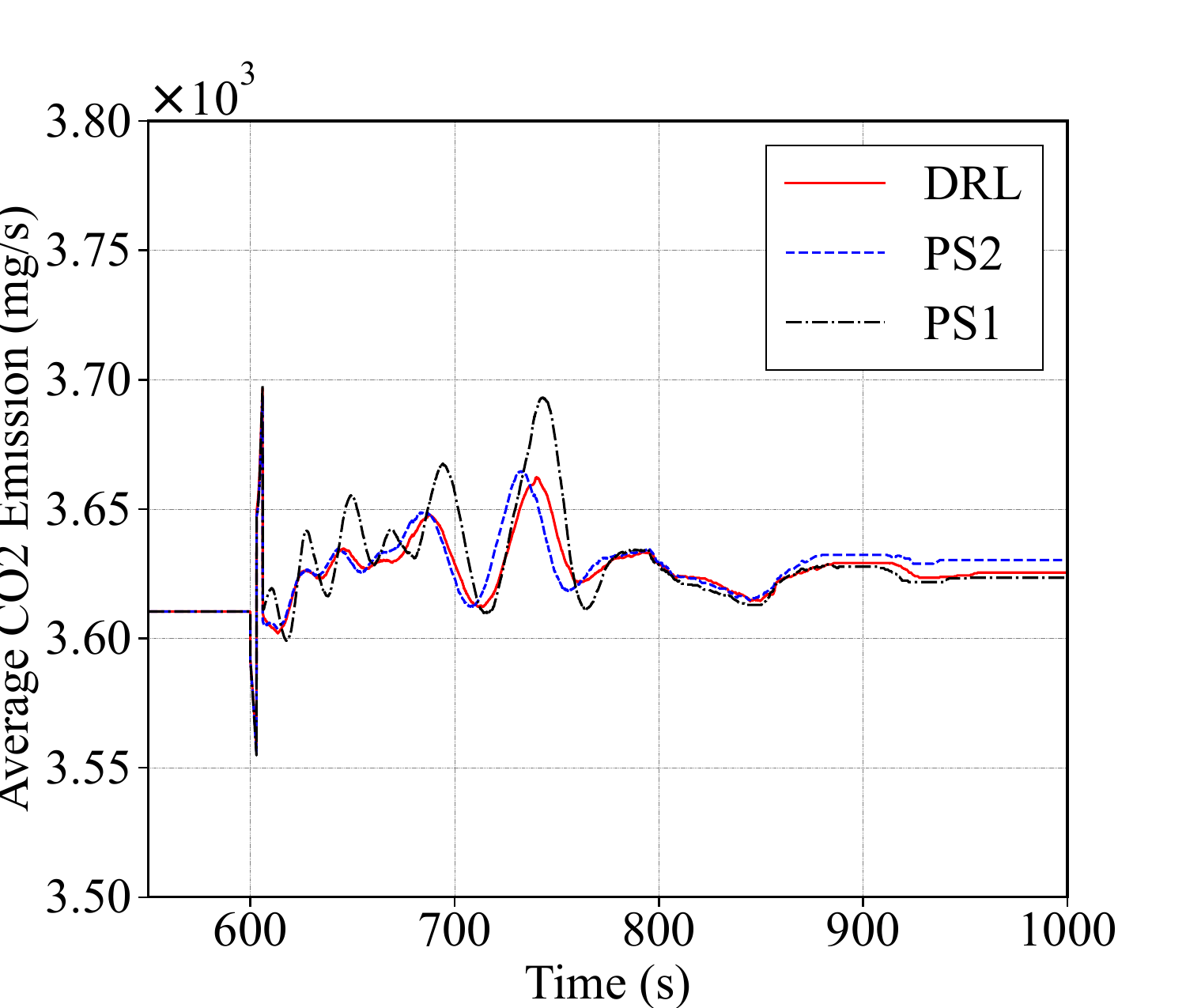}
}
\caption{Average $\mathrm{CO_{2}}$ emissions under different penetration rates $P_{\text{CAV}}$. 
%From top left to bottom right, $P_{\text{CAV}} = 0.1$, $0.2$, $0.3$, and $0.4$, respectively.
}
\label{fig:chap4_avg_co2}
\end{figure}

\begin{figure}[!t]
\centering
\subfloat[$P_{\text{CAV}} = 0.1$]{
    \includegraphics[width=0.45\linewidth]{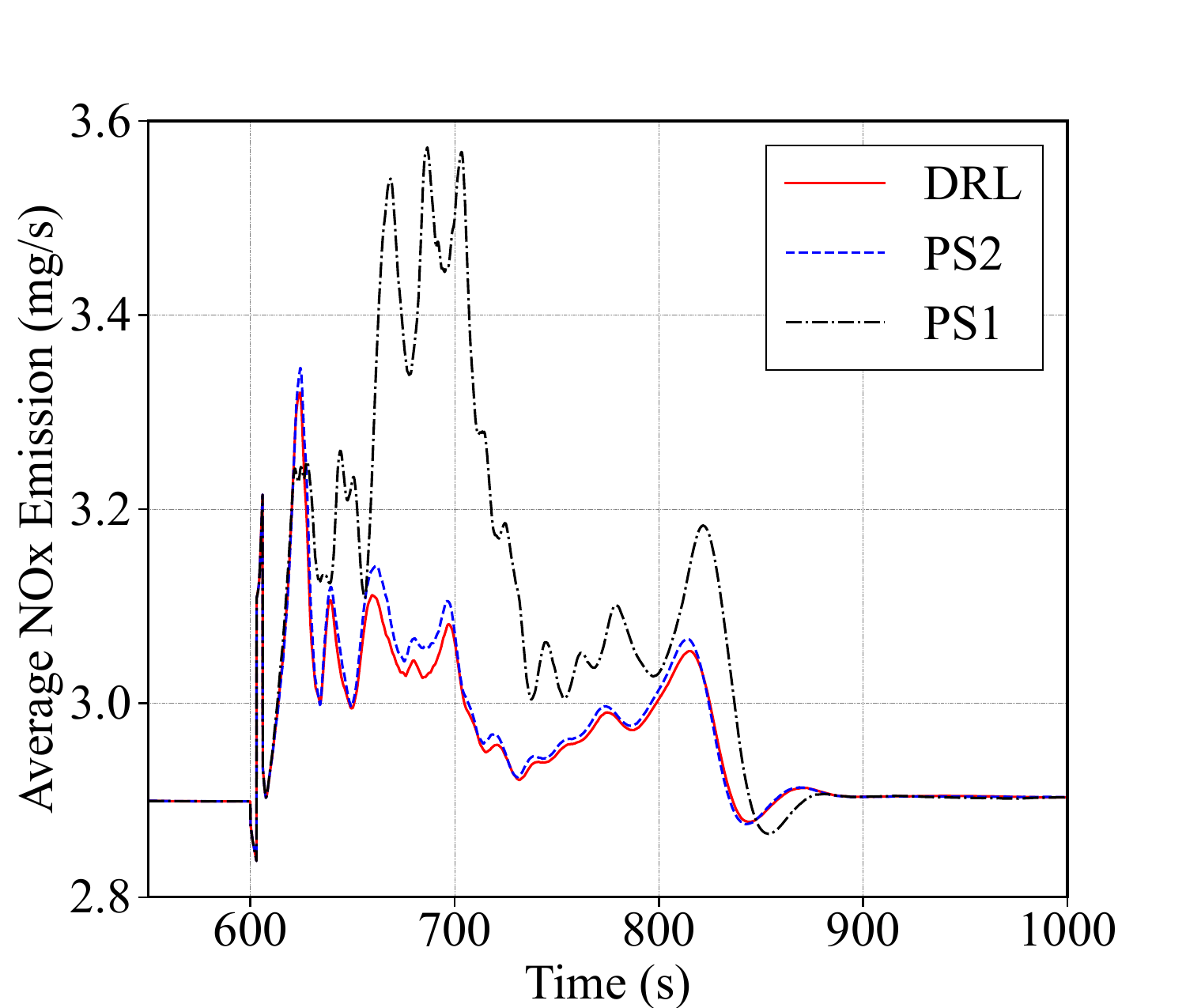}
}
\subfloat[$P_{\text{CAV}} = 0.2$]{
    \includegraphics[width=0.45\linewidth]{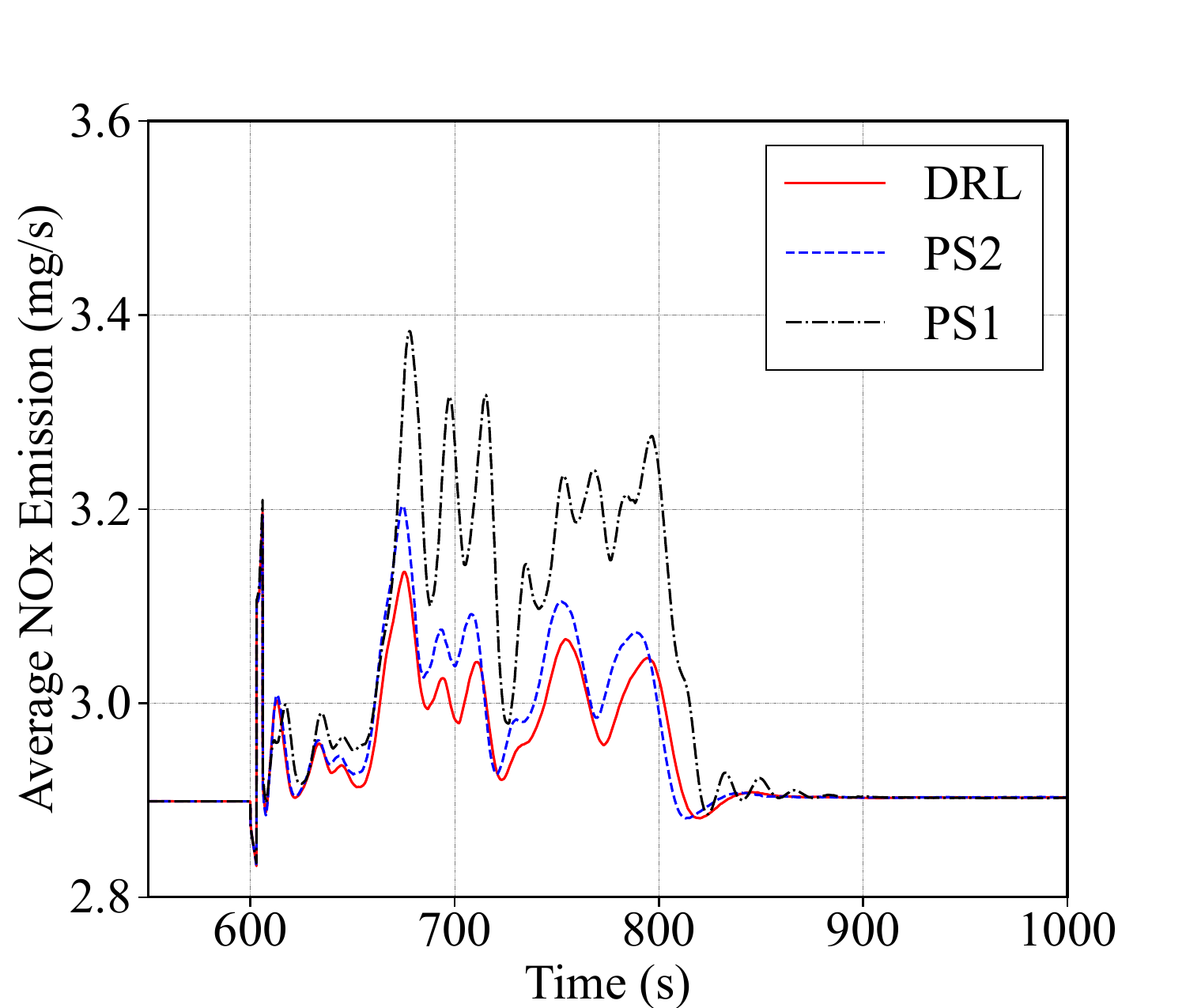}
}\\
\subfloat[$P_{\text{CAV}} = 0.3$]{
    \includegraphics[width=0.45\linewidth]{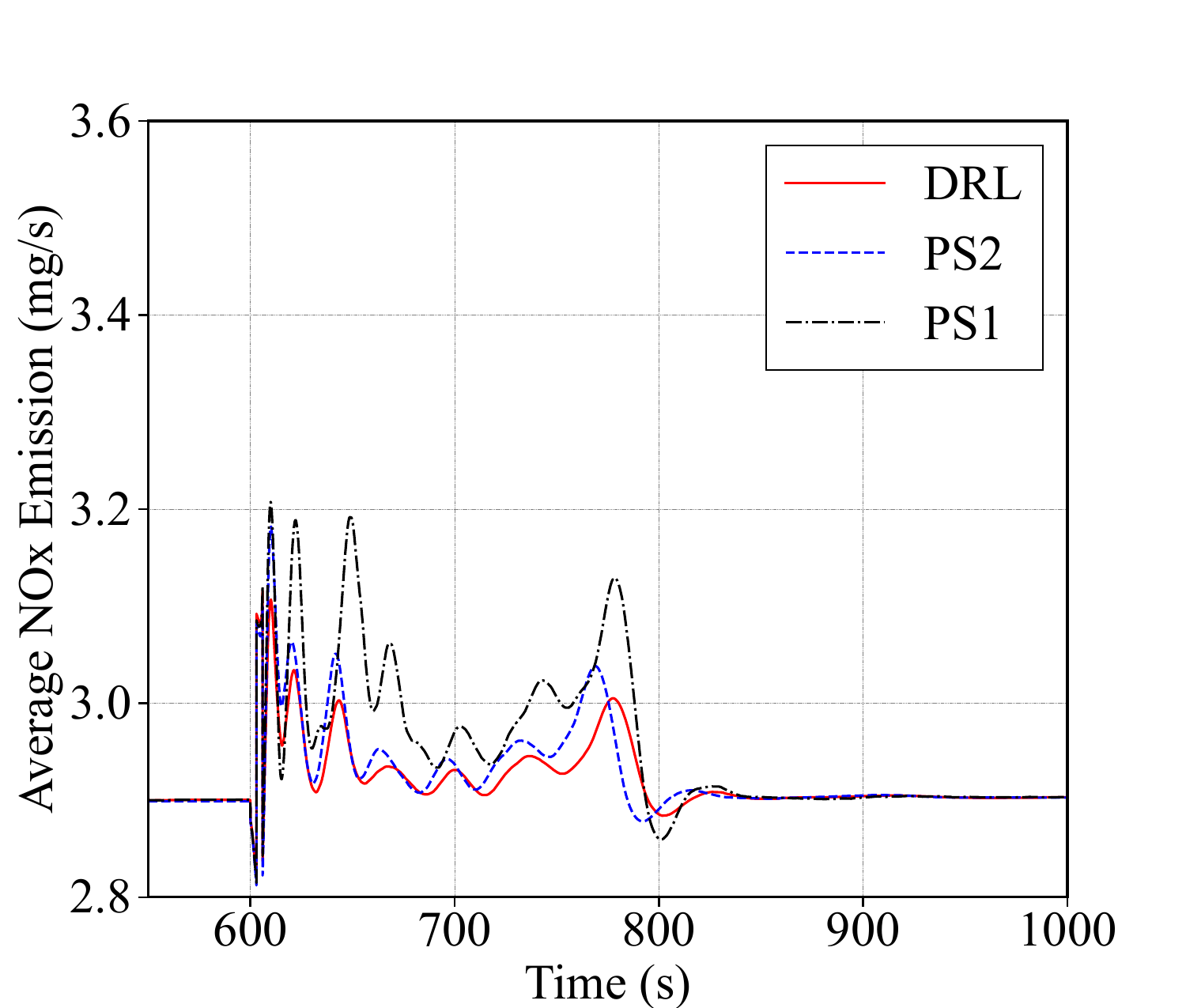}
}
\subfloat[$P_{\text{CAV}} = 0.4$]{
    \includegraphics[width=0.45\linewidth]{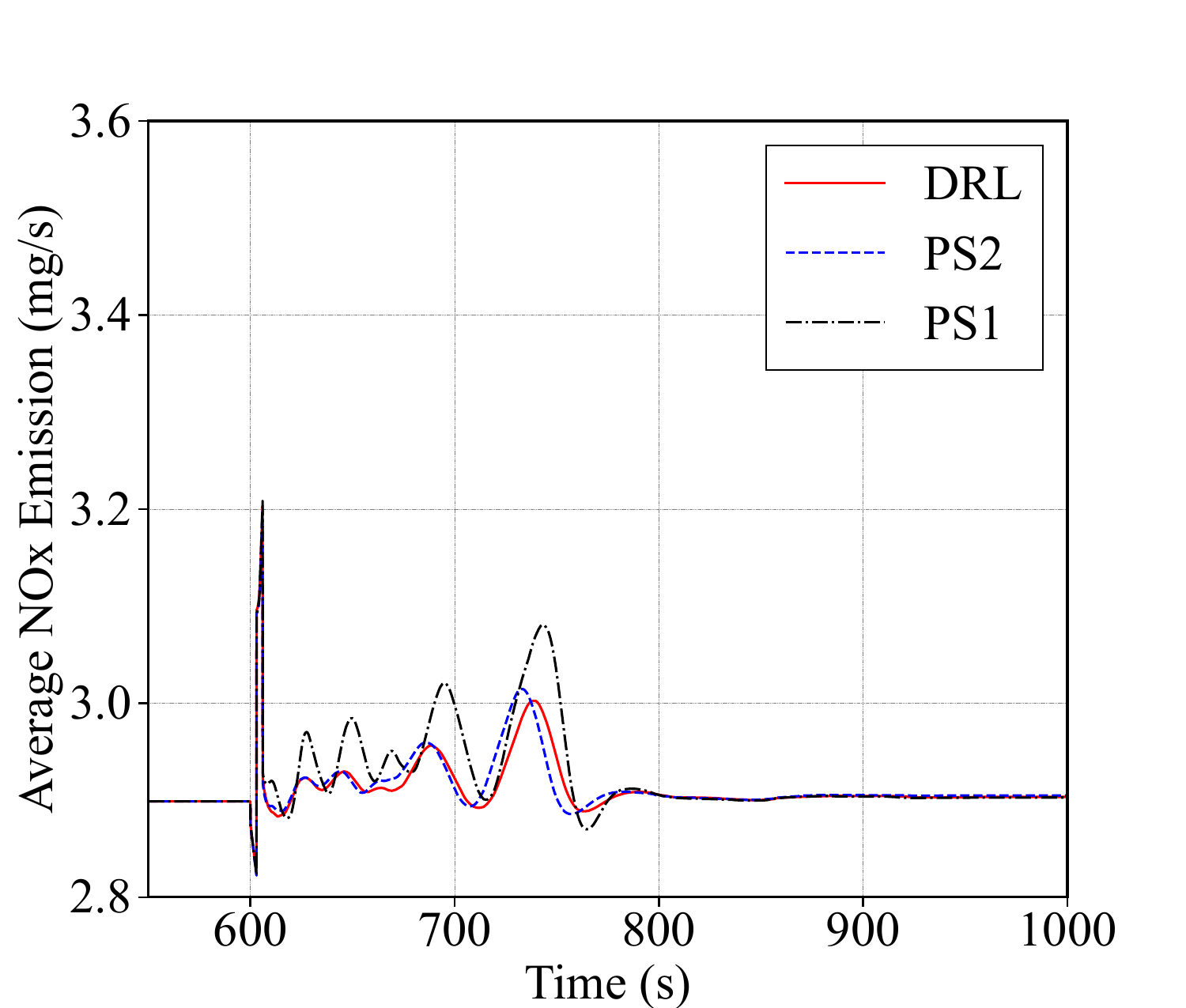}
}
\caption{Average $\mathrm{NO_{x}}$ emissions under different penetration rates $P_{\text{CAV}}$. 
%From top left to bottom right, $P_{\text{CAV}} = 0.1$, $0.2$, $0.3$, and $0.4$, respectively.
}
\label{fig:chap4_avg_nox}
\end{figure}

Building upon the disturbance propagation scenario in Section \ref{subsec:chap4_traffic_stability}, Figs. \ref{fig:chap4_avg_fuel}-\ref{fig:chap4_avg_nox} delineate the evolution of energy consumption and emission under different platooning strategies. 
Experimental results indicate that at low CAV penetration rates ($P_{\mathrm{CAV}} \leq 0.3$), the DRL-based strategy reduces peak average energy consumption by up to $1.90\%$ and $0.37\%$ compared to conventional and hybrid platooning schemes ($\text{PS}1$ and $\text{PS}2$) respectively, effectively mitigating incomplete fuel combustion caused by frequent acceleration/deceleration.

Analysis of emission characteristics reveals divergent response mechanisms between CO$_2$ and NO$_x$. For CO$_2$ emissions, the DRL-based platooning strategy demonstrates performance consistent with its energy consumption profile: under low CAV penetration scenarios, it reduces CO$_2$ emissions by over $1.2\%$ compared to conventional platooning $\text{PS}1$, while achieving up to $7.1\%$ reduction in NO$_x$ emissions. 
Relative to hybrid platooning $\text{PS}2$, the DRL-based strategy yields reductions of $0.37\%$ in CO$_2$ and $2.1\%$ in NO$_x$ emissions. 
These experimental findings substantiate the energy efficiency and environmental benefits of DRL-based strategy from both fuel consumption and emission perspectives.

\section{Conclusions and future research}
\label{sec:Conclusions}

To overcome the limitations of traditional CAV-exclusive platooning in mixed traffic, we propose a flexible platoon formation approach that incorporates HVs/AVs, controlled via a novel DRL framework. 
We further develop a DRL-based platooning strategy for hybrid platoons in mixed traffic. 
Our method leverages a POMDP formulation with multi-level state representations to enable distributed control of heterogeneous platoons. 
Experimental results show that the proposed strategy improves traffic stability mainly by regulating platoon structures rather than simply maximizing platoon formation. 
In the training evaluation, the learned policy improves the traffic stability discriminant by $27.9\%$ and reduces the maximum platoon length by $28.3\%$ compared with the conventional hybrid platooning strategy. 
In the disturbance-propagation evaluation, the trained policy reduces peak average fuel consumption by up to $1.90\%$ and CO$_2$ emissions by over $1.2\%$ compared with the conventional connected platooning strategy, while reducing NO$_x$ emissions by up to $7.1\%$ under low CAV penetration conditions.
The strategy demonstrates robust generalization across varying penetration rates $P_{\mathrm{CAV}}\in(0.1,0.45)$, effectively balancing capacity-stability tradeoffs while suppressing disturbance propagation throughout the traffic flow.
While these findings highlight the effectiveness of the proposed framework, several practical issues remain to be further investigated. The current study adopts a simplified mixed-traffic setting to focus on heterogeneous platoon organization. In real-world traffic, multi-lane interactions, uncertain events, and dynamic platoon splitting and re-forming may jointly affect platoon stability, safety, and efficiency. Future work will extend the framework by incorporating these complex operational conditions and robustness considerations.

\bibliographystyle{model2-names}
\bibliography{Ref.bib}

@article{Lan_2023_Data_Driven_Robust,
    author={Lan, Jianglin and Zhao, Dezong and Tian, Daxin},
    journal={IEEE Transactions on Intelligent Transportation Systems},
    title={Data-Driven Robust Predictive Control for Mixed Vehicle Platoons Using Noisy Measurement},
    year={2023},
    volume={24},
    number={6},
    pages={6586-6596}
}

@ARTICLE{shi_2025_Mixed_Vehicle_Platoon,
    author={Shi, Yujie and Dong, Haoxuan and He, Chaozhe R. and Chen, Yuxiao and Song, Ziyou},
    journal={IEEE Internet of Things Journal},
    title={Mixed Vehicle Platoon Forming: A Multi-Agent Reinforcement Learning Approach},
    year={2025},
    volume={12},
    number={11},
    pages={16886-16898},
}

@article{MAITI_2023_Ad_hoc_platoon,
    title = {Ad-hoc platoon formation and dissolution strategies for multi-lane highways},
    journal = {Journal of Intelligent Transportation Systems},
    volume = {27},
    number = {2},
    pages = {161-173},
    year = {2023},
    issn = {1547-2450},
    author = {Santa Maiti and Stephan Winter and Lars Kulik and Sudeshna Sarkar}
}

@article{feng_2021_Robust_Platoon_Control,
    author={Feng, Shuo and Song, Ziyou and Li, Zhaojian and Zhang, Yi and Li, Li},
    journal={IEEE Transactions on Intelligent Vehicles},
    title={Robust Platoon Control in Mixed Traffic Flow Based on Tube Model Predictive Control},
    year={2021},
    volume={6},
    number={4},
    pages={711-722}
}

@article{yao2024impact,
    title={Impact of the heterogeneity and platoon size of connected vehicles on the capacity of mixed traffic flow},
    author={Yao, Zhihong and Ma, Yuqin and Ren, Tingting and Jiang, Yangsheng},
    journal={Applied Mathematical Modelling},
    volume={125},
    pages={367--389},
    year={2024},
    publisher={Elsevier}
}

@article{guan2023markov,
    title={Markov chain-based traffic analysis on platooning effect among mixed semi-and fully-autonomous vehicles in a freeway lane},
    author={Guan, Hao and Wang, Hua and Meng, Qiang and Mak, Chin Long},
    journal={Transportation Research Part B: Methodological},
    volume={173},
    pages={176--202},
    year={2023},
    publisher={Elsevier}
}

@article{jiang2023platoon,
    author = {Jiang, Yangsheng and Zhu, Fangyi and Yao, Zhihong and Gu, Qiufan and Ran, Bin},
    title = {Platoon Intensity of Connected Automated Vehicles: Definition, Formulas, Examples, and Applications},
    journal = {Journal of Advanced Transportation},
    volume = {2023},
    number = {1},
    pages = {3325530},
    year = {2023}
}

@article{qin2023stability,
    title={Stability analysis and connected vehicles management for mixed traffic flow with platoons of connected automated vehicles},
    author={Qin, Yanyan and Luo, Qinzhong and Wang, Hua},
    journal={Transportation Research Part C: Emerging Technologies},
    volume={157},
    pages={104370},
    year={2023},
    publisher={Elsevier}
}

@article{xin2021modeling,
    title={Modeling and impact analysis of connected vehicle merging accounting for mainline random length tight-platoon},
    author={Xin, Qi and Fu, Rui and Ukkusuri, Satish V and Yu, Shaowei and Jiang, Rui},
    journal={Physica A: Statistical Mechanics and its Applications},
    volume={563},
    pages={125452},
    year={2021},
    publisher={Elsevier}
}

@article{huang2018path,
    title={Path planning and cooperative control for automated vehicle platoon using hybrid automata},
    author={Huang, Zichao and Chu, Duanfeng and Wu, Chaozhong and He, Yi},
    journal={IEEE Transactions on Intelligent Transportation Systems},
    volume={20},
    number={3},
    pages={959--974},
    year={2018},
    publisher={IEEE}
}

@article{sala2020macroscopic,
    title={Macroscopic modeling of connected autonomous vehicle platoons under mixed traffic conditions},
    author={Sala, Marcel and Soriguera, Francesc},
    journal={Transportation Research Procedia},
    volume={47},
    pages={163--170},
    year={2020},
    publisher={Elsevier}
}

@article{yao2023analysis,
    title={Analysis of the impact of maximum platoon size of CAVs on mixed traffic flow: An analytical and simulation method},
    author={Yao, Zhihong and Wu, Yunxia and Wang, Yi and Zhao, Bin and Jiang, Yangsheng},
    journal={Transportation Research Part C: Emerging Technologies},
    volume={147},
    pages={103989},
    year={2023},
    publisher={Elsevier}
}

@article{chang2020analysis,
    title={Analysis on traffic stability and capacity for mixed traffic flow with platoons of intelligent connected vehicles},
    author={Chang, Xin and Li, Haijian and Rong, Jian and Zhao, Xiaohua and others},
    journal={Physica A: Statistical Mechanics and Its Applications},
    volume={557},
    pages={124829},
    year={2020},
    publisher={Elsevier}
}

@article{ghiasi2020lane,
    title={Lane management with variable lane width and model calibration for connected automated vehicles},
    author={Ghiasi, Amir and Hussain, Omar and Qian, Zhen and Li, Xiaopeng},
    journal={Journal of Transportation Engineering, Part A: Systems},
    volume={146},
    number={3},
    pages={04019075},
    year={2020},
    publisher={American Society of Civil Engineers}
}

@article{ghiasi2017mixed,
    title={A mixed traffic capacity analysis and lane management model for connected automated vehicles: A Markov chain method},
    author={Ghiasi, Amir and Hussain, Omar and Qian, Zhen Sean and Li, Xiaopeng},
    journal={Transportation Research Part B: Methodological},
    volume={106},
    pages={266--292},
    year={2017},
    publisher={Elsevier}
}

@article{yao2022fundamental,
    title={Fundamental diagram and stability of mixed traffic flow considering platoon size and intensity of connected automated vehicles},
    author={Yao, Zhihong and Gu, Qiufan and Jiang, Yangsheng and Ran, Bin},
    journal={Physica A: Statistical Mechanics and its Applications},
    volume={604},
    pages={127857},
    year={2022},
    publisher={Elsevier}
}

@article{zhou2021impact,
    title={Impact of CAV platoon management on traffic flow considering degradation of control mode},
    author={Zhou, Linjie and Ruan, Tiancheng and Ma, Ke and Dong, Changyin and Wang, Hao},
    journal={Physica A: Statistical Mechanics and its Applications},
    volume={581},
    pages={126193},
    year={2021},
    publisher={Elsevier}
}

@article{liu2018modeling,
    title={Modeling impacts of cooperative adaptive cruise control on mixed traffic flow in multi-lane freeway facilities},
    author={Liu, Hao and Kan, Xingan David and Shladover, Steven E and Lu, Xiao-Yun and Ferlis, Robert E},
    journal={Transportation Research Part C: Emerging Technologies},
    volume={95},
    pages={261--279},
    year={2018},
    publisher={Elsevier}
}

@article{liu2018impact,
    title={Impact of cooperative adaptive cruise control on multilane freeway merge capacity},
    author={Liu, Hao and Kan, Xingan and Shladover, Steven E and Lu, Xiao-Yun and Ferlis, Robert E},
    journal={Journal of Intelligent Transportation Systems},
    volume={22},
    number={3},
    pages={263--275},
    year={2018},
    publisher={Taylor \& Francis}
}

@article{van2006impact,
    title={The impact of cooperative adaptive cruise control on traffic-flow characteristics},
    author={Van Arem, Bart and Van Driel, Cornelie JG and Visser, Ruben},
    journal={IEEE Transactions on intelligent transportation systems},
    volume={7},
    number={4},
    pages={429--436},
    year={2006},
    publisher={IEEE}
}

@article{zhou2021analytical,
    title={Analytical analysis of the effect of maximum platoon size of connected and automated vehicles},
    author={Zhou, Jiazu and Zhu, Feng},
    journal={Transportation Research Part C: Emerging Technologies},
    volume={122},
    pages={102882},
    year={2021},
    publisher={Elsevier}
}

@article{seraj2018modeling,
    author = {  Mudasser Seraj and     Jiangchen Li and Zhijun Qiu},
    title = {Modeling Microscopic Car-Following Strategy of Mixed Traffic to Identify Optimal Platoon Configurations for Multiobjective Decision-Making},
    journal = {Journal of Advanced Transportation},
    volume = {2018},
    pages = {1-15},
    year = {2018},
}

@article{jin2020analysis,
    author={Jin, Li and Čičić, Mladen and Johansson, Karl H. and Amin, Saurabh},
    journal={IEEE Transactions on Automatic Control},
    title={Analysis and Design of Vehicle Platooning Operations on Mixed-Traffic Highways},
    year={2021},
    volume={66},
    number={10},
    pages={4715-4730}
}

@article{JIN_2021_Spatial_Distribution,
    title = {The Impact of Spatial Distribution of Heterogeneous Vehicles on Performance of Mixed Platoon: A Cyber-Physical Perspective},
    journal = {KSCE Journal of Civil Engineering},
    volume = {25},
    number = {1},
    pages = {303-315},
    year = {2021},
    issn = {1226-7988},
    author = {Shuang Jin and Dihua Sun and Zhongcheng Liu}
}

@ARTICLE{li_2023_Cooperative_Formation,
    author={Li, Keqiang and Wang, Jiawei and Zheng, Yang},
    journal={IEEE Transactions on Intelligent Transportation Systems},
    title={Cooperative Formation of Autonomous Vehicles in Mixed Traffic Flow: Beyond Platooning},
    year={2022},
    volume={23},
    number={9},
    pages={15951-15966}
}

@ARTICLE{wu_2024_Multi_Lane_Unsignalized,
    author={Wu, Ruiyi and Jia, Hongfei and Huang, Qiuyang and Tian, Jingjing and Gao, Heyao and Wang, Guanfeng},
    journal={IEEE Transactions on Intelligent Transportation Systems},
    title={Multi-Lane Unsignalized Intersection Cooperation Strategy Considering Platoons Formation in a Mixed Connected Automated Vehicles and Connected Human-Driven Vehicles Environment},
    year={2024},
    volume={25},
    number={2},
    pages={1569-1585}
}

@article{treiber2013traffic,
    title={Traffic flow dynamics},
    author={Treiber, Martin and Kesting, Arne},
    journal={Traffic Flow Dynamics: Data, Models and Simulation, Springer-Verlag Berlin Heidelberg},
    pages={983--1000},
    year={2013},
    publisher={Springer}
}

@article{milanes2014modeling,
    title={Modeling cooperative and autonomous adaptive cruise control dynamic responses using experimental data},
    author={Milan{\'e}s, Vicente and Shladover, Steven E},
    journal={Transportation Research Part C: Emerging Technologies},
    volume={48},
    pages={285--300},
    year={2014},
    publisher={Elsevier}
}

@article{milanes2013cooperative,
    title={Cooperative adaptive cruise control in real traffic situations},
    author={Milan{\'e}s, Vicente and Shladover, Steven E and Spring, John and Nowakowski, Christopher and Kawazoe, Hiroshi and Nakamura, Masahide},
    journal={IEEE Transactions on intelligent transportation systems},
    volume={15},
    number={1},
    pages={296--305},
    year={2013},
    publisher={IEEE}
}

@article{Ngoduy_2013_analytical_studies_nonlinear_sci,
    title = {Analytical studies on the instabilities of heterogeneous intelligent traffic flow},
    journal = {Communications in Nonlinear Science and Numerical Simulation},
    volume = {18},
    number = {10},
    pages = {2699-2706},
    year = {2013},
    issn = {1007-5704},
    author = {D. Ngoduy}
}

@inproceedings{Bernstein_2000_ctde,
    author = {Bernstein, Daniel S. and Zilberstein, Shlomo and Immerman, Neil},
    title = {The complexity of decentralized control of Markov decision processes},
    year = {2000},
    isbn = {1558607099},
    publisher = {Morgan Kaufmann Publishers Inc.},
    address = {San Francisco, CA, USA},
    booktitle = {Proceedings of the Sixteenth Conference on Uncertainty in Artificial Intelligence},
    pages = {32-37},
    numpages = {6},
    location = {Stanford, California},
    series = {UAI'00}
}

@article{Schulman_2015_gae,
    author = {Schulman, John and Moritz, Philipp and Levine, Sergey and Jordan, Michael and Abbeel, Pieter},
    year = {2015},
    month = {06},
    pages = {},
    title = {High-Dimensional Continuous Control Using Generalized Advantage Estimation}
}

@inproceedings{Lin_1991_data_memory,
    author = {Lin, Long Ji},
    title = {Self-improvement based on reinforcement learning, planning and teaching},
    year = {1991},
    isbn = {1558602003},
    publisher = {Morgan Kaufmann Publishers Inc.},
    address = {San Francisco, CA, USA},
    booktitle = {Proceedings of the Eighth International Conference on Machine Learning},
    pages = {323-327},
    numpages = {5},
    location = {Evanston, Illinois, USA},
    series = {ML'91}
}

@article{LI_2024_safety_Enhancing,
    title = {Enhancing vehicular platoon stability in the presence of communication Cyberattacks: A reliable longitudinal cooperative control strategy},
    journal = {Transportation Research Part C: Emerging Technologies},
    volume = {163},
    pages = {104660},
    year = {2024},
    issn = {0968-090X},
    author = {Zihao Li and Yang Zhou and Yunlong Zhang and Xiaopeng Li}
}

@article{HUA_2023_sd_Impact,
    title = {Impact of multi-class stochastic cyberattacks on vehicle dynamics and rear-end collision risks for heterogeneous traffic},
    journal = {Physica A: Statistical Mechanics and its Applications},
    volume = {626},
    pages = {129095},
    year = {2023},
    issn = {0378-4371},
    author = {Xuedong Hua and Weijie Yu and Wei Wang and De Zhao}
}

@article{LUO_2024_safety_Modeling,
    title = {Modeling and analyzing self-resistance of connected automated vehicular platoons under different cyberattack injection modes},
    journal = {Accident Analysis \& Prevention},
    volume = {198},
    pages = {107494},
    year = {2024},
    issn = {0001-4575},
    author = {Dongyu Luo and Jiangfeng Wang and Yu Wang and Jiakuan Dong}
}

@mastersthesis{Ahn1998,
    author  = {Ahn, Kyoungho},
    title   = {Microscopic Fuel Consumption and Emission Modeling},
    school  = {Virginia Polytechnic Institute and State University},
    year    = {1998},
    address = {Blacksburg, VA, USA},
    type    = {Master's thesis}
}

@article{LI_2014_Stop_and_go,
    title = {Stop-and-go traffic analysis: Theoretical properties, environmental impacts and oscillation mitigation},
    journal = {Transportation Research Part B: Methodological},
    volume = {70},
    pages = {319-339},
    year = {2014},
    issn = {0191-2615},
    author = {Xiaopeng Li and Jianxun Cui and Shi An and Mohsen Parsafard}
}

@article{Wu2023DistributedDataDrivenModel,
  title = {Distributed Data-Driven Model Predictive Control for Heterogeneous Vehicular Platoon With Uncertain Dynamics},
  author = {Wu, Yanhong and Zuo, Zhiqiang and Wang, Yijing and Han, Qiaoni and Hu, Chuan},
  year = {2023},
  month = aug,
  journal = {IEEE Transactions on Vehicular Technology},
  volume = {72},
  number = {8},
  pages = {9969--9983},
  issn = {0018-9545, 1939-9359},
  
  urldate = {2024-03-14},
  langid = {english}
}

@article{Wang2025DistributedNonlinearModel,
  title = {Distributed Nonlinear Model Predictive Control of Vehicular Platoon Orienting Practical Driving Conditions},
  author = {Wang, Shenyi and Yang, Xiujian and Chen, Zheng and Zhang, Yuanjian},
  year = {2025},
  journal = {IEEE Transactions on Transportation Electrification},
  volume = {11},
  number = {1},
  pages = {2684--2695},
  issn = {2332-7782},
  
  urldate = {2024-07-20},
  langid = {english}
}

@article{Yang2023DistributedModelPredictive,
  title = {Distributed Model Predictive Control for Heterogeneous Platoon With Leading Human-Driven Vehicle Acceleration Prediction},
  author = {Yang, Junru and Chu, Duanfeng and Yin, Jianhua and Pi, Dawei and Wang, Jinxiang and Lu, Liping},
  year = {2023},
  journal = {IEEE Transactions on Intelligent Transportation Systems},
  pages = {1--16},
  issn = {1558-0016},
  
  urldate = {2023-12-18},
  langid = {english}
}

@article{SUN2024,
title = {Enhancing stability and flexibility of platoon control with an adaptive spacing policy},
journal = {Transportmetrica A Transport Science},
year = {2024},
issn = {2324-9935},
author = {Dengjiang Sun and Dong-Fan Xie and Xiao-Mei Zhao and Bin Jia},
}

@article{Zhang2025PlatoonFormation,
   author = {Mengqi Zhang and Chunyan Wang and Wanzhong Zhao and Jinqiang Liu and Ziyu Zhang},
   issn = {15580016},
   issue = {3},
   journal = {IEEE Transactions on Intelligent Transportation Systems},
   pages = {4002-4018},
   publisher = {Institute of Electrical and Electronics Engineers Inc.},
   title = {A Multi-Vehicle Self-Organized Cooperative Control Strategy for Platoon Formation in Connected Environment},
   volume = {26},
   year = {2025},
}

@article{Du2024heuristicRL,
   author = {Guodong Du and Yuan Zou and Xudong Zhang and Jie Fan and Wenjing Sun and Zirui Li},
   issn = {23274662},
   issue = {23},
   journal = {IEEE Internet of Things Journal},
   pages = {38273-38290},
   publisher = {Institute of Electrical and Electronics Engineers Inc.},
   title = {Efficient Motion Control for Heterogeneous Autonomous Vehicle Platoon Using Multilayer Predictive Control Framework},
   volume = {11},
   year = {2024},
}

@article{Ma2024ModelFree,
   author = {Yong Sheng Ma and Wei Wei Che and Chao Deng and Zheng Guang Wu},

   issn = {15580016},
   issue = {3},
   journal = {IEEE Transactions on Intelligent Transportation Systems},
   month = {3},
   pages = {2373-2381},
   publisher = {Institute of Electrical and Electronics Engineers Inc.},
   title = {Data-Driven Distributed Vehicle Platoon Control for Heterogeneous Nonlinear Vehicle Systems},
   volume = {25},
   year = {2024},
}

@article{Zong2025He,
   author = {Fang Zong and Sheng Yue and Meng Zeng and Zhengbing He and Dong Ngoduy},
   issn = {09600779},
   journal = {Chaos, Solitons and Fractals},
   month = {2},
   pages = {115850},
   publisher = {Elsevier Ltd},
   title = {Platoon or individual: An adaptive car-following control of connected and automated vehicles},
   volume = {191},
   year = {2025},
}

@article{Liu2025ZhengWang,
   author = {Gongzhe Liu and Nan Zheng and Hao Wang},
   issn = {0968090X},
   journal = {Transportation Research Part C: Emerging Technologies},
   month = {11},
   pages = {105353},
   title = {Cooperative control method for connected and automated vehicle platoon based on arbitrary time headway switched system},
   volume = {180},
   year = {2025},
}

@article{Liang2025Jun,
   author = {Jun Liang and Luyang Liu and Xinqi Yu and Wensa Wang and Chaofeng Pan and Long Chen},
   issn = {03784371},
   journal = {Physica A: Statistical Mechanics and its Applications},
   month = {10},
   pages = {130844},
   publisher = {Elsevier B.V.},
   title = {Cooperative control of mixed CAV-HDV platoons with hierarchical optimization for oscillation mitigation},
   volume = {675},
   year = {2025},
}

@article{Shi2023DRLDistributedControl,
   author = {Haotian Shi and Danjue Chen and Nan Zheng and Xin Wang and Yang Zhou and Bin Ran},
   issn = {0968090X},
   journal = {Transportation Research Part C: Emerging Technologies},
   month = {3},
   pages = {104019},
   publisher = {Elsevier Ltd},
   title = {A deep reinforcement learning based distributed control strategy for connected automated vehicles in mixed traffic platoon},
   volume = {148},
   year = {2023},
}

@article{Lyu2025ConflictRisk,
  title = {A Real-Time Conflict Risk Prediction Modeling Based on Causal Inference and Graph-Model: {{Applied}} to Multi-Configuration Expressway Diversion Zones},
  shorttitle = {A Real-Time Conflict Risk Prediction Modeling Based on Causal Inference and Graph-Model},
  author = {Lyu, Nengchao and Xie, Tian and Wen, Jiaqiang and Wang, Yugang},
  year = 2025,
  month = sep,
  journal = {Transportation Research Part C: Emerging Technologies},
  volume = {178},
  pages = {105219},
  issn = {0968090X},
  doi = {10.1016/j.trc.2025.105219},
  urldate = {2026-03-26},
  langid = {english}
}

@article{Song2026VSL,
  title = {Exploring Mechanisms of Integrating Global Perception Prediction for Connected Vehicles with Lane-Specific Reinforcement Learning-Based Variable Speed Limits},
  author = {Song, Li and Li, Shijie and Chen, Guojun and Zhao, Xin and Lyu, Nengchao and Fan, Wei (David)},
  year = 2026,
  month = mar,
  journal = {Expert Systems with Applications},
  volume = {299},
  pages = {129958},
  issn = {09574174},
  doi = {10.1016/j.eswa.2025.129958},
  urldate = {2026-03-26},
  langid = {english}
}

@article{Liu2026StringStability,
  title = {A {{Deep Reinforcement Learning Based Cooperative Control}} for {{Mixed Platoon}} with {{Guaranteed String Stability}}},
  author = {Liu, Guichuan and Li, Yongfu and Liu, Ling},
  year = 2026,
  month = apr,
  journal = {IEEE Transactions on Industrial Electronics},
  volume = {73},
  number = {4},
  pages = {6324--6334},
  issn = {0278-0046, 1557-9948},
  doi = {10.1109/TIE.2025.3637325},
  urldate = {2026-05-14},
  copyright = {https://ieeexplore.ieee.org/Xplorehelp/downloads/license-information/IEEE.html},
  langid = {english}
}

@article{Shi2025MixedVehiclePlatoonForming,
  title = {Mixed {{Vehicle Platoon Forming}}: {{A Multiagent Reinforcement Learning Approach}}},
  shorttitle = {Mixed {{Vehicle Platoon Forming}}},
  author = {Shi, Yujie and Dong, Haoxuan and He, Chaozhe R. and Chen, Yuxiao and Song, Ziyou},
  year = 2025,
  month = jun,
  journal = {IEEE Internet of Things Journal},
  volume = {12},
  number = {11},
  pages = {16886--16898},
  issn = {2327-4662, 2372-2541},
  doi = {10.1109/JIOT.2025.3535732},
  urldate = {2026-03-22},
  copyright = {https://ieeexplore.ieee.org/Xplorehelp/downloads/license-information/IEEE.html},
  langid = {english}
}

@article{Hua2025Multi-agentDRL,
   author = {Min Hua and Dong Chen and Kun Jiang and Fanggang Zhang and Jinhai Wang and Bo Wang and Quan Zhou and Hongming Xu},
   issn = {19399359},
   issue = {4},
   journal = {IEEE Transactions on Vehicular Technology},
   pages = {6076-6087},
   publisher = {Institute of Electrical and Electronics Engineers Inc.},
   title = {Communication-Efficient MARL for Platoon Stability and Energy-Efficiency Co-Optimization in Cooperative Adaptive Cruise Control of CAVs},
   volume = {74},
   year = {2025},
}

@article{Yang2025MixedPlatoonRL,
   author = {Zhiwei Yang and Zuduo Zheng and Jiwon Kim and Hesham Rakha},
   issn = {13619209},
   journal = {Transportation Research Part D: Transport and Environment},
   month = {5},
   pages = {104658},
   publisher = {Elsevier Ltd},
   title = {Eco-cooperative adaptive cruise control for platoons in mixed traffic using single-agent and multi-agent reinforcement learning},
   volume = {142},
   year = {2025},
}

@article{Wijnbergen2025NonlinearSystems,
   author = {Paul Wijnbergen and Mark Jeeninga and Redmer de Haan and Erjen Lefeber},
   issn = {00051098},
   journal = {Automatica},
   month = {7},
   pages = {112340},
   publisher = {Elsevier Ltd},
   title = {Longitudinal and lateral control of vehicle platoons: A unifying framework to prevent corner cutting},
   volume = {177},
   year = {2025},
}

@article{Han2025Model-freeAdaptiveControl,
   author = {Qiaoni Han and Jianguo Ma and Zhiqiang Zuo and Xiaocheng Wang and Xinping Guan},
   issn = {24751456},
   journal = {IEEE Control Systems Letters},
   pages = {33-37},
   publisher = {Institute of Electrical and Electronics Engineers Inc.},
   title = {Data-Driven Finite-Time Platooning Control for Heterogeneous Nonlinear Vehicle Systems},
   volume = {9},
   year = {2025},
}

@ARTICLE{Lin2024Multi-AgentRLPlatoon,
  author={Lin, Hongyi and Lyu, Cheng and He, Yixu and Liu, Yang and Gao, Kun and Qu, Xiaobo},
  journal={IEEE Transactions on Vehicular Technology}, 
  title={Enhancing State Representation in Multi-Agent Reinforcement Learning for Platoon-Following Models}, 
  year={2024},
  volume={73},
  number={8},
  pages={12110-12114},
}

@ARTICLE{Qiang2023DMPC,
  author={Qiang, Zhiwen and Dai, Li and Chen, Boli and Xia, Yuanqing},
  journal={IEEE Transactions on Intelligent Transportation Systems}, 
  title={Distributed Model Predictive Control for Heterogeneous Vehicle Platoon With Inter-Vehicular Spacing Constraints}, 
  year={2023},
  volume={24},
  number={3},
  pages={3339-3351},
}

@ARTICLE{Xie2019MixedTraffic,
  author={Xie, Dong-Fan and Zhao, Xiao-Mei and He, Zhengbing},
  journal={IEEE Transactions on Intelligent Transportation Systems}, 
  title={Heterogeneous Traffic Mixing Regular and Connected Vehicles: Modeling and Stabilization}, 
  year={2019},
  volume={20},
  number={6},
  pages={2060-2071},
}

\newpage

\appendix

\section{Definitions and Mappings of Time Headway Patterns}
\label{app:headway_patterns}

Applying the CAV demotion criterion and CHV demotion criterion established, these 32 car-following types can be consolidated into 6 distinct time headway patterns.
In a car-following pair written as 
$\mathrm{X-Y}$ (or $\Delta T_\mathrm{X,Y}$), 
$X$ denotes the leading vehicle and 
$Y$ denotes the following vehicle; the time headway pattern is primarily determined by the following vehicle’s capability and whether cooperation is available.

\subsection{Fundamental autonomous headway ($\Delta T_\mathrm{A}$)}

Applicable when the following vehicle is an AV or demoted CAV. The corresponding time headway is expressed as:
\begin{equation}
\begin{aligned}
\Delta T_\mathrm{A}
& = \Delta T_\mathrm{AV,CAV} = \Delta T_\mathrm{HV,CAV} = \Delta T_\mathrm{CAV,AV} \\
& = \Delta T_\mathrm{CAV_{pla},AV_{pla}} = \Delta T_\mathrm{CHV,AV} = \Delta T_\mathrm{CHV_{pla},AV_{pla}} \\
& = \Delta T_\mathrm{AV,AV} = \Delta T_\mathrm{AV_{pla},AV_{pla}} = \Delta T_\mathrm{HV,AV} \\
& = \Delta T_\mathrm{HV_{pla},AV_{pla}}
\end{aligned}
\label{eq:chap3_delta_T_A}
\end{equation}

\subsection{Manual driving headway ($\Delta T_\mathrm{H}$)}

Corresponds to cases where the following vehicle is an HV or demoted CHV. 
The corresponding time headway is expressed as:
\begin{equation}
\begin{aligned}
\Delta T_H
& = \Delta T_\mathrm{AV,CHV} = \Delta T_\mathrm{HV,CHV} = \Delta T_\mathrm{CAV,HV} \\
& = \Delta T_\mathrm{CAV_{pla},HV_{pla}} = \Delta T_\mathrm{CHV,HV} = \Delta T_\mathrm{CHV_{pla},HV_{pla}} \\
& = \Delta T_\mathrm{AV,HV} = \Delta T_\mathrm{AV_{pla},HV_{pla}} = \Delta T_\mathrm{HV,HV} \\
& = \Delta T_\mathrm{HV_{pla},HV_{pla}}
\end{aligned}
\label{eq:chap3_delta_T_H}
\end{equation}

\subsection{Connected autonomous headway outside platoons ($\Delta T_\mathrm{CA}$)}

Applicable when CHV/CAV engage in cooperative control without platoon coordination. %Specifically, this occurs when:
%Both leading and following vehicles possess communication capabilities; the following vehicle features autonomous driving functionality; and the two vehicles belong to different platoons.
This pattern comprises the following configurations: CAV-CAV and CHV-CAV.
The corresponding time headway is expressed as:
\begin{equation}
\Delta T_\mathrm{CA} = \Delta T_\mathrm{CAV,CAV} = \Delta T_\mathrm{CHV,CAV}
\label{eq:chap3_delta_T_CA}
\end{equation}

\subsection{Connected manual headway outside platoons ($\Delta T_\mathrm{CH}$)}

Applies to manual control scenarios involving CHVs/CAVs. 
%Specifically, this occurs when: both leading and following vehicles possess communication capabilities; the following vehicle lacks autonomous driving functionality; and the two vehicles belong to different platoons.
This pattern comprises the configurations as CAV-CHV and CHV-CHV.
The corresponding time headway is expressed as:
\begin{equation}
\Delta T_\mathrm{CH} = \Delta T_\mathrm{CAV,CHV} = \Delta T_\mathrm{CHV,CHV}
\label{eq:chap3_delta_T_CH}
\end{equation}

\subsection{Intra-platoon cooperative autonomous headway ($\Delta T_\mathrm{PA}$)}

Applies to autonomous cooperative driving scenarios within a platoon. 
%Specifically, this occurs when: both leading and following vehicles belong to the same platoon; and the following vehicle features autonomous driving functionality.
This pattern comprises the following configurations:
$\text{CAV}{\text{pla}}\text{-CAV}{\text{pla}}$, $\text{CHV}{\text{pla}}\text{-CAV}{\text{pla}}$, $\text{AV}{\text{pla}}\text{-CAV}{\text{pla}}$, and $\text{HV}{\text{pla}}\text{-CAV}{\text{pla}}$.
The corresponding time headway is expressed as:
\begin{equation}
\begin{aligned}
\Delta T_\mathrm{PA} & = \Delta T_\mathrm{CAV_{pla}, CAV_{pla}} = \Delta T_\mathrm{CHV_{pla}, CAV_{pla}} \\
& = \Delta T_\mathrm{AV_{pla}, CAV_{pla}} = \Delta T_\mathrm{HV_{pla}, CAV_\mathrm{pla}}
\end{aligned}
\label{eq:chap3_delta_T_PA}
\end{equation}

\subsection{Intra-platoon cooperative manual headway ($\Delta T_{PH}$)}

Applies to manual following scenarios within a platoon. %Specifically, this occurs when: both leading and following vehicles belong to the same platoon; the following vehicle lacks autonomous driving functionality.
This pattern comprises the following configurations:
$\text{CAV}{\text{pla}}\text{-CHV}{\text{pla}}$, $\text{CHV}{\text{pla}}\text{-CHV}{\text{pla}}$, $\text{AV}{\text{pla}}\text{-CHV}{\text{pla}}$, and $\text{HV}{\text{pla}}\text{-CHV}{\text{pla}}$.
The corresponding time headway is expressed as:
\begin{equation}
\begin{aligned}
\Delta T_\mathrm{PH} &= \Delta T_\mathrm{CAV_{pla}, CHV_{pla}} = \Delta T_\mathrm{CHV_{pla}, CHV_{pla}} \\
& = \Delta T_\mathrm{AV_{pla}, CHV_{pla}} = \Delta T_\mathrm{HV_{pla}, CHV_\mathrm{pla}}
\end{aligned}
\label{eq:chap3_delta_T_PH}
\end{equation}

\section{Car-Following Model Formulations and Parameters}
\label{app:cf_models}

To accurately characterize heterogeneous driving behaviors in mixed traffic flows, this study employs three car-following models for distinct vehicle types. 

\subsection{The intelligent driver model for human-driven vehicles}

The intelligent driver model (IDM) \citep{treiber2013traffic}, featuring minimal parameters with clear physical interpretations, is employed to simulate human-driven vehicle. It is formulated as:
\begin{equation}
\begin{aligned}
\dot{v}_{n}(t) = &
A\Bigg(1-\left(\frac{v_n(t)}{v_f}\right)^4 \\
& -\Bigg(\frac{s_{\min}+ T_{\mathrm{des}}\, v_n(t)-\dfrac{v_n(t)\,\Delta v_n(t)}{2\sqrt{AB}}}{s_n(t)}\Bigg)^2\Bigg)
\end{aligned}
\label{eq:chap2_idm}
\end{equation}
where $A$ and $B$ are the maximum acceleration and desired deceleration, respectively; $s_{\min}$ represents the jam gap, that is the minimum spacing between two adjacent vehicles;  $T_\mathrm{des}$ denotes the safe time headway; and $v_f$ is the free-flow velocity.

\subsection{The cooperative adaptive cruise control model for CAVs}

When CAVs operate within platoons, they move in cooperative adaptive cruise control (CACC) mode. This study employs the CACC model calibrated by PATH Laboratory through field tests \citep{milanes2013cooperative,milanes2014modeling}, mathematically expressed as follows:
\begin{equation}
v_n(t+\Delta t) =v_n(t)+k_p e_n(t)+k_d \dot{e_n}(t)
\label{eq:chap2_cacc_1}
\end{equation}
\begin{equation}
e_n(t) = s_n(t) - T_\mathrm{des} v_n(t)
\label{eq:chap2_cacc_2}
\end{equation}
where $\Delta t$ denotes length for each time step; $k_p$ and $k_d$ represent control gains; and $e_n(t)$ quantifies the deviation between actual and desired spacing for vehicle $n$ at time $t$.

To ensure consistency with the general car-following model form in Equation (\ref{eq:chap2_stable_analysis_general_cf2}), the CACC model is approximated through first-order Taylor expansion:
\begin{equation}
\dot{v}_{n}(t) = \frac{k_p (s_n(t) - T_\mathrm{des} v_n(t)) + k_d \Delta v_n(t)}{k_d T_\mathrm{des} + \Delta t}
\label{eq:chap2_cacc_3}
\end{equation}

\subsection{The adaptive cruise control model for AVs}

For AVs and degraded CAVs without connection, this study employs the adaptive cruise control (ACC) model developed by PATH Laboratory \citep{milanes2013cooperative,milanes2014modeling} to characterize their car-following behavior. The model is mathematical formulated as:
\begin{equation}
\dot{v}_{n}(t) = k_1 (s_n(t) - T_\mathrm{des} v_n(t)) + k_2 \Delta v_n(t)
\label{eq:chap2_acc}
\end{equation}
where $k_1$ denotes the spacing error control gain, and $k_2$ represents the velocity difference control gain.

Based on the three car-following models, specific behaviors are defined for each scenario in Appendix~\ref{app:headway_patterns}:
Car-following patterns (1)-(2) occur in non-communication environments where following behavior depends solely on the follower type: IDM for human-driven vehicles (HVs), ACC for AVs.
Car-following patterns (3)-(6) feature V2V communication enabling vehicle state sharing: CACC for autonomous-capable followers, IDM for human-driven followers.
Time headway parameters and corresponding model coefficients are listed in Table \ref{tab:cf_params} \citep{yao2023analysis,qin2023stability,jiang2023platoon,yao2024impact}.

\begin{table}[H]
    \caption{Parameters of car-following models.}
    \label{tab:cf_params}
    \centering
    \begin{tabularx}{\columnwidth}{p{0.6\columnwidth}>{\centering\arraybackslash}p{0.125\columnwidth}>{\centering\arraybackslash}p{0.125\columnwidth}}
        \toprule
        \textbf{Parameter} & \textbf{Value} & \textbf{Unit} \\
        \midrule
        Free-flow velocity, $v_f$ & 33.3 & m/s \\
        Minimum spacing, $s_{\min}$ & 2.0 & m \\
        Maximum acceleration, $A$ & 1.0 & m/s$^2$ \\
        Desired deceleration, $B$ & 2.0 & m/s$^2$ \\
        Gain, $k_p$ & 0.45 & s$^{-1}$ \\
        Gain, $k_d$ & 0.25 &  -\\
        Simulation time step, $\Delta t$ & 0.01 & s \\
        Spacing error control gain, $k_1$ & 0.23 & s$^{-1}$ \\
        Velocity difference control gain, $k_2$ & 0.07 & s$^{-1}$ \\
        $\Delta T_\mathrm{H}$ & 1.8 & s \\
        $\Delta T_\mathrm{CH}$ & 1.2 & s \\
        $\Delta T_\mathrm{PH}$ & 0.8 & s \\
        $\Delta T_\mathrm{CA}$ & 1.0 & s \\
        $\Delta T_\mathrm{A}$ & 1.4 & s \\
        $\Delta T_\mathrm{PA}$ & 0.6 & s \\
        \bottomrule
    \end{tabularx}
\end{table}

\end{spacing}
\end{document}